\documentclass[11pt]{cmuthesis}

\usepackage{placeins}
\usepackage{afterpage}

\usepackage{multirow}
\usepackage[ruled]{algorithm2e} 

\usepackage{amsmath,amssymb}  
\usepackage{amsfonts}

\usepackage{array}
\usepackage{mathtools}

\usepackage{xspace}

\newcommand{\etal}{et al.~}
\newcommand{\ie}{\textit{i}.\textit{e}.,}
\newcommand{\eg}{\textit{e}.\textit{g}.}

\newcommand*\Laplace{\mathop{}\!\mathbin\bigtriangleup}

\newcommand{\mycolor}[1]{\textcolor{black}{#1}}

\begin{document}
\frontmatter
\pagestyle{plain}

\title{Computational Design and Evaluation Methods for Empowering Non-Experts in Digital Fabrication}

\author{Nurcan Gecer Ulu}

\bsdegree{Mechanical Engineering, Middle East Technical University}
\msdegree{Mechanical Engineering, Bilkent University}

\Month{May}

\Year{2018}

\permission{\textit{All Rights Reserved}}

\maketitle

\begin{abstract}

Despite the increasing availability of personal fabrication hardware and services, the true potential of digital fabrication remains unrealized due to lack of computational techniques that can support 3D shape design by non-experts. This work develops computational methods that address two key aspects of content creation:(1) Function-driven design synthesis, (2) Design assessment. 

For design synthesis, a generative shape modeling algorithm that facilitates automatic geometry synthesis and user-driven modification for non-experts is introduced. A critical observation that arises from this study is that the most geometrical specifications are dictated by functional requirements.  To support design by high-level functional prescriptions, a physics based shape optimization method for compliant coupling behavior design has been developed. In line with this idea, producing complex 3D surfaces from flat 2D sheets by exploiting the concept of buckling beams has also been explored. Effective design assessment, the second key aspect, becomes critical for problems in which computational solutions do not exist. For these problems, this work proposes crowdsourcing as a way to empower non-experts in esoteric design domains that traditionally require expertise and specialized knowledge.

\end{abstract}


\begin{acknowledgments}

I would like to thank my advisor Prof. Levent Burak Kara for giving me this opportunity. Your guidance and support during my PhD has been invaluable.
\hfill \break

Special thanks go to the committee Prof. Jessica Zhang, Prof. Kenji Shimada and Prof. Stelian Coros. Prof. Zhang, your input has been very helpful in shaping up this thesis. Prof. Shimada, you have been a great example for me to look up to in your classes as well as an expert in the field. Stelian, you have been an integral part of my PhD. I owe a great deal of my success to your involvement in my research. 
\hfill \break

Thank you labmates and collaborators, this thesis would not come together without you.
\hfill \break

I am eternally grateful to my family for their endless support and love. My dear friends, you made my time here enjoyable.
\hfill \break

Very few people get to work with their spouses and even fewer like the idea of it. Erva, I consider myself the luckiest person in the world for sharing the whole experience with you and having your support along the way.
\hfill \break

The work was supported by NSF CMMI1235427, America Makes Project \#4058, Siemens Corporate Research, and Carnegie Mellon Manufacturing Futures Initiative.

\end{acknowledgments}
\tableofcontents
\listoffigures
\listoftables
\mainmatter


\chapter{Introduction}
\label{chap:intro}

\section{Motivation}

Digital fabrication is expected to bring a new revolution~\cite{gerald2012fa}. Recent advances in 3D printing and scanning technologies are the main drivers of these expectations. Despite the increasing availability of fabrication equipment and services, the dissemination of digital fabrication to large populations is yet to remain a challenge.

Achieving the real revolution of digital fabrication is about developing the fundamental concepts, such as design process and design assessment, as much as developing the fabrication techniques. For a successful realization of digital fabrication, the effective  engagement of large populations is of utmost importance. The key objective is to foster creativity of a more general public rather than limited number of experts. To date, we have digitally native generations, widely accessible fabrication techniques and depth sensors incorporated in our phones that support a general access. Surrounded by these developments, here, we see an opportunity to bridge the gap between these advances and content creation through novel computational approaches.

Current computer-aided design technology is largely tailored towards expert users and require extensive knowledge and training. For this reason, the current content creation does not cater to needs of non-expert users. If we want to expand access to these cheap and effective tools, we need to make them easier to use for non-experts. Additionally, as the technology develops, the complexity is increasing for non-experts and experts alike; is not meeting the needs of the experts themselves.

\begin{figure}
  \centering  
  \includegraphics[trim = 0in 0in 0in 0in, clip, width=\textwidth]{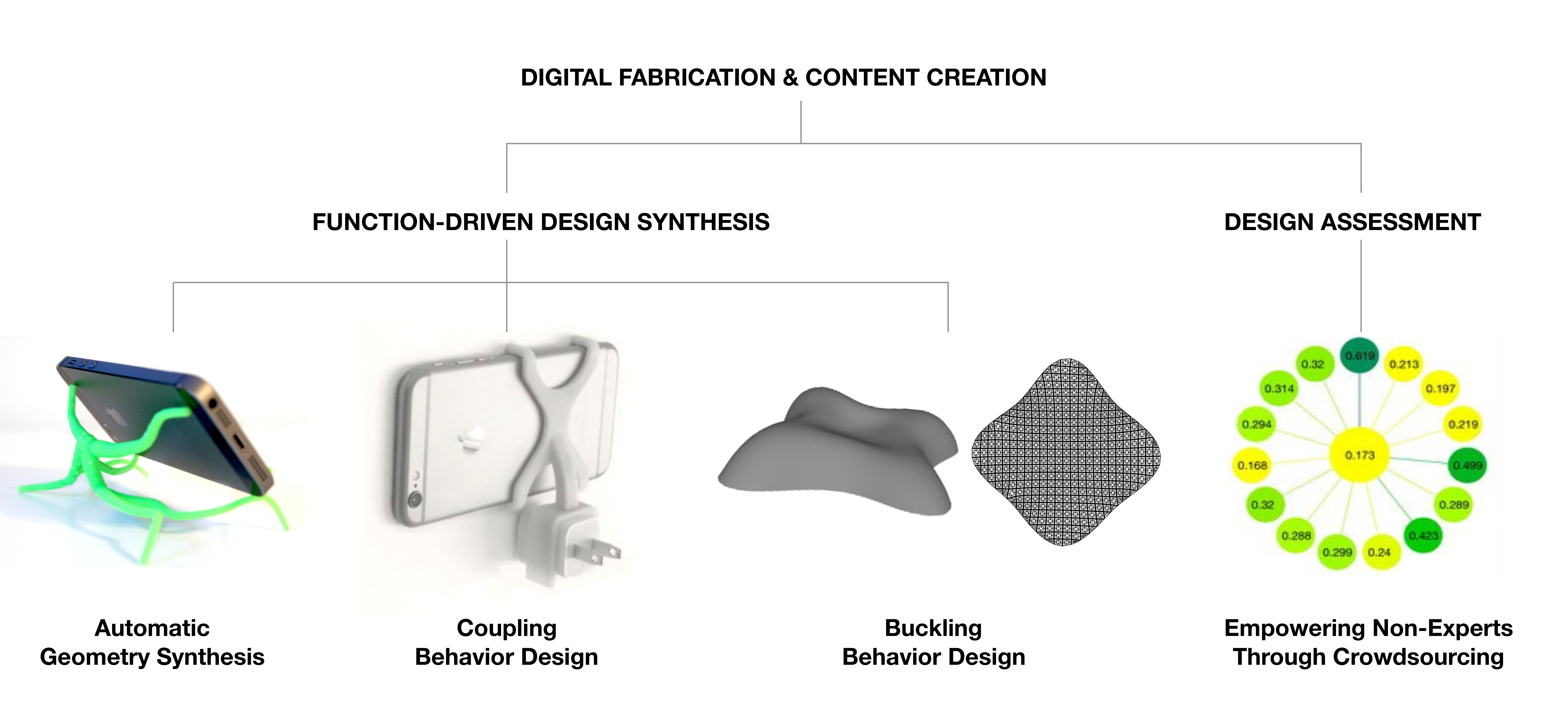}
  \caption{We investigate content creation for digital fabrication through four computational techniques from the two key aspects.}
  \label{fig:intro:digitalfab}
\end{figure}

\mycolor{The overarching goal of this thesis is to explore and develop computational tools for function-driven design by simple geometry specification and directly prescribing high level functions. Both synthesis and assessment aspects of function-driven design are investigated. As depicted in Figure~\ref{fig:intro:digitalfab}, this thesis embodies four computational methods that explore these two key aspects of content creation to enhance the overall digital fabrication experience and catalyze its widespread dissemination.}

\section{Research Questions and Scientific Challenges}

In general, widespread dissemination of digital fabrication is hindered hindered by the lack of computational tools that can support the engagement of the general population in content creation. Under the overall aim of enhancing computational fabrication experience, the following research questions and challenges are addressed in this thesis.

\begin{enumerate}
\item Can we make geometry modeling easier? Effective use of current 3D modeling and design software is still grueling for most novice users. Discovering core design intentions and developing practical design tools yet to remain a challenge. Another challenge in 3D modeling originate from the increasing capabilities of the machines to produce complex geometries. Current modeling software requires tedious design process as the geometries become more detailed. (Chapter \ref{chap:growth}) 

\item Geometry design is usually hard and indirect, can we directly prescribe function instead of geometry? As we seek to step into not only understanding but also to enabling design by functional specifications, we choose one particular example, mechanical coupling behavior. Here, we specifically address the challenges in extending appropriate coupling behavior to arbitrary object pairs as opposed to limited prevalent uses (\eg snap fits). (Chapter \ref{chap:coupling})

\item Can we fabricate complex 3D surfaces in simpler and more convenient ways? 3D printing thin surface objects may turn into a very costly process due to support material requirements and production time. We present 2D laser cut patterns that deploy into 3D surfaces after the addition of rigid inserts. Specific challenge addressed here stem from understanding the highly non-linear underlying mechanism in the deployment of flat patterns into 3D surfaces. (Chapter \ref{chap:inserts})

\item How can we empower non-experts in problem solving? More sorecifically, can we employ the power of crowds in digital fabrication process? While crowdsourcing has been proven to be effective in many applications, esoteric design domain has not been taking advantage of these advances. An open challenge here is determining under what conditions crowdsourcing can be utilized for esoteric problems. (Chapter \ref{chap:crowd})

\end{enumerate}

\section{Methodology}

This thesis presents methods for facilitating prevalent use of digital fabrication. The presented methods span a variety of aspects in content creation that are open to significant advancement in order to catch up with the coveted rise of fabrication techniques. In Chapter \ref{chap:relatedwork}, we go over the works that are foundational to this thesis. We describe how we build on them and extend their methodologies.

In Chapter \ref{chap:growth}, we introduce a shape modeling system to increase engagement of novice users in design. To make the modeling process easier and more convenient for novice users, we developed a generative design system where the shape is automatically created by only specifying a few key points. To better facilitate expression of design intentions, we utilize modifier sketches on the automatically generated content. The main idea of the presented generative modeling system is to enable the design flexibility to generate structures that could serve different functions as opposed to parameterized designs of specific objects with certain functions.

A critical observation that arises from this study is that the most geometrical specifications are dictated by functional requirements. This observation lends itself naturally to the immediate insight of prescribing function instead of geometry. In line with this idea, Chapter \ref{chap:coupling} presents a computational method for designing mechanical coupling behavior between a rigid object and a compliant enclosure based on high-level functional specifications. In particular, we introduce a method that maps the geometrical parameters of the compliant object onto sequentially observed coupling behaviors, such as the grip, insertion and removal forces that develop as the rigid object is engaged.

Although digital fabrication has enabled us to make almost anything, there are still many factors that need to be considered such as material and time cost, product quality and even manufacturability. Moreover, depending on the application, some fabrication techniques become more desirable. For example, laser cutting is faster than 3D printing but 3D printing can generate more complex objects. In Chapter \ref{chap:inserts}, we present a shape parameterization method to map 3D complex surfaces into 2D flat patterns such that the 3D surfaces can be realized by installing rigid inserts. In this work, we exploit the concept of buckling beams to achieve 3D surfaces from flat layouts. This approach will pave the way for simpler, faster manufacturing and convenient flat shipping of 3D surface objects.

In Chapter \ref{chap:crowd},  we explore the use of crowdsourced populations in esoteric engineering design domain. Understanding the underlying mechanisms of crowdsourcing in esoteric problems, we aim to solve challenging design problems, that can only be solved by experts, via crowdsourced populations. In this work, we gather empirical data by conducting a crowdsourcing experiments with varying difficulty and intuitiveness levels on crowds with ranging expertise. We analyzed this empirical data using a computational Bayesian framework. We observe that crowd estimations exceed the accuracy of individuals in the vast majority of instances indicating a wisdom of crowd effect. Our observations also show that the wisdom of crowd effect is maintained for micro-crowds of practitioners, 4-15 individuals, for less intuitive problems suggesting that the challenging engineering design problems can vastly benefit from crowdsourcing.

\section{Contributions}

In line with the aforementioned motivations and challenges,  main contributions of thesis are:

\begin{itemize}

\item A generative shape modeling framework that facilitates easy geometry specification and modification for novice users.
\item	The use of deformation profiles to describe and optimize mechanical behavior.
\item	A physics based shape optimization method for compliant coupling behavior design involving two-part interactions.
\item A practical insertion simulation based on collision elimination for computing deformation profiles.
\item Showing that wisdom of crowds effect is valid for esoteric engineering problems where crowd estimations exceed the accuracy of individuals in the vast majority of instances
\item Finding the wisdom of crowd effect is maintained for micro-crowds of practitioners, 4-15 individuals, for less intuitive problems.

\end{itemize}

\section{Publications}

\mycolor{This thesis work has lead to a number of publications:}

\begin{itemize}

\item \mycolor{ \textbf{Nurcan Gecer Ulu}, Stelian Coros and Levent Burak Kara, 
"Designing Coupling Behaviors Using Compliant Shape Optimization",
 \textit{Computer-Aided Design}, 2018. \cite{ulu2018coupling} }

\item \mycolor{ \textbf{Nurcan Gecer Ulu}, Michael Messersmith, Kosa Goucher-Lambert, Jonathan Cagan and Levent Burak Kara, 
"Wisdom of Micro-Crowds in Evaluating Solutions to Esoteric Engineering Problems",
 \textit{Journal of Mechanical Design}, 2018. (Under review) \cite{ulu2018crowdsourcing} }
 
\item \mycolor{ \textbf{Nurcan Gecer Ulu} and Levent Burak Kara, 
"Generative Interface Structure Design for Supporting Existing Objects",
 \textit{Journal of Visual Languages and Computing}, 2015. \cite{ulu2015jvlc} }
 
\item \mycolor{ \textbf{Nurcan Gecer Ulu} and Levent Burak Kara, 
"Generative Interface Structure Design for Supporting Existing Objects",
 \textit{International Conference on Distributed Multimedia Systems Workshop on Visual Languages and Computing (VLC)}, 2015. \cite{ulu2015vlc} }
 
\item \mycolor{ Guanyun Wang, Humohrey Yang, Zeyu Yan, \textbf{Nurcan Gecer Ulu}, Ye Tao,Jianzhe Yu, Levent Burak Kara and Lining Yao, "4DMesh Inverse Design Tools for Self-Assembling Non-Developable Mesh Surfaces via 4D Printing",
 \textit{UIST}, 2018. (Under review) \cite{wang20184DMesh} }
 
\item \mycolor{Wentai Zhang, Jonelle Z. Yu, Fangcheng Zhu, Yifang Zhu,, \textbf{Nurcan Gecer Ulu}, Erhan Batuhan Arisoy and Levent Burak Kara,
"High Degree of Freedom Hand Pose Tracking Using Limited Strain Sensing and Optimal Training",
 \textit{ASME International Design Engineering Technical Conferences and Computers and Information in Engineering Conference}, 2018. \cite{zhang2018high} }
 
 \item \mycolor{ Erhan Batuhan Arisoy, Guannan Ren, Erva Ulu, \textbf{Nurcan Gecer Ulu} and Suraj Musuvathy,
"A Data-Driven Approach to Predict Hand Positions for Two-Hand Grasps of Industrial Objects",
 \textit{ASME International Design Engineering Technical Conferences and Computers and Information in Engineering Conference}, 2016. \cite{arisoy2016data} }
 
\item \mycolor{ Suraj Musuvathy, George Allen, Lucia Mirabella, Louis Komzsik and \textbf{Nurcan Gecer Ulu}, 
"System and Method for Modeling of Parts with Lattice Structures",
\textit{Patent Publication Number:WO2017088134 A1}, 2017. \cite{musuvathy2017methods}  }
 
\item \mycolor{ Erhan Batuhan Arisoy, Suraj Musuvathy, Erva Ulu and \textbf{Nurcan Gecer Ulu}, 
"Methods and System to Predict Hand Positions for Multi-hand Grasps of Industrial Objects",
\textit{Patent Publication Number:WO2017132134 A1}, 2017. \cite{arisoy2017methods} }

\end{itemize}

\chapter{Background and Related work}
\label{chap:relatedwork}

This thesis is related to digital fabrication which is making things using an all digital process. More specifically, digital fabrication is studied through geometry design, functional behavior design and crowdsourcing esoteric problems. These approaches are discussed in detail in this chapter.

\section{Generative Interface Structure Design}

One objective of this thesis is investigating how geometry specification can be made easier for non-experts. Below we cover the background on non-expert design tools, generative design and bio-inspired design algorithms.

\subsection{Design Tools for Non-Expert Users} 

Recently, design tools for non-expert users have received significant attention. In \cite{sketCHair_11}, a chair design tool is proposed to create balanced chairs from extruded 2D profile sketches. To enable informed exploration, Umetani et al.\cite{umetani_sigg12} proposed a suggestive design tool for plank-based furniture. In that work, the user adds planks and edits their positions, orientations or size. A data-driven approach to interactive design of model airplanes is proposed in \cite{Umetani:Pter:2014} where the user creates free-flight gliders with 2D sketches. In this work, we focus on creating a large range of products instead of one specific group such as chairs or gliders. While all three systems are notable interactive tools, components of the resulting designs are limited to 2D laser cut pieces. Our system generates organic 3D geometries that can take advantage of the opportunities in 3D printing. 

Much of recent research on design for 3D printing addresses modifications of existing digital models by optimizing physical properties, such as balance and structural strength. For this purpose, inner carving with deformations \cite{Prevost:2013, Bacher:2014} and thickening of thin sections \cite{Stava_2012} have been used. Here, we focus on geometric shape \emph{generation} whereas their focus is on shape modification.

\subsection{Generative Design} 

Generative design methods are recognized as significant technologies to rapidly generate different design alternatives. Fabrication of generatively produced designs have been examined illustrating how geometrically complex shapes can be physically created in \cite{Wang2002, Sass}. In computer graphics, generative design methods have been used to create architectural models such as buildings \cite{Muller_2006}, virtual cities \cite{Parish_2001} and trees \cite{Prusinkiewicz_sigg94,Runions_sigg2005}. In this work, we are inspired by a specific generative design method developed to simulate the tree growth process for automatic shape generation.

\subsection{Topology Optimization as Generative Design} 

\mycolor{ In structural design, an ubiquitous generative design approach is topology optimization. One major drawback of the topology optimization is its high computational cost. In the current state, 3D topology optimization with even a fairly small grid resolution may take hours to days depending on the complexity of the chosen finite element types, \ie~ linear or nonliner material models, linear or higher order shape functions~\cite{Bendsoe:2004}. In contrast, our aim in this work is to develop a practical, easy to use geometry modeling tool for non-experts. Therefore, an interactive framework with small computational cost is desirable for our purposes. In addition, our work puts an emphasis on user's design intentions with sketch modifications. This presents another challenge in a topology optimization based method since maintaining structural optimality under user modifications such as sketch strokes would require an additional optimization step. In addition, we believe one-of-a-kind property of the designs that are produced with our framework is an important characteristic for casual designers and mass customization that are not prevalent in conventional topology optimization approaches.}

\subsection{Bio-inspired Growth Algorithms} 

Tree Growth methods have been widely used in computer graphics for urban modeling and computer animation. As such, tree growth has been vastly investigated in the literature. L-systems have been used to generate trees in \cite{Prusinkiewicz_sigg94}. Runions et al.~\cite{Runions_sigg2005} proposed a space colonization algorithm to mimic open and closed venation process in the leaf formation. Later, the space colonization algorithm has been extended to grow trees in \cite{Runions_07} and \cite{Palubicki_sigg09}. Tree generation inside various geometries is investigated using spatial attractor distribution in \cite{Longay2012}. In this work, our automatic interface generation process has been inspired by the space colonization algorithm. We use its spatial attractor distribution feature to enable the interaction with the design space. Moreover, interaction of the tree model with the obstacles in the environment has been studied in \cite{Pirk_sigg12}. In that, a fully grown tree model is placed around an object and the colliding branches are removed. For our purposes, this approach cannot be used since the grown structure has to be connected to the target points and a branch connected to a target can not be removed. Hence, we utilize obstacle avoidance during the growth process.

\section{ Designing Coupling Behaviors}
Fundamentally, our approach involves mechanical behavior control through shape optimization. Below, we review the works that are foundational to our work.

\subsection{Deformation Behavior Control} 

Deformation control through shape and structure optimization has been addressed in various ways including (1) Material distribution optimization ~\cite{Bickel_sigg2010,Panetta_sigg2015,Schumacher_sigg2015}, (2) Multi-material distributions ~\cite{Skouras_sigg2013}, (3)  Wireframe thickness optimization ~\cite{Perez_sigg2015}, and (4) Nonlinear material design through prescribed stress-strain curves ~\cite{Xu:2015:nonlinearMaterial}. Xu et~al.~\cite{Xu:2015:materialmodes} introduce model reduction to design heterogeneous deformable materials to achieve prescribed displacements and forces. Chen et~al.~\cite{Chen:2014} explore rest shape optimization to account for the deformations due to prescribed forces to obtain desired deformed states. 

We extend these works to scenarios involving a coupling process with part contacts between a compliant and a rigid object rather than relying on forces known a priori. We formulate a broader problem where the compliant object acquires  its final (steady) state through a progression of contacts where neither the location of the contact, nor the resulting contact forces can be known in advance. Additionally, it is not possible to prescribe the final deformed configuration in advance, as the contact forces deforming the object cannot be known explicitly a priori. Finally, each new hypothesis for the compliant object during its design likely produces new contact configurations. This necessitates shape design and contact analysis to be performed conjointly.

\subsection{Computational Design for Fabrication} 

There exists a large body of work for structure design to enable prescribed functional objectives such as kinematic goals ~\cite{Prevost:2013,Bacher:2014,Musialski:2015}, strength improvements ~\cite{Stava:2012,Wang:2013,Lu:2014,ulu2017lightweight}, or other physical qualities of interest ~\cite{Bharaj:2015,Umetani:Pter:2014,Umetani:duduk:2016}. Closely related to our work ~\cite{Koyama:2015} create automatic connectors between object pairs involving parameterized primitive geometries such as cylinders and rectangular prisms using a data-driven approach informed by a battery of physical experiments, or use partitioned rigid connectors to accommodate free-form objects.  Our work extends their work by formulating compliant mechanical coupling design as a conjoint shape optimization and physics-based contact simulation. This allows our method to transform arbitrary free-form objects into pairs that can be made attachable to one another.

\subsection{Compliant Mechanisms}

Compliant mechanisms exploit flexible and continuous joint structures~\cite{Bendsoe:2004}. Typically, compliant mechanisms are structurally optimized for input/output displacement or force transfer ratios~\cite{Lu:2003,Lu:2005},  for matching the displacement path of a compliant mechanism for an input actuation~\cite{pedersen2001}, or for enabling gripping behavior through known input force points~\cite{gripper2003topology,gripperfreeform,gripper98,Bruns2004,Ohsaki2005}. Our compliant structures are not externally activated through prescribed contact points. Instead, deformations are generated through part interactions that are unknown a priori. 

Bruns et al.~\cite{Bruns2004} present a designer guided topology optimization method for generating a snap-fit mechanism to mount onto walled openings. However, contacts are deterministic as they follow imposed boundaries such as continuous sliding across a line. Lawry et al.~\cite{lawry2017level} present a topology optimization method that produces  a  snap fit pair starting from objects with perfectly matching boundaries (\ie one object is a complement of the other). The aim is to optimize harmonic separation forces without considering grip. Optimization of connectors for simple pin geometries has been shown in~\cite{yunConnector,hsu2000shape}. Our approach builds on these ideas to make arbitrary geometries attachable to one another rather than fine-tuning existing snap fit configurations. With our formulation, engagement and disengagement forces as well as grip tightness can be designed in a decoupled way, thereby enabling the creation of couplings that require weak engagement forces but result in tight grips. Moreover, our work extends the above works in 2D to 3D.

\subsection{Contact Simulation} 

Our approach seeks to optimize the compliant object so as to produce the desired deformation behavior in the form of deformation profiles. The deformation, however, depends exclusively on the interaction between the current shape of the compliant object and the rigid object, thus necessitates a heavy use of contact simulations throughout the shape optimization process. Kloosterman~\cite{Kloosterman:2002:contact} provides a detailed review of the large body of research in contact simulations. Voxmap Point Shell~\cite{McNeely:1999:boeing}  models the environment as a map of voxels for penetration calculations and computes virtual penalty forces to eliminate penetrations. This method works for rigid object contacts but it is also extended to deformable objects in~\cite{Barbic:2008}. Kaufman et~al.~\cite{Kaufman:2008} presents a method that can model the frictional contact between deformable objects. Complex contact scenarios in dynamic simulations are studied in~\cite{Harmon:2009}. Continuous penalty force approach is presented in~\cite{Tang:2012}. Based primarily on these works,  we formulate our insertion simulation as a friction-free penetration elimination problem using distance fields.

\section{Wisdom of Micro-Crowds in Esoteric Design Problems}

Amazon Mechanical Turk (AMT)~\cite{AMT}, CrowdFlower (CF)~\cite{crowdflower} and Prolific Academic (ProA)~\cite{prolificAcademic} are among the most prominent crowdsourcing platforms.  These platforms have a wide reach and are designed to be representative of the general public consisting of diverse crowds~\cite{berinsky2012evaluating}.  While AMT allows the surveys to be targeted toward specific demographics, it is difficult to identify crowds that share a prescribed technical background. By contrast, our work focuses on solving esoteric problems via micro-crowds that consist of practitioners.  

Previous studies have developed task design and response quality detection methods as a way to maximize the useful information  content in crowdsourcing~\cite{kittur2008crowdsourcing,kittur2013future}. Example methods include the use of explicitly verifiable questions to identify malicious users and to encourage honest responses, and task fingerprinting to monitor completion time, mouse movements, key presses, and scroll movements, which can all be used as indicator attributes for detecting suspect responses~\cite{rzeszotarski2011instrumenting}. The presented work uses response speed as one such indicator to vet data quality.

Consensus through collaboration is a widely used approach in engineering~\cite{summers2010mechanical,takai2010game}. However, driven by the previous observations that there is a danger of expert collaboration to result in a singular thought pattern that could be outperformed by diverse groups~\cite{hong2004groups}, this work explores WoMC with individuals who remain independent and  form crowds that are more diverse than collaborating experts. Surowiecki~\cite{surowiecki2005wisdom} argues that one requirement for a good crowd judgement is that people's decisions remain independent of one another. This was further validated by Lorenz \etal~\cite{lorenz2011social} where individuals were observed to produce collectively more accurate crowd estimations over cases where the same individuals were informed by others' estimates. 

Burnap \etal~\cite{burnap2015crowdsourcing,burnap_identify} explored the use of crowdsourcing in engineering design assessment as well as techniques for identifying the experts in a crowd. These studies do not assume an apriori knowledge of the individuals' background and are thus greatly suited for  studies involving large crowds. Our work builds on and complements these studies by focusing on a small group of  practitioners, none of whom may be an expert but whose technical familiarity with the problem domain is significantly higher and more homogeneous compared to crowds extracted from the general population.

Crowdsourcing has also been used in  design for identifying customer preferences to balance style with brand recognition~\cite{burnap2015balancing} or to study the relationship between product geometry and consumer judgment of style~\cite{orbay2015deciphering}. While these works primarily focus on eliciting subjective judgments of preference and perception, the main focus of the presented work is to crowdsource solutions to engineering problems where an objectively true solution must exist (albeit unknown).

Another popular use of crowdsourcing involves the discovery of diverse solutions to complex technical problems involving very high-dimensional design spaces, such as the GE bracket design challenge~\cite{morgan2014ge}. While the generation of  solutions is typically the core challenge (hence crowdsourced), candidate solutions can be rather easily assessed using computational analysis tools. However, the main hypothesis and the utility of our work is that further crowdsourcing to \emph{assess} candidate designs may in fact produce successful outcomes, which would be critically important in cases where no appropriate computational evaluation technologies exist. 

Another open problem within the engineering design research community where crowdsourcing could provide value relates to the consistent evaluation of conceptual designs. In contrast to engineering problems with known solutions (\ie~structural mechanics), conceptual design problems have no \textit{true} solution. When studying the conceptual design process, researchers often utilize cognitive studies to explore specific process characteristics, such as the impact of analogical stimuli on solution output~\cite{fu2013meaning,murphy2014function,moreno2014fundamental}. Typically, design output from such studies is evaluated qualitatively; trained experts rate defined metrics, such as the novelty or quality, across a wide design space~\cite{shah2000evaluation}. Unsurprisingly, the process of both training and rating design solutions can be incredibly time consuming and costly. This is particularly true for cognitive studies requiring hundreds of design concepts to be evaluated at a given time~\cite{goucher2017using}. Another challenge with the current approach to evaluating conceptual design solutions is that when multiple experts are used, they do not always agree upon the particular merits of a given design concept. This can lead to low inter-rater reliability metrics, and require researchers to retrain experts prior to having them re-evaluate designs. With this in mind, a combined human-computational framework that removes the necessity of training experts could greatly improve and expedite the conceptual design evaluation process. In this work, we also explore WoMC for evaluation of conceptual designs.

\chapter[Generative Interface Structure Design]{Generative Interface Structure Design for Supporting Existing Objects}
\label{chap:growth}
\blindfootnote{This chapter is based on Ulu and Kara, 2015 \cite{ulu2015jvlc,ulu2015vlc}.}

Increasing availability of high quality 3D printing devices and services now enable ordinary people to create, edit and repair products for their custom needs. However, an effective use of current 3D modeling and design software is still a challenge for most novice users. In this work, we introduce a new computational method to automatically generate an organic interface structure that allows existing objects to be statically supported within a prescribed physical environment. Taking the digital model of the environment and a set of points that the generated structure should touch as an input, our biologically inspired growth algorithm automatically produces a support structure that when physically fabricated helps keep the target object in the desired position and orientation. The proposed growth algorithm uses an attractor based form generation process based on the space colonization algorithm and introduces a novel target attractor concept. Moreover, obstacle avoidance, symmetrical growth, smoothing and sketch modification techniques have been developed to adapt the nature inspired growth algorithm into a design tool that is interactive with the design space. We present the details of our technique and illustrate its use on a collection of examples from different categories.

\section{Introduction} 
\label{sec:growth:intro}

The customization and personalization of products started to compete with traditional mass production principles with the contribution of maker movement and DIY (Do-It-Yourself) culture. DIY commonly refers to any fabrication, modification or repair event that is outside of one's professional expertise \cite{Mota_2011}. With the rise of DIY culture, there is a growing interest for design and fabrication tools tailored towards non-expert users.

Recent advances in 3D design and manufacturing technologies now have made content creation accessible to novice users. Besides the basic consumer level 3D printers, online on-demand 3D printing services (e.g. Shapeways, i.Materialise) have enabled ordinary people to access high quality machines. 3D modeling software, such as Autodesk 123D and Tinkercad, allow consumers to create 3D shapes using simplified geometric interaction methods. However, current commercial design software do not take advantage of capabilities of 3D printing. While \textit{almost anything} can be fabricated using 3D printing, these design software limit potential design outputs by mimicking features of traditional manufacturing and assembly methods. In this work, we extend the design possibilities by taking a generative design approach to create organic looking branching shapes that would be challenging to design and fabricate with traditional methods.

\begin{figure}[t]
  \centering  
  \includegraphics[trim = 0in 4.2in 5in 0in, clip, width=3.5in]{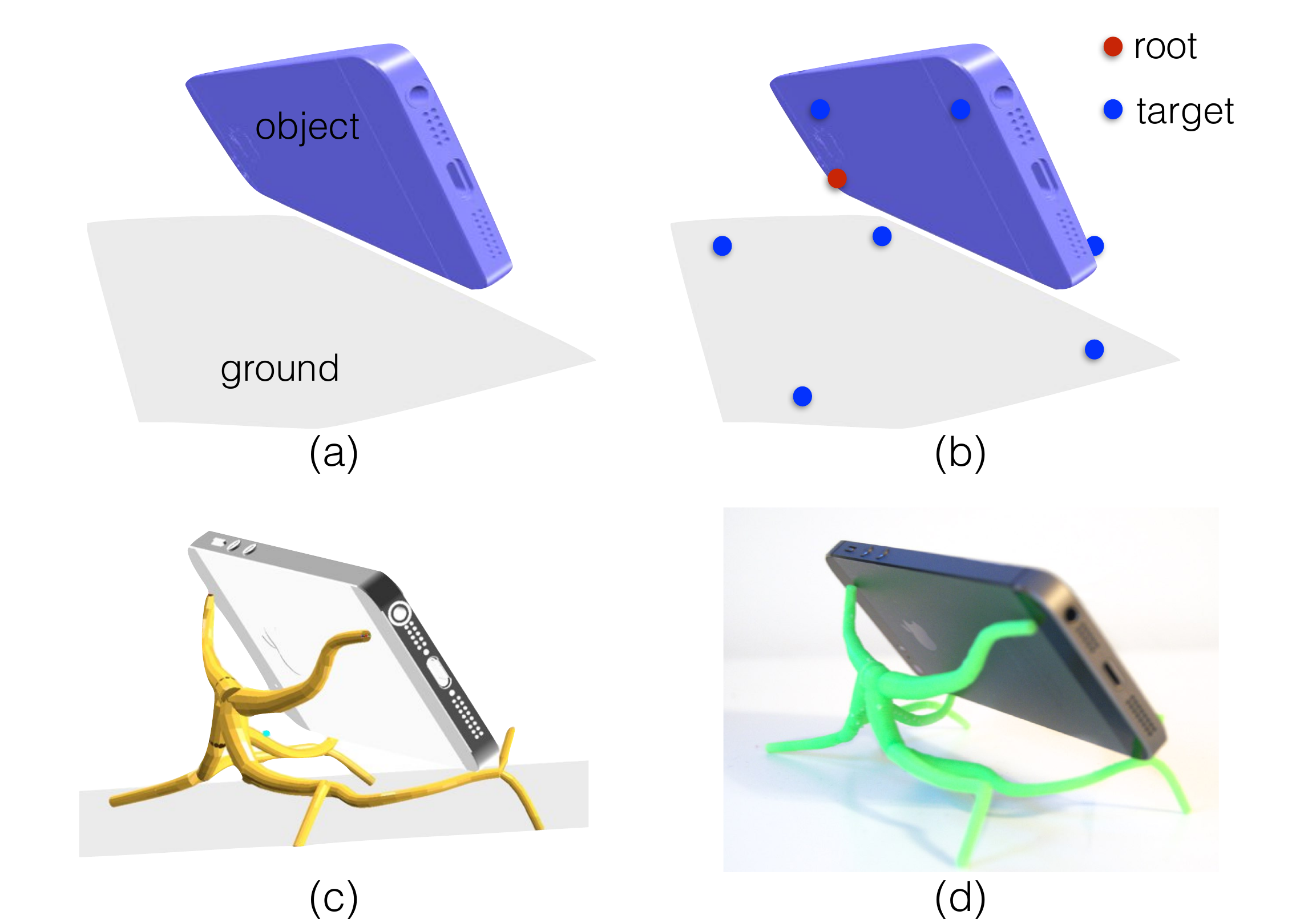}
  \caption{Example problem to generate a phone stand (a) given object and environment configuration (b) user defined target and root points (c) generated interface structure (d) 3D printed result.}
  \label{growth:teaser}
\end{figure}

We propose a framework that automatically generates interface structures under prescribed constraints. The input to our algorithm is a surface mesh for the object to support and a mesh to represent the ground surface with target and root points to create a shape in between (Fig.\ref{growth:teaser}). Then, automatic interface structure generation is achieved by a nature inspired growth mechanism. Users can control the design by changing target-root combinations at the input phase as well as by using sketch modifications after the shape is created. Moreover, the stochastic nature of the growth algorithm lets users design one of a kind pieces by generating different outputs for the same problem on each run. The main contribution of this work is the novel application of a nature inspired growth algorithm for automatic product generation. This is accomplished by the introduction of target attractor and pruning concepts, embedding product design considerations and user interaction.

\begin{figure*}[t]
  \centering
  \includegraphics[trim = 0in 3.0in 0in 3.5in, clip, width=\textwidth]{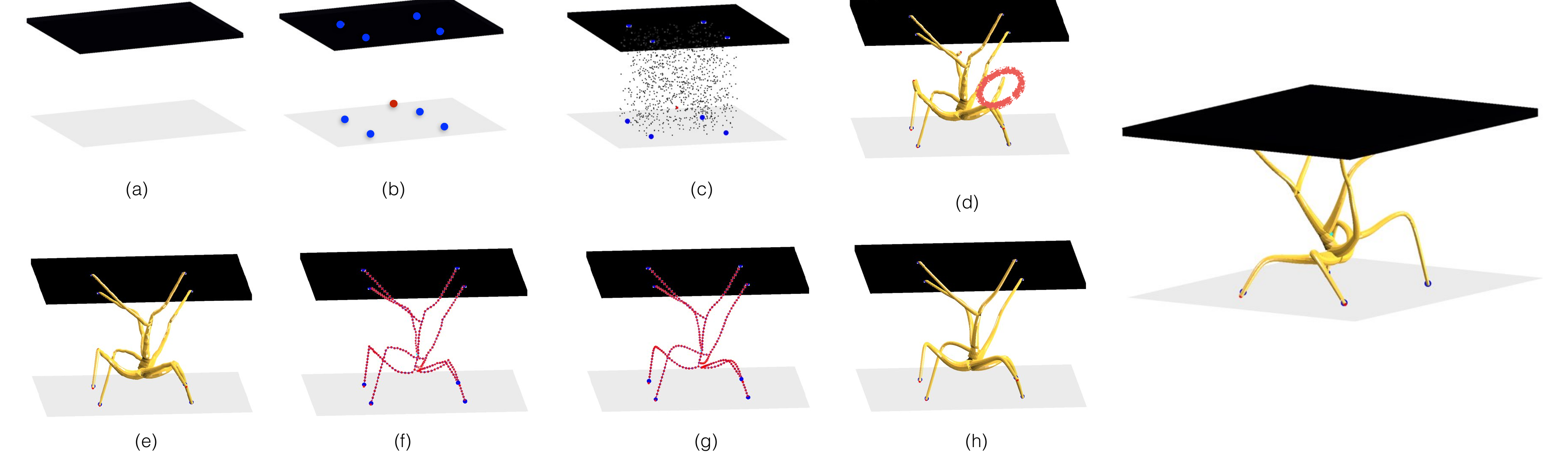}         \caption{Interface structure generation process. First, user defines the environment configuration (a) and selects the root and target points shown as red and blue, respectively (b). Attractors are generated randomly in the design space (c) and interface structure is grown automatically (d). Then, unnecessary branches are removed automatically (d-e) and the skeleton of the interface structure is smoothed as desired (f-h). Resulting shape is shown on the right.}
  \label{growth:process}
\end{figure*}

\section{Methods}
\label{sec:growth:methods}

In this work, the aim is to automatically create an interface structure between given objects. This process is illustrated in Fig.\ref{growth:process} starting with the user input and the steps of the automatic shape creation performed on the background. First, the user supplies the input geometries as 3D mesh models (a). Then, a set of target points are selected by the user to define where the interface structure should be in contact with the input models (b). Then, a root point or points are provided by the user to start the growth process (b). Attractors are randomly generated inside the design space (c). The structure is generated akin to a tree originating at its root and growing in 3D space to reach the targets (d). Branches that are not connected to the input objects through target points are removed from the structure (e). We also refer to this step as pruning or unnecessary branch removal. Finally, the skeleton of the structure (f) is smoothed (g-h). In the following sections, the details of these steps are described.

\begin{figure}
\centering
\includegraphics[trim = 0in 4.0in 0in 0in, clip, width=\textwidth]{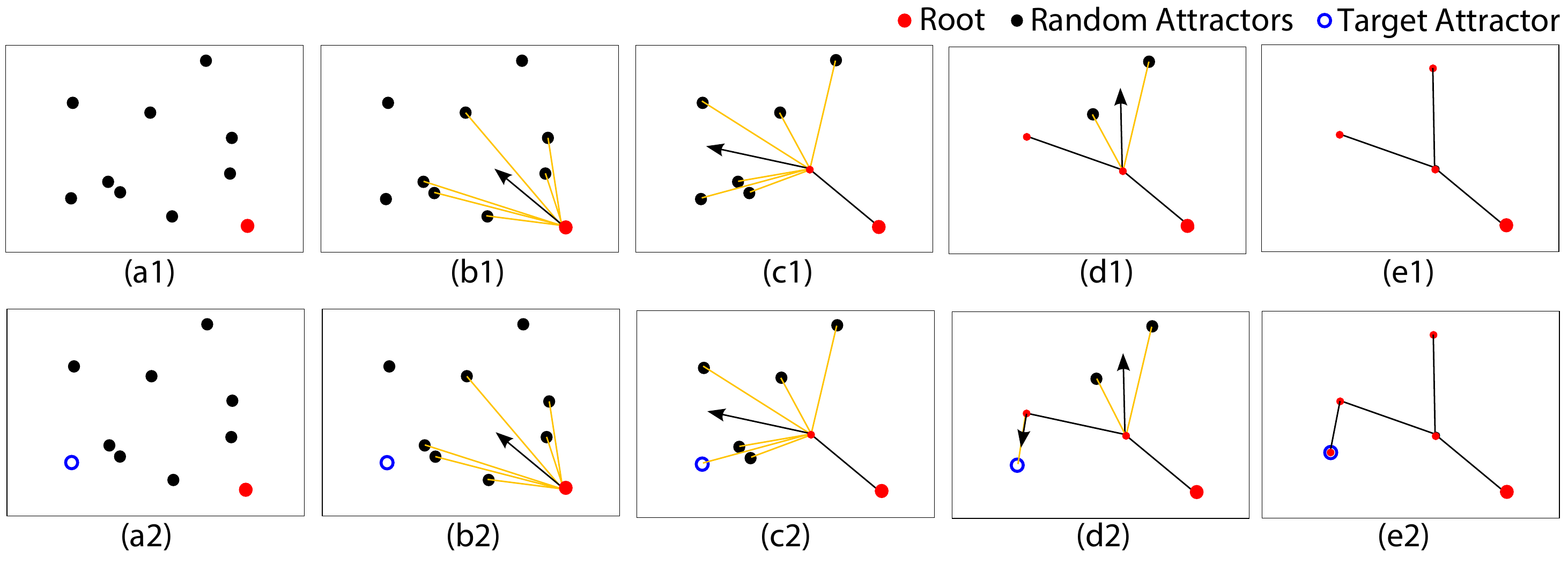} 
\caption{Space colonization algorithm with (a1-e1) and without (a2-e2) target attractors.}
\label{growth:colonization}
\end{figure} 

\subsection{Growth Algorithm}

\begin{figure}
\centering
\includegraphics[trim = 0in 2in 2.3in 0in, clip, width=4in]{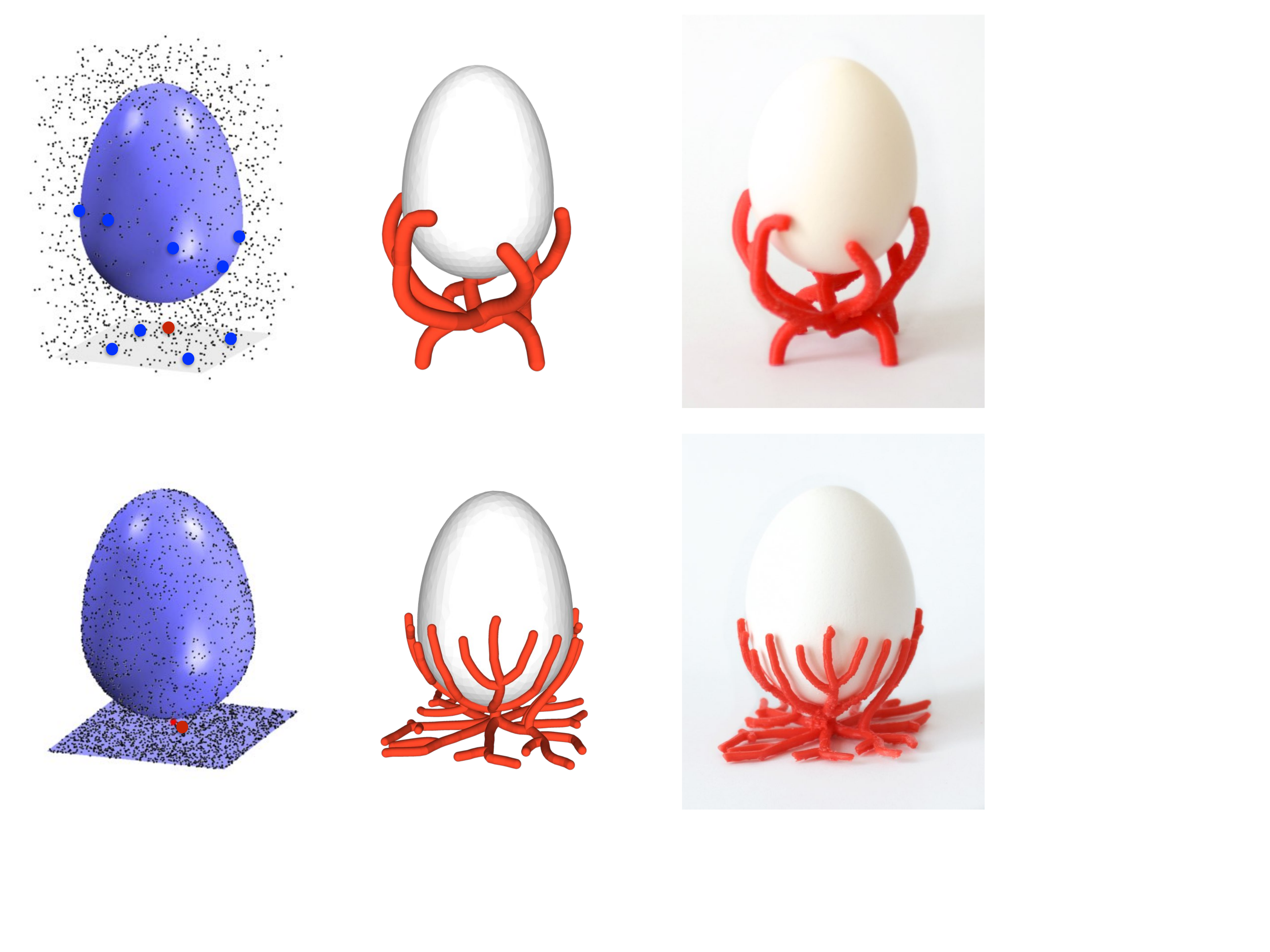} 
\caption{Egg holder generated using volume (top) and surface (bottom) hormones. Left: problem setup, middle: digital model, right: 3D printed result.}
\label{growth:egg}
\end{figure}

The proposed method uses an attractor based growth approach of space colonization algorithm given in \cite{Runions_sigg2005}. Space colonization algorithm creates a branching tree structure in space as demonstrated in Fig.\ref{growth:colonization}.a1-e1. The tree structure grows without the guarantee that it would touch any specific point in the design environment. In this work, we need to create shapes between objects ensuring that the generated interface would be in contact with the target objects to support them. For this reason, we introduce a novel target attractor concept to create branching structures that grow to the required target positions (Fig.\ref{growth:colonization}.a2-e2).

The target based growth process starts with the definition of the design space, e.g.~rectangle in (a2) and target-root point selections. Then, random attractors are sampled uniformly inside the design space. These random attractors have an \emph{influence distance} that they can pull a branch to themselves as well as a \emph{kill distance} that makes them inactive when they get too close to a branch in the growing structure. At every growth step, depending on the influence and kill distance, each attractor is associated with the tree node that is closest to it (yellow lines) if the node is within the influence distance. Then, normalized vectors from the node to the attractors are created and their average (black arrow) is calculated and used as the growth direction for the node (b2). The new node is added in the growth direction in the distance of branch length. All attractors are checked if they are in the kill distances of nodes. In other words, an attractor is killed if it is close enough to the tree (c2). This process iterates until all attractors are killed. While target attractors also pull the branches towards them, they are a special type of attractor with \emph{zero} kill distance. If an attractor is a target attractor, it does not get \textit{killed} until a tree node reaches it (notice difference in d1 and d2). The position of a new node is calculated as follows 

\begin{equation}
\label{growth:growthEq}
\vec{v}^{'} = 
    \begin{cases}
      \vec{t}, & \text{if}\ \text{reaching target}\\
      \vec{v} + L\hat{n}, & \text{otherwise}
    \end{cases}
\end{equation}

\begin{equation}
\label{growth:growthEq}
\;\;\;\;\vec{n} = \dfrac{\vec{a}-\vec{v}}{||\vec{a}-\vec{v}||}
\end{equation}

\begin{equation}
\label{growth:growthEq}
\hat{n} = \dfrac{\vec{n}}{||\vec{n}||} \\
\end{equation}

\noindent $\vec{v}^{'},\; \vec{v},\; \vec{t}, \; L,\; \hat{n},\; \vec{a} $ are the position vector of a new node, the position vector of node in the tree set, the position vector of the target attractor, branch length, unit growth direction vector and position vector of attractor in the set, respectively.

\begin{figure}
\centering
\includegraphics[trim = 0in 4.5in 5in 0in, clip, width=4in]{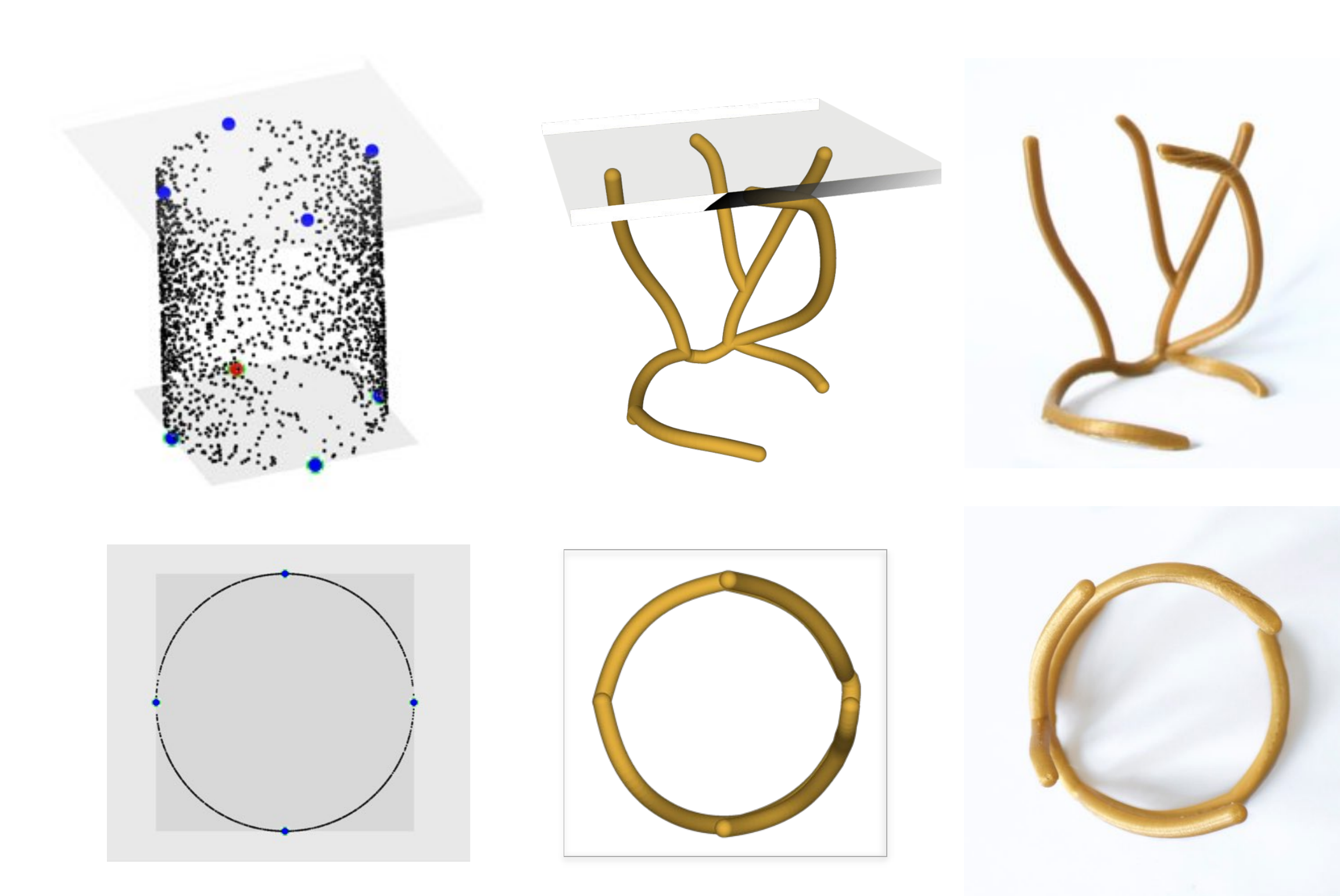} 
\caption{Example projection growth. Generating the interface structure. Left: problem setup with root \& target points illustrated, middle: digital model, right: 3D printed result. Orthogonal (top) and top (bottom) views of the same part are given.}
\label{growth:projection}
\end{figure} 
 
\subsection{Random Attractor Placement}

Placement of random attractors is a crucial step in our algorithm, especially to create variations for the same problem. Since the placement of the attractors defines the virtual design space in which our structure can grow, how attractors are placed in 3D drastically affects the resulting geometry. This effect is demonstrated for three distinct cases in Fig.\ref{growth:egg} and Fig.\ref{growth:projection}. Figure~\ref{growth:egg} compares the use of volume and surface attractors to generate an interface structure between the same objects. In the first one, we use the bounding box volume of the two objects to generate the attractors inside of the volume. On the other hand, in the second one, attractors are sampled on the surface of these objects. From this figure, it can be seen that the resulting interface geometries with very distinct characteristics can be obtained by only changing the distribution of the attractors even for the same problem setting. Here, an important distinction between these two cases is that we do not require target attractors for the surface growth case simply because we are guaranteed to touch the surfaces of both objects in this case. Another distinct case for attractor distribution is illustrated in Fig.~\ref{growth:projection}. Here, the aim is to generate an interface structure that would give a desired 2D profile when viewed from a certain direction. For this purpose, we sample the attractors on a surface that is created by extruding the desired 2D profile in the viewing direction. Hence, this specific attractor generation case can be classified as a subset of the surface growth explained previously. Moreover, any 3D swept/curved surface can be used to create 3D profiles. However, the main distinction here is that we require target attractors to be defined in this case, to ensure the resulting structure touches and supports the objects in the problem setup. This is mainly because the surface which the attractors generated on is a virtual one rather than the actual surface of the objects.

Apart from the aforementioned cases, we also enable symmetry in the resulting geometries. In product design, symmetry is considered to be a critical feature for everyday objects \cite{de2003design}. In our shape creation algorithm, we can ensure symmetry by simply placing the attractors in the design space symmetrically. Hence, the addition of a symmetry feature does not add any computational cost in our algorithm. However, the only case that may need special attention is the one where the root point is placed on the symmetry plane. In such cases, growth only happens on the symmetry plane because of the equal attraction from both sides. We solve this problem by moving the user defined root point slightly in both directions orthogonal to the symmetry plane by duplicating the root point.

\subsection{Obstacle Avoidance}

\begin{figure}
\centering
\includegraphics[trim = 0in 5.3in 0in 0in, clip, width=4.5in]{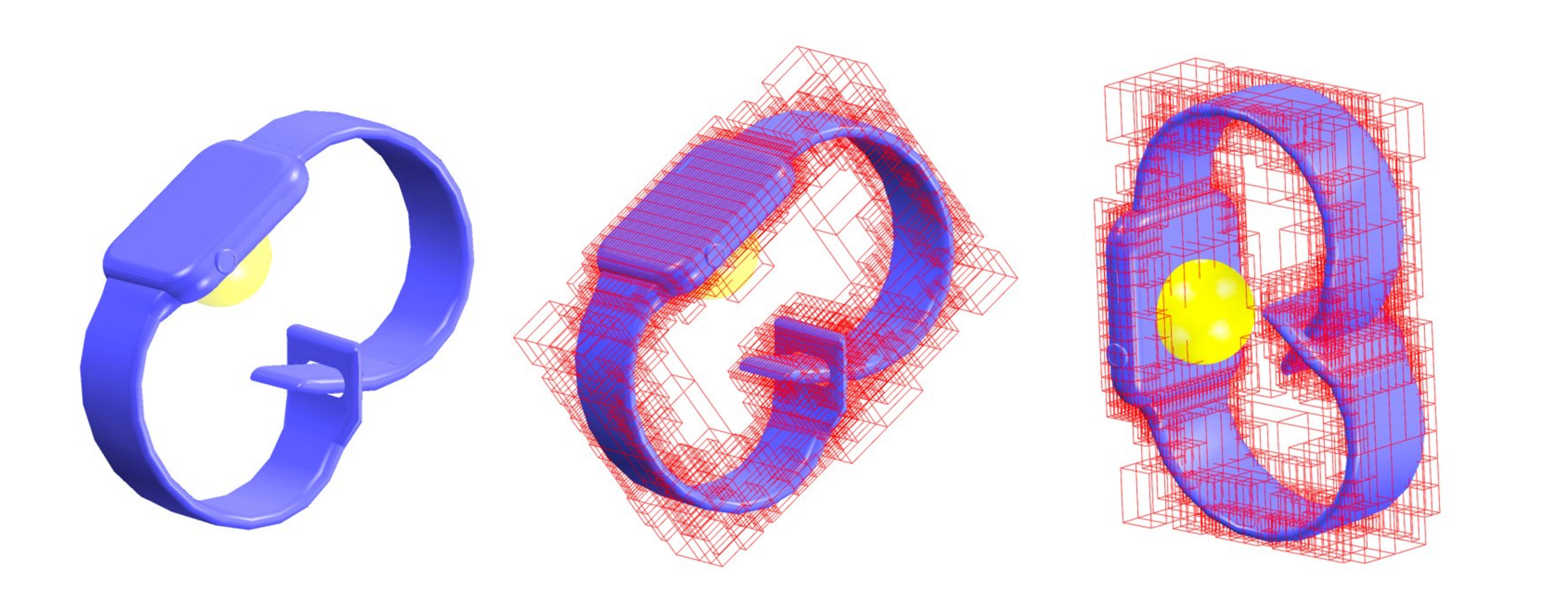} 
\caption{Obstacle avoidance is used to restrict growth in specific parts in the space. The restricted regions may be the objects represented as octrees or user defined spheres.}
\label{growth:Obstacle}
\end{figure} 

When designing an object, its interaction with the environment is important. For this reason, structure growth may be restricted in some parts of the design space. First of all, the interface structure should not intersect with the objects that it is intended to support. For this purpose, we utilize mesh representations of the objects for collision detection. In addition, users may define geometric obstacles in the form of spheres to restrict the growth. As an example for the use of spheres for functional purposes, a sphere is placed under the smart watch to limit the growth of the interface structure on the magnetic touch charging area in Fig~\ref{growth:Obstacle}.

During growth, the intersection of the new branch and obstacles are tested at each step. If there is a collision between the obstacle and the branch, a random direction is chosen for growth until collision is eliminated or maximum number of trials is reached. Intersection tests are conducted using a parametric representation of a line segment and an implicit representation of spherical and triangular objects. Details of the intersection test can be found in~\cite{ShirleyGraphicsBook}. To increase the efficiency of collision detection for triangular meshes, we use octree representation~\cite{Revelles00}.

\subsection{Smoothing}

\begin{figure}
\centering
\includegraphics[trim = 0in 0.5in 0.5in 0in, clip, width=4in]{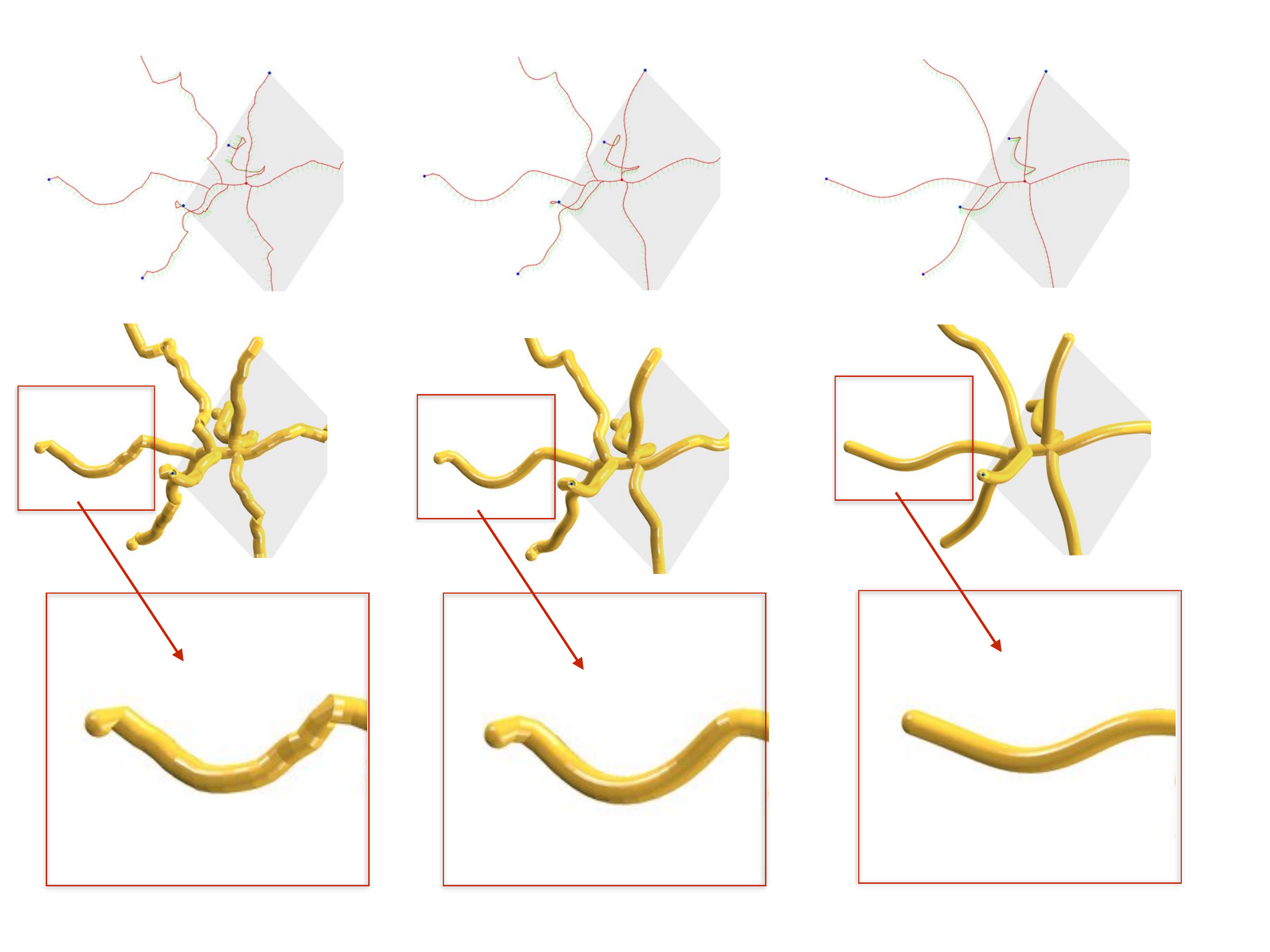} 
\caption{Effect of Laplace and Biharmonic operators on smoothing is illustrated for skeleton (top) and skin (middle-bottom) of the resulting geometries. Left: the original, middle: after smoothing with Laplace \& Biharmonic operators together, right: after smoothing with Laplace operator only. Note that the use of the combined Laplace \& Biharmonic operators allows smoothing without significant shrinkage.}
\label{growth:smooth}
\end{figure} 

While jagged transitions between biological branches look realistic for trees, smooth transitions are usually more appealing in product design. For this purpose, we apply curve smoothing to the tree structure as shown in Fig.\ref{growth:smooth}. Here, the position of each node on the tree skeleton is updated based on the positions of the neighboring nodes using Laplacian and Biharmonic operators as follows~\cite{mesh_processing_book, arisoy2012}. 

\begin{equation}
\label{growth:smoothingEq}
\vec{v}_i = \vec{v}_i + \lambda_1 \Laplace \vec{v}_i + \lambda_2 \Laplace(\Laplace \vec{v}_i)
\end{equation}

\noindent where Laplace and Biharmonic operators can be defined as:

\begin{equation}
\label{growth:smoothingEq2}
\Laplace \vec{v}_i = \nabla^2 \vec{v}_i = \dfrac{1}{2}(\vec{v}_{i+1} - \vec{v}_i) + \dfrac{1}{2}(\vec{v}_{i-1} - \vec{v}_i) 
\end{equation}

\begin{equation}
\label{growth:smoothingEq3}
\Laplace(\Laplace \vec{v}_i) = \nabla^4 \vec{v}_i
\end{equation}

\noindent $\vec{v}_{i-1}$ and $\vec{v}_{i+1}$ denote two neighbors of the node, $\vec{v}_i$.

In order to achieve smooth curves, we linearly combined Laplace and Biharmonic operators. Although it is possible to accomplish smoothing with only the Laplacian term, Biharmonic term is included to suppress the shrinking behavior arising from the Laplace operator when used alone (Fig.\ref{growth:smooth}). Here, we select the coefficients $\lambda_1$, $\lambda_2$ as 0.2 and 0.1, respectively.

\subsection{Branch Pruning}

As can be observed from Fig.\ref{growth:process} and Fig.\ref{growth:colonization}, our approach creates many branches that may not serve a structural function on the interface (i.e., branches that do not touch a target point). Hence, branches that are not connected to the target object or ground are automatically detected and removed form the tree graph (Fig.\ref{growth:process}(d)-(e)).

\subsection{Modifications and Variation In The Design}
Although we produce the interface structures automatically, we enable users to control many aspects of the geometry generation. The user control starts by importing 3D models of the objects and the selection of root-target attractor configurations. Then, another significant control comes from the placement of random attractors as explained previously. In addition to these inputs, there are four factors that affect the growth process (1) influence distance, (2) kill distance, (3) branch length and (4) number of random attractors. These factors are very important to generate variations in the space colonizations algorithm for tree generation such as trees with dense or sparse branch structures \cite{Runions_07}. On the other hand, results of our proposed growth algorithm are not affected by the changes in those parameters primarily due to the target attractor and branch removal concepts. As long as the parameters are \emph{suitable}, results do not change significantly. There is a wide range of suitable parameters for a given problem. Any suitable parameter set has the following properties:

\begin{itemize}
\item Influence distance is greater than kill distance.
\item Branch length, which can be considered as step size, is small compared to the environment dimensions but it is large enough to facilitate efficient growth computation. 
\item The number of random attractors is high enough to create uniform distribution in the design space, we used 2000 attractors for the examples in this work.
\item Influence distance is high enough to enable attraction of a node for the created uniform distribution.
\end{itemize}

\noindent In this work, we choose default values using the given guidelines. For each problem setup, we use the default values by scaling them with the dimensions of the bounding box of the system.

Another set of important controls comes into play after the interface structure is generated automatically. At this point, users can control the radius variation in the branches of the interface structure as well as modify the skeleton of the structure by sketched strokes. Now that the skeleton of the structure is obtained, a 3D skin is created by covering each branch with a truncated cone and taking the union of all cones. The radius at each node is calculated based on its age as

\begin{equation}
\label{growth:radiusEq}
r = r_{min} + (r_{max}-r_{min})*e^{-k\alpha}
\end{equation}

\noindent where $r,\;r_{min},\;r_{max},\;\alpha,\;k$ is the radius, minimum radius, maximum radius, age and decay of radius, respectively. Here, age of each node is determined in such a way that every node starts with age of 0 and the age increases by 1 at each growth step. Decay of radius defines how fast the radius changes from the root to the targets between maximum radius and minimum radius and is set by the user. Also, $r_{min},\;r_{max}$ are set by the user.

Sketch modifications are performed through modifier sketches performed by the user to specify the new shape of the skeleton curve as it would occur from the current viewpoint. To do this, a surface is created by the rays emanating from the user's eyes, passing through the strokes and extends into the page. In theory,  there are infinitely many candidate solutions on this surface. The best 3D configuration is thus found by computing the minimum distance projection of the original curve onto the surface. For the details of the sketch modifications please refer to \cite{sketch2007}.

\begin{figure}
\centering
\includegraphics[trim = 0.9in 2.5in 0in 0in, clip, width=4.5in]{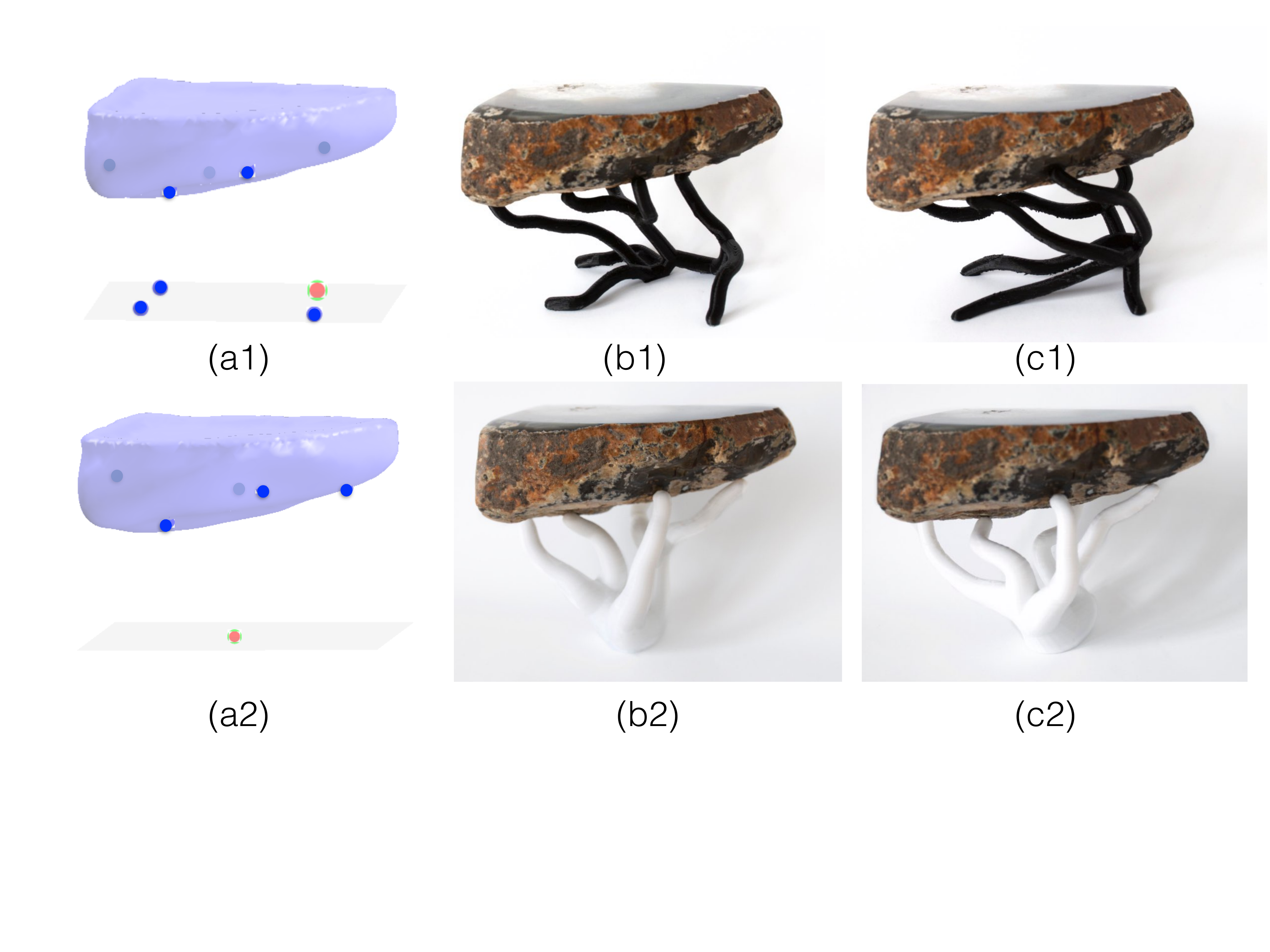} 
\caption{Effect of target-root selection and stochastic growth demonstrated on two different target-root configurations (a1-c1, a2-c2). For the same problem setup (a), various stochastic growth results (b and c) can be obtained.}
\label{growth:rock} 
\end{figure}

\section{Results}
\label{sec:growth:results}

Our approach enables the automatic generation of interface structures for 3D printing. The user may control the geometry generation through target-root configurations, random attractor placement, skin radius selection and sketch modifications. We applied our algorithm to a variety of examples including gadget accessories, decoration and restoration of existing objects and furniture. In order to transform the existing objects into the digital design environment, we utilized 3D scanning using a Kinect device. We downloaded the 3D digital models through the stock 3D model websites like GrabCAD and Google 3D Warehouse for the common objects.

The latest trends in decorating and modern furniture design include hybrid design approaches where natural materials with imperfections are combined with machine-made parts to create innovative and original designs. In Figure~\ref{growth:rock}, an example hybrid design created using our system is shown. Here, we take a natural rock piece and design a support structure that complements its organic geometrical features. One important control that our system provides is the target-root placement. We generated two different target-root configurations (a1, a2) for the same problem to demonstrate the significant variance in the resulting geometries (b1, b2). Moreover, we would like to draw attention to the stochastic nature of our algorithm that comes from the random sampling of attractors inside the design space where different results are obtained for distinct set of random attractors. However, the effect of the stochastic nature on the resulting geometries (b1-c1 and b2-c2) are subtle compared to the effect of target-root selection.

\begin{figure}
\centering
\includegraphics[trim = 0in 3.5in 0in 0in, clip, width=4.0in]{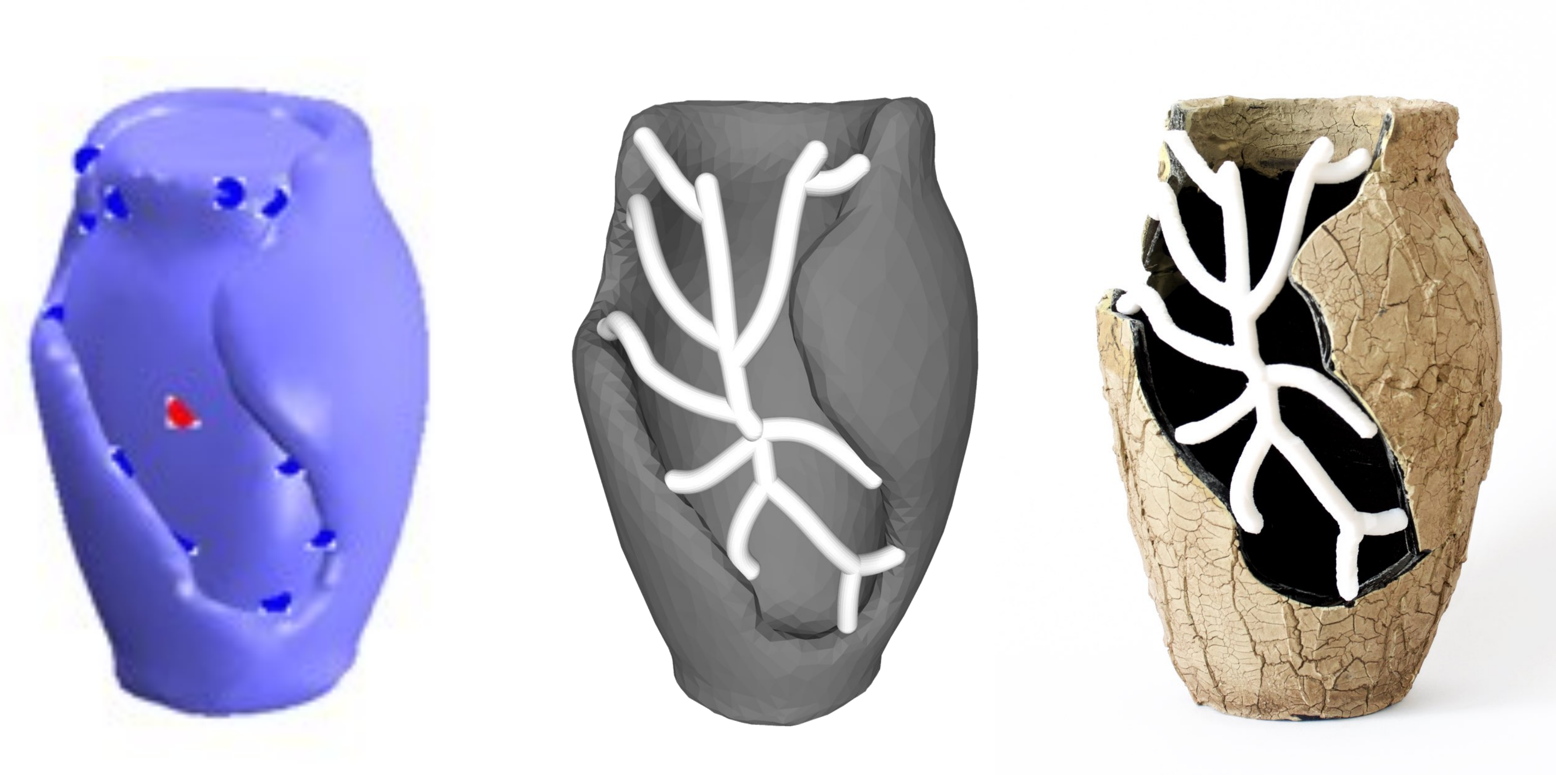} 
\caption{Form completion: A broken vase is restored using a scanned model. Left: problem setup, middle: digital model, right: 3D printed result.}
\label{growth:vase}
\end{figure}

\begin{figure*}[t]
  \centering
  \includegraphics[trim = 0in 0in 0in 0in, clip, width=\textwidth]{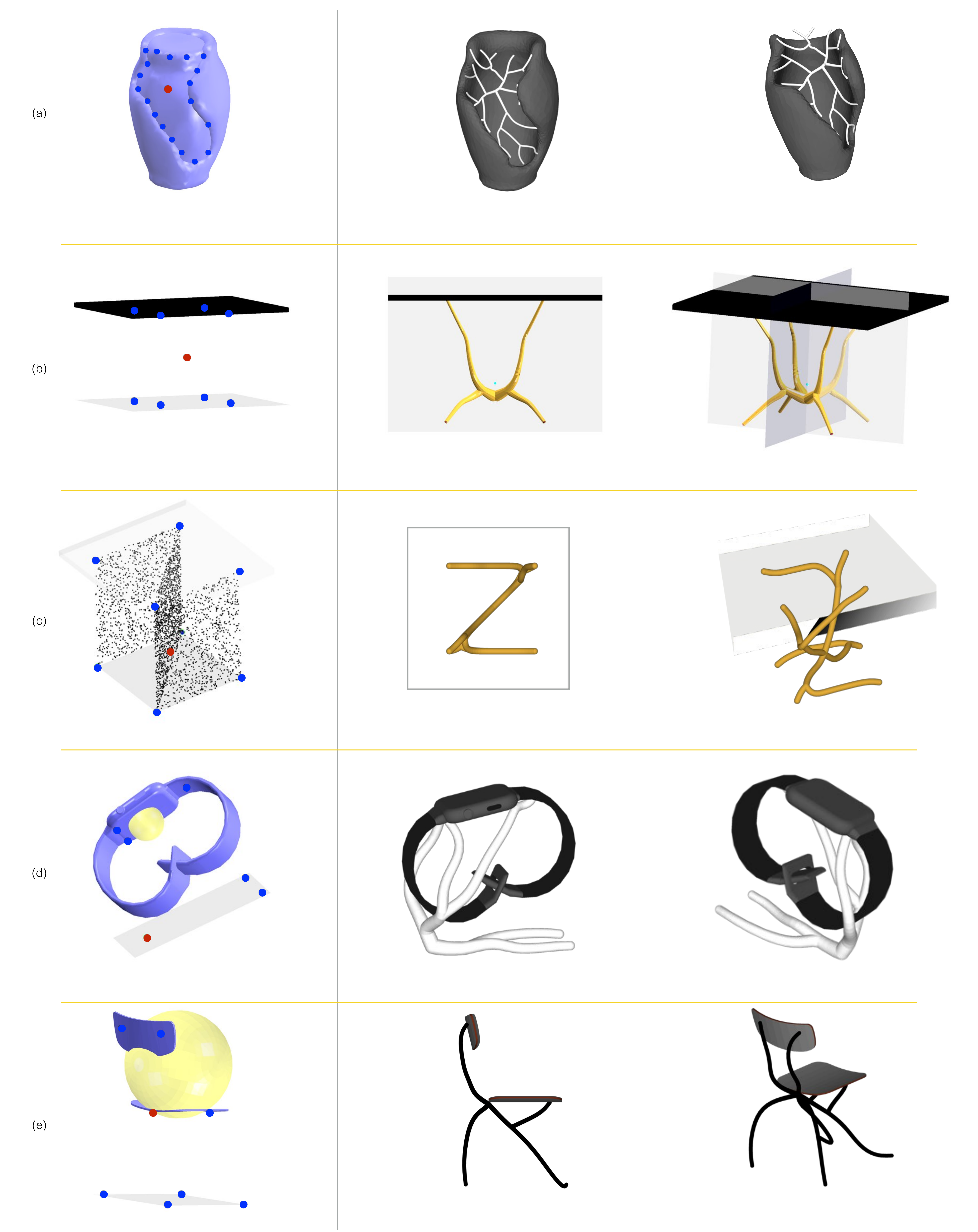}         \caption{Designs created with our system. Left: problem setup, right: resulting design from two different views.}
  \label{growth:final}
\end{figure*}

Another direction for craft, arts and design is the restoration of broken objects through 3D printing to obtain new artistic expressions rather than restoring the original object~\cite{zoran2013hybrid}. In that, the motivation is not to restore the initial function of the object, but rather use it to function as a memorial. In Fig.\ref{growth:vase}, we show that our method can be used for similar purposes. Here, a missing part of a broken vase is restored with the generated interface shape. For this, we first scanned the broken vase and placed desired target-root points. Then, the resulting piece to complete the broken part is grown using our algorithm. We also 3D printed and assembled the resulting part to the broken vase. Another alternative interface structure for this example can also be seen in Fig.\ref{growth:final}.a.

\begin{figure}
\centering
\includegraphics[trim = 0in 3.6in 5.3in 0in, clip, width=4.0in]{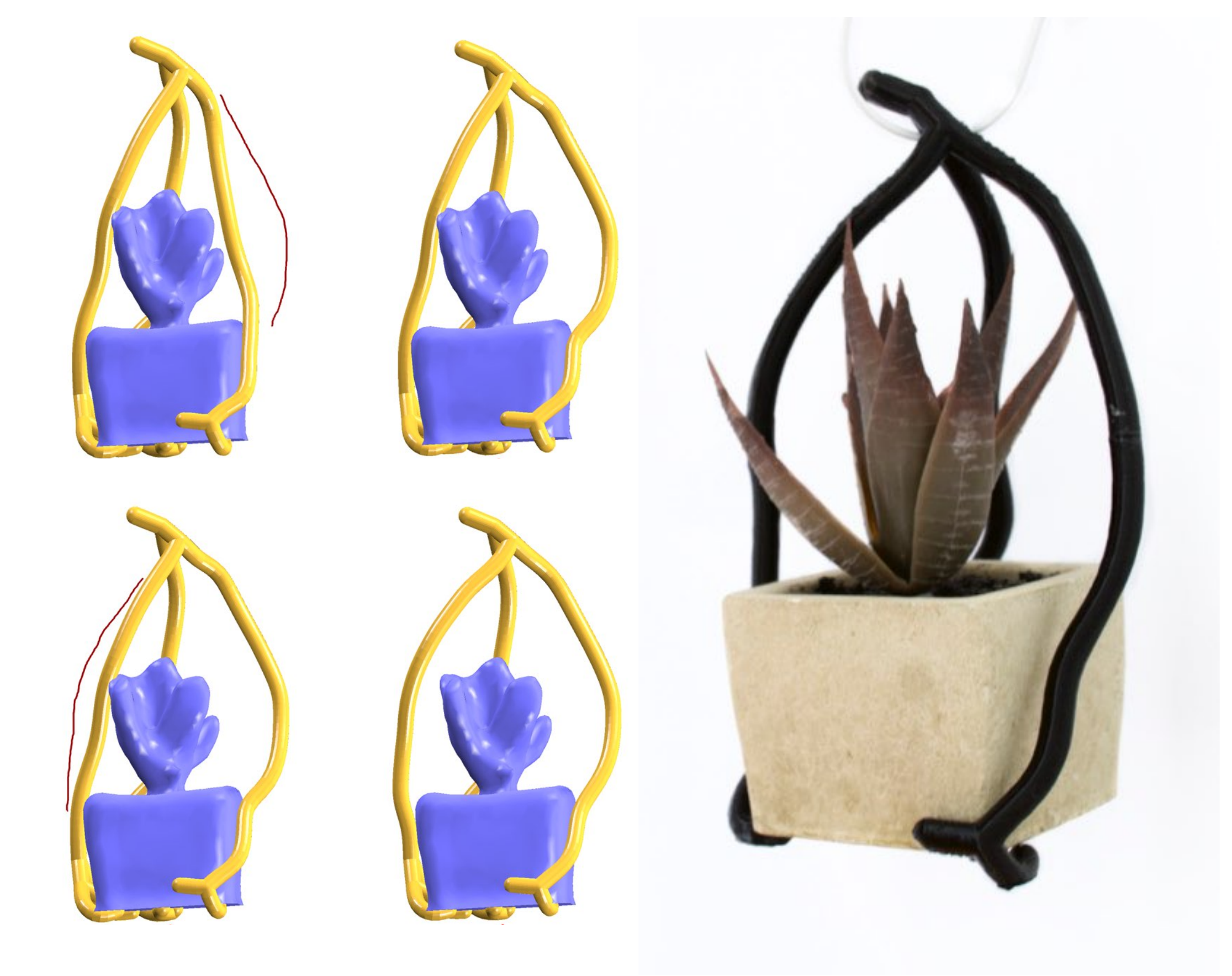} 
\caption{Use of sketch modifications for a functional purpose. Top part of the planter holder is enlarged to be able to insert the planter. Left: Sketch modification steps, Right: 3D printed result with the inserted planter.}
\label{growth:planterSketch}
\end{figure}

In addition to the aesthetic needs, the need for sketch modifications may arise from functional requirements. Use of sketch modifications for a functional purpose is illustrated in Fig.\ref{growth:planterSketch}. In this example, a hanger is designed to suspend the planter. However, the planter can not be inserted into this automatically generated structure. For this reason, sketch modifications are applied on the skeleton of the structure to enlarge the top part of the hanger so that the planter can be inserted.

\begin{figure*}[t]
  \centering
  \includegraphics[trim = 0in 0.0in 0in 0in, clip, width=\textwidth]{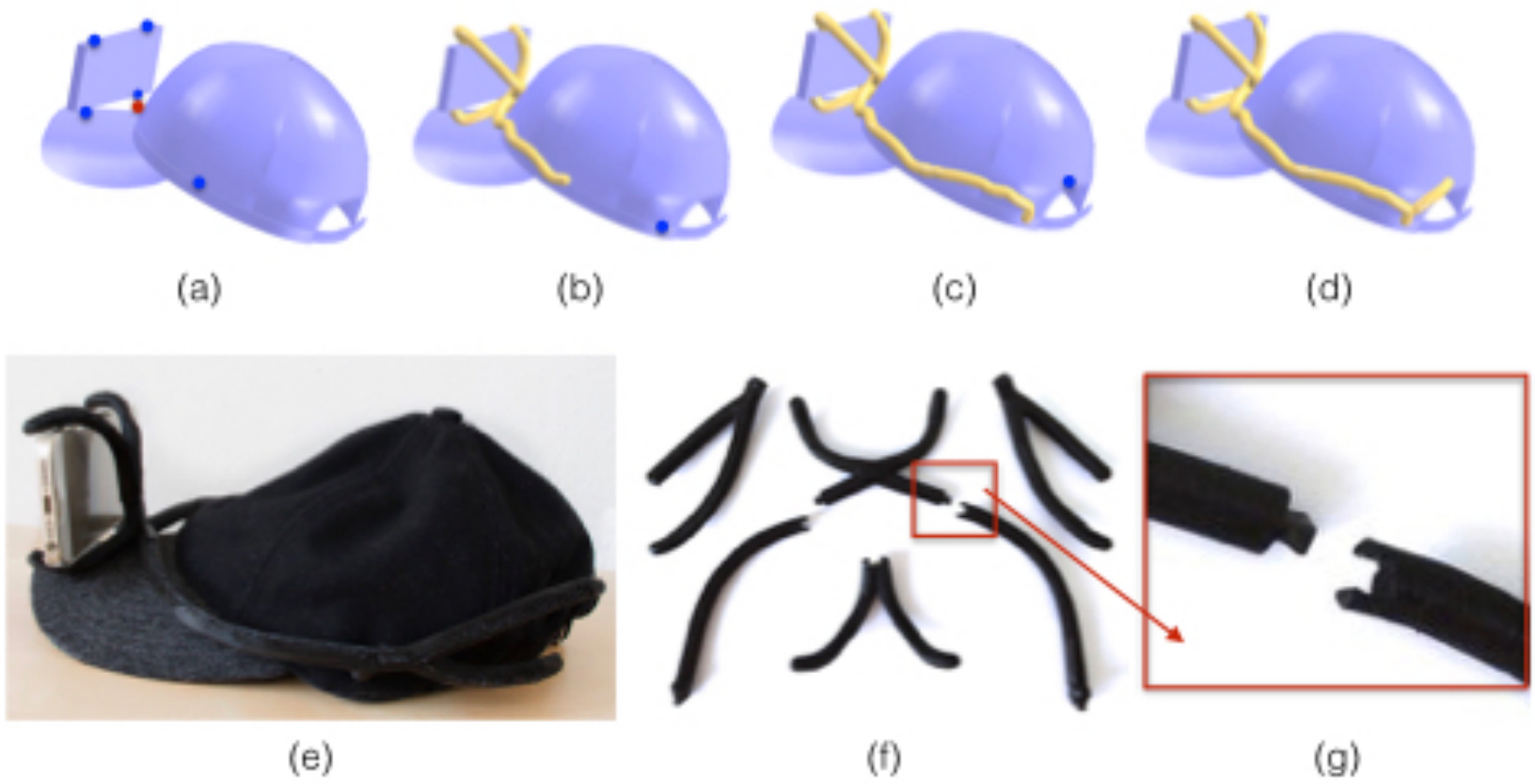}         \caption{Guided progressive growth (top) is shown on a baseball cap example to attach a phone for first-person  view recording. Use of assembly structures to 3D print larger designs (bottom) have been demonstrated with the zoomed in dovetail joint detail (g).}
  \label{growth:guided}
\end{figure*}

In some configurations, users might need to control the growth process more strictly to achieve a geometry with particular desired properties. In such cases, our geometry creation process can be guided by progressively manipulating the problem setup. To do this, instead of defining all target points at once, we start with a subset of targets and progressively add the remaining ones as we grow the structure. Figure~\ref{growth:guided} demonstrates this on an example to attach the phone to a baseball cap for first person view camera shots. Here, the aim is to guide the growth on the side of the cap instead of any other possible outcome. First, only five of the targets are defined (a) and the growth process is completed (b). Then, a new target point is added (b) and another growth process is accomplished. This process is iterated (c) until the final desired shape is created (d). Since we are using a consumer level 3D printer with a limited build volume, we partitioned the resulting object into smaller pieces to be able to 3D print. For the assembly, we manually added dovetail structures on the assembly surfaces~(f).

There are many communities that promote reuse of materials through community engagement, resource conservation and creativity \eg Pittsburgh Center for Creative Reuse and Lancaster Creative Reuse. Since our design framework is developed to work with existing objects, users can easily utilize our algorithm for creative reuse purposes. A virtual example of material reuse is shown in Fig.~\ref{growth:final}.e. Here, the usage of a seat and back from a broken chair to design a new support and legs is demonstrated. In the example, while we have virtual models for the elements to be reused, as mentioned earlier any object can be scanned and used to create the interface structures. Although for the previous examples, we focus on creating attachment structures that hold the object in place without fixing or gluing, this example requires the interface structure to be fixed to the supported objects.

Since our framework is tailored towards non-expert users, we fabricated all our examples with a consumer level low-cost 3D printer, PrintrBot Simple 1405, to study the printability of our results. However, more advanced 3D printers can be used to fabricate resulting geometries with higher qualities using various material options. 

We recorded the computation time for automatic shape generation for a number of examples. Since our method has a stochastic nature, computation time changes as the random attractor set changes. Thus, the results are reported for three different random attractor sets for each example in Table~\ref{growth:table_performance}. One reason computation time changes for each example is the change in the complexity of the objects that increases the time for collision checks. Another, important factor is how easy or difficult it is to reach the targets inside the design space.

\begin{table}[!t]
\renewcommand{\arraystretch}{1.3}
\caption{Runtime Performance of Our Generative Design Algorithm}
\label{growth:table_performance}
\centering
\begin{tabular}{m{0.48in}|>{\centering\arraybackslash}m{0.39in}|>{\centering\arraybackslash}m{0.39in}|>{\centering\arraybackslash}m{0.39in}|>{\centering\arraybackslash}m{0.39in}|>{\centering\arraybackslash}m{0.39in}|}
\cline{2-6}
&\multicolumn{5}{c|}{Total Run Time [s]}\\
\cline{2-6}
  & Run 1 &Run 2 &Run 3 &Run 4&Run 5\\
\hline
\multicolumn{1}{|c|}{ Fig.\ref{growth:final}.a}& 11 & 12 & 10 & 10 & 9 \\
\hline
\multicolumn{1}{|c|}{ Fig.\ref{growth:final}.b}& $<1$ & $<1$ & $<1$ & $<1$ & $<1$\\
\hline
\multicolumn{1}{|c|}{ Fig.\ref{growth:final}.c}& $<1$ & $<1$ & $<1$ & $<1$ & $<1$\\
\hline
\multicolumn{1}{|c|}{ Fig.\ref{growth:final}.d}& 2 & 2 & 2 & 2 & 2\\
\hline
\multicolumn{1}{|c|}{ Fig.\ref{growth:final}.e}& 4 & 2 & 4 & 3 & 2\\
\hline
\end{tabular}
\end{table}

\noindent \textbf{User Study:} 

\begin{figure}
\centering
\includegraphics[trim = 0in 3.6in 0in 0in, clip, width=4.5in]{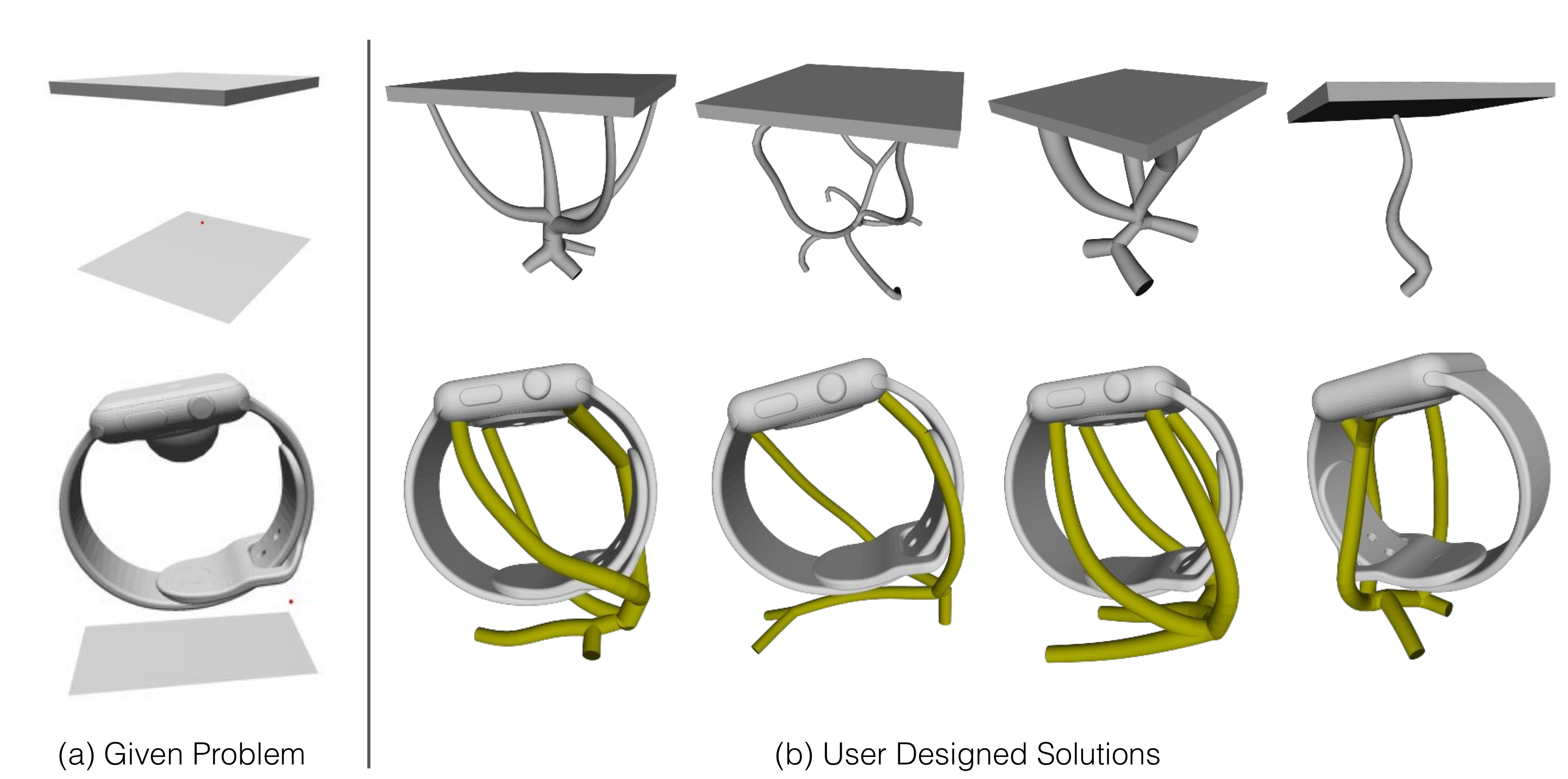} 
\caption{Users were given two design problems: designing legs for a table top (a-top row), and designing a stand for a smart watch (a-bottom row). Some designs created by the users are shown (b). The users configure the root and target nodes for the given problem and the software produces the final shape. The users may further modify the shapes using sketch input.}
\label{growth:userresults}
\end{figure}

\noindent We  conducted a user study to evaluate the usability of our system. 25 users who had no prior knowledge of our software participated in the user study. Each participant was given 30 minutes to finish all the tasks including software introduction, two modeling assignments and completing out a post survey. The same two modeling tasks were assigned to all participants as shown in Fig. \ref{growth:userresults}. Some example designs generated by the users are also shown.

\begin{figure}
\centering
\includegraphics[trim = 4.34in 0in 0in 7.3in, clip, width=4.5in]{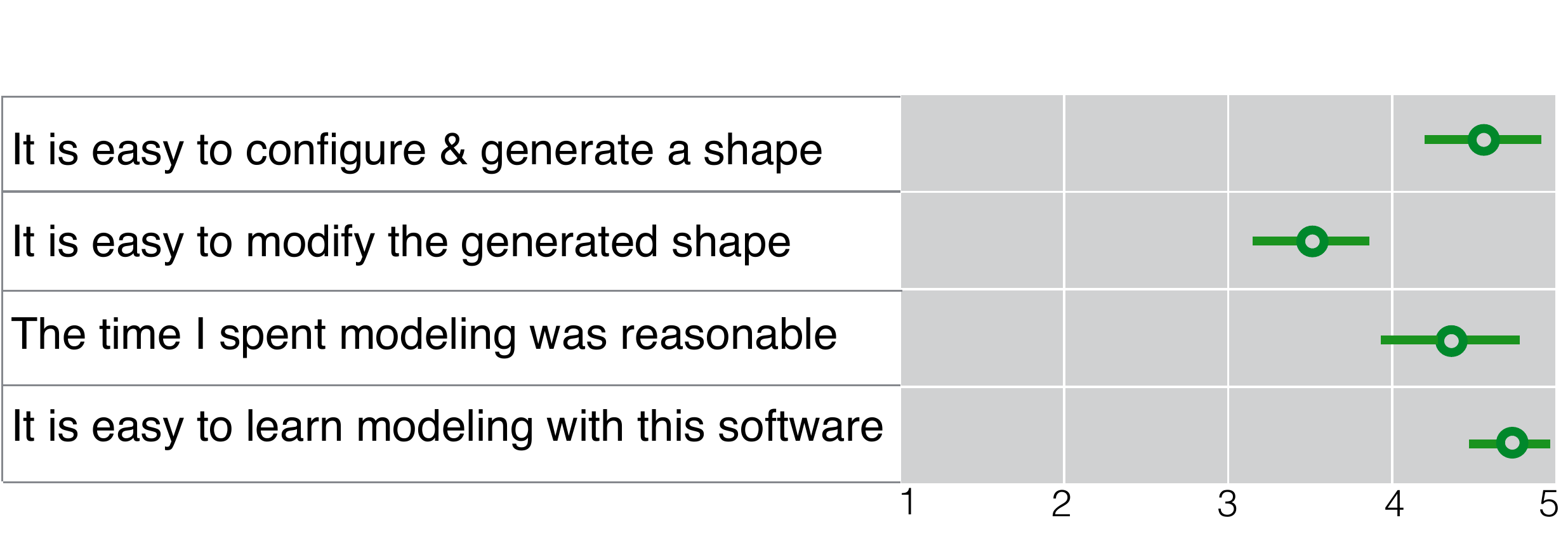} 
\caption{Survey results collected from all 25 participants of the user study (1: Strongly Disagree, 5: Strongly Agree). Circles and error bars represent the mean value and one standard deviation, respectively.}
\label{growth:userstudy}
\end{figure}

All users were able to complete the tasks in 30 minutes or less. We believe this indicates that users were able to learn the software easily and use it efficiently. The survey results also support this with a strong agreement in questions 1, 3 and 4 (Fig.~\ref{growth:userstudy}). During the user study, we observed some issues with shape  modification. Some participants had a hard time figuring out how sketch modification works. This observation also explains the weaker agreement in question 2 in the survey. However, our observations showed that if the participants experimented more with the sketching part of the software, they were able to understand and efficiently use sketch  modifications.

In addition to the Likert scale questions (Fig.~\ref{growth:userstudy}) on the survey, we asked the participants to write comments if they had any. One important conclusion we drew from some of the comments was that some of the participants tended to imagine a design and tried to generate that exact solution. Since, our system automatically creates shape  solutions to a given problem, the software is not intended to be used to produce a specific shape the user has in mind. It was interesting to see that people felt compelled to control every  aspect of the shape with the conventional design approaches even when the results were automatically generated for them. We believe increasing the expressive power of such a design system is very important even when the results are automatically created.

\section{Limitations and Future Work}
\label{sec:growth:limitations}

Our focus has been on generating tree-like  structures on the skeletons we grow that make the resulting designs resemble biological trees. We expect the proposed formulation to be readily applicable with different building blocks instead of our current truncated cones to achieve a richer variation in form. Moreover, since our obstacle avoidance is achieved through random search directions, our algorithm may not converge to a solution within predefined maximum number of trials. While we have observed this issue rarely, increasing the maximum trial number for complex problem settings may be required. Finally, in this work, we do not consider structural performance of the resulting shapes. Yet, our algorithm can be extended to ensure structural soundness for a given problem configuration. This may require finite element analysis during the shape generation process.

\section{Conclusions}
\label{sec:growth:conclusions}

We present a generative design framework to create interface structures to support existing objects. The proposed method enables novice users to automatically generate geometries and edit them once the shape is created. Our approach introduces a novel application of a nature inspired growth algorithm with embedded product design considerations. Our current studies indicate that the approach works well for a variety of design problems with the presented actual 3D printed results alongside their digital models. Also, the user study supports the practical usage of the proposed system. We consider this work as a step towards future customized design software where users only define functional constraints and the CAD system automatically creates a solution.

\chapter[Designing Coupling Behaviors]{Designing Coupling Behaviors Using Compliant Shape Optimization}
\label{chap:coupling}
\blindfootnote{This chapter is based on Ulu et al.2018 \cite{ulu2018coupling}.}

A wide set of assembly blocks such as attachments, connectors, joints, and supports rely on the principle of passively coupling two objects using structural compliance. However, only a limited variety of configurations are prevalent in daily use (\eg snap fits) due to the challenge of extending the appropriate mechanical behavior to arbitrary object pairs. In this work, we present a method for computationally designing the mechanical coupling behavior between a rigid object and a compliant enclosure based on high-level specifications such as the ease of engagement and disengagement. At the heart of our approach is the use of deformation profiles as the means to describe and optimize physical coupling characteristics. In particular, we introduce a method that maps the shape parameters of the compliant object onto sequentially observed coupling descriptors such as the grip, insertion and removal forces that develop as the rigid object is engaged. Using this formulation, we present a method for optimizing the rest shape of the compliant object to produce the desired coupling behavior. We demonstrate our approach through a variety of designs and validate it with 3D printed physical prototypes.

\section{Introduction}
\label{sec:coupling:intro}

\begin{figure}
   \includegraphics[width=\textwidth]{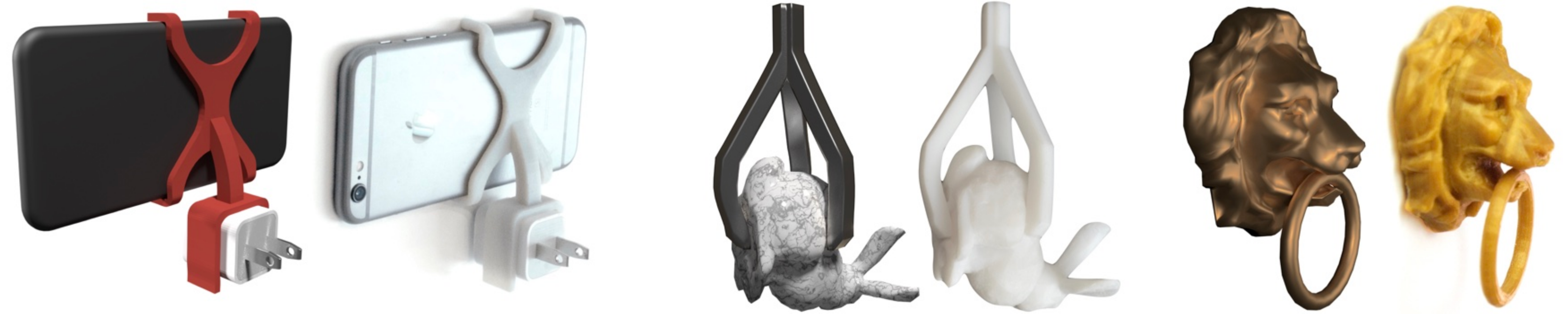}
   \caption{We introduce a method for designing coupling behaviors between an arbitrary compliant structure and an arbitrary rigid object. Resulting structures exhibit the desired coupling behavior such as ease of engagement/disengagement and grip.}
   \label{fig:teaser}
\end{figure}

Mechanical coupling, defined as attaching two objects to one another, is a fundamental notion that underpins the realization of all types of connectors, joints, fixtures and attachments, which enable the creation of complex assemblies and mechanisms~\cite{Steif:2012}. In daily use, a large class of couplings are intended to be \emph{readily separable} with as few parts as possible as a way to facilitate temporary affixing, quick assembly and maintenance, and general ease of use. To this end, passive coupling of objects through structural compliance is a widely used method involving minimal number of parts  and mechanical complexity, as part engagement is primarily enabled by elastic body deformations over coupling rigid objects. For designing monolithic compliant structures, topology optimization is a widely used approach that allows a tuning of force-displacement characteristics at prescribed end  states~\cite{Lu:2003,Lu:2005}, or to achieve structures that satisfy strength or compliance requirements under specified load configurations~\cite{bendsoe1989optimal}.

However, only a limited variety of configurations are prevalent in daily use (\eg snap fits) due to the challenge of extending the appropriate mechanical behavior to arbitrary object pairs. In particular, the compliant structures and their rigid counterparts are typically tailored such that either  there exists known and permanent contact points that do not change during coupling~\cite{casals2004dynamic}, or the contact points involving the maximally deformed state can be known a priori~\cite{Chen:2014}. Realizing these limitations, Koyama et~al.~\cite{Koyama:2015} in an inspiring work present a data-driven approach for designing compliant attachments using parameterized basic geometries such as cylinders and rectangular prisms. However, the analysis does not extend  to arbitrary free-form objects, necessitating rigid, multi-part solutions for such instances.

\begin{figure*}[t]
  \centering  
  \includegraphics[trim = 0in 0in 0in 0in, clip, width = \textwidth]{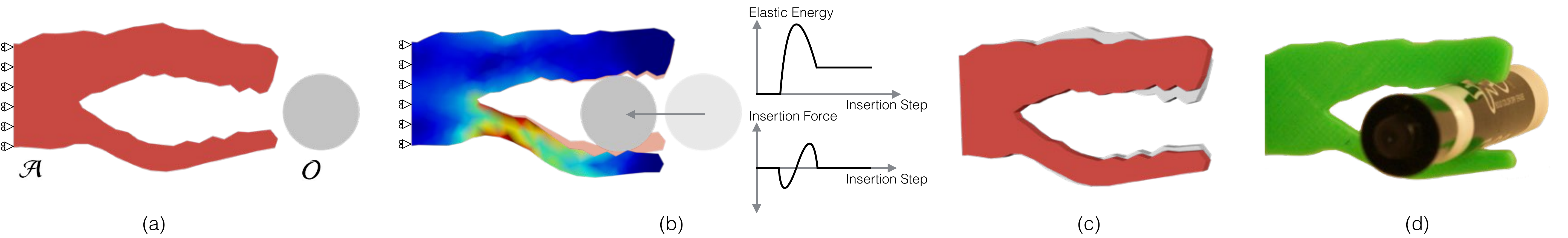}
  \caption{Given a compliant structure $\mathcal{A}$ and a rigid object $\mathcal{O}$ (a), our algorithm optimizes the rest shape of $\mathcal{A}$  based on the deformation behavior obtained through insertion simulations (b). The resulting structure (c) exhibits the desired compliant behavior when coupled with $\mathcal{O}$  (d). Grey silhouette in (c) is the original unmodified shape.}
  \label{fig:overview}
\end{figure*}

In this work, we present a physics-based method for designing the mechanical coupling behavior between a rigid and a compliant object such that the engagement and disengagement forces during the process of coupling, as well as the grip forces that lock the object pair together can all be customized by optimizing the shape of the compliant object (Figure~\ref{fig:teaser}). Given an arbitrary rigid and  compliant object, we use deformation profiles as a means to describe and optimize physical coupling. We introduce a method that maps the compliant object's shape parameters onto sequentially observed coupling descriptors such as the grip, insertion and removal forces that develop when the compliant object engages the rigid object. Using this formulation, we present a method for optimizing the rest shape of the compliant object to produce the desired coupling behavior. 

A distinguishing feature of our approach is that it allows coupling behaviors to be designed for part interactions that may not be known a priori. In particular, our approach does not rely on the knowledge of known contact points or deformed states, thereby extending prior work on compliant attachments to scenarios involving arbitrary object pairs.

Our main contributions are
\begin{itemize}
\item	The use of deformation profiles to describe and optimize mechanical behavior.
\item	A physics based shape optimization method for compliant coupling behavior design involving two-part interactions.
\item A practical insertion simulation based on collision elimination for computing deformation profiles.
\end{itemize}

\section{Fundamentals and Overview}
\label{FundamentalsAndOverview}

\begin{figure}
  \centering  
  \includegraphics[trim = 0in 0in 0in 0in, clip, width = \textwidth]{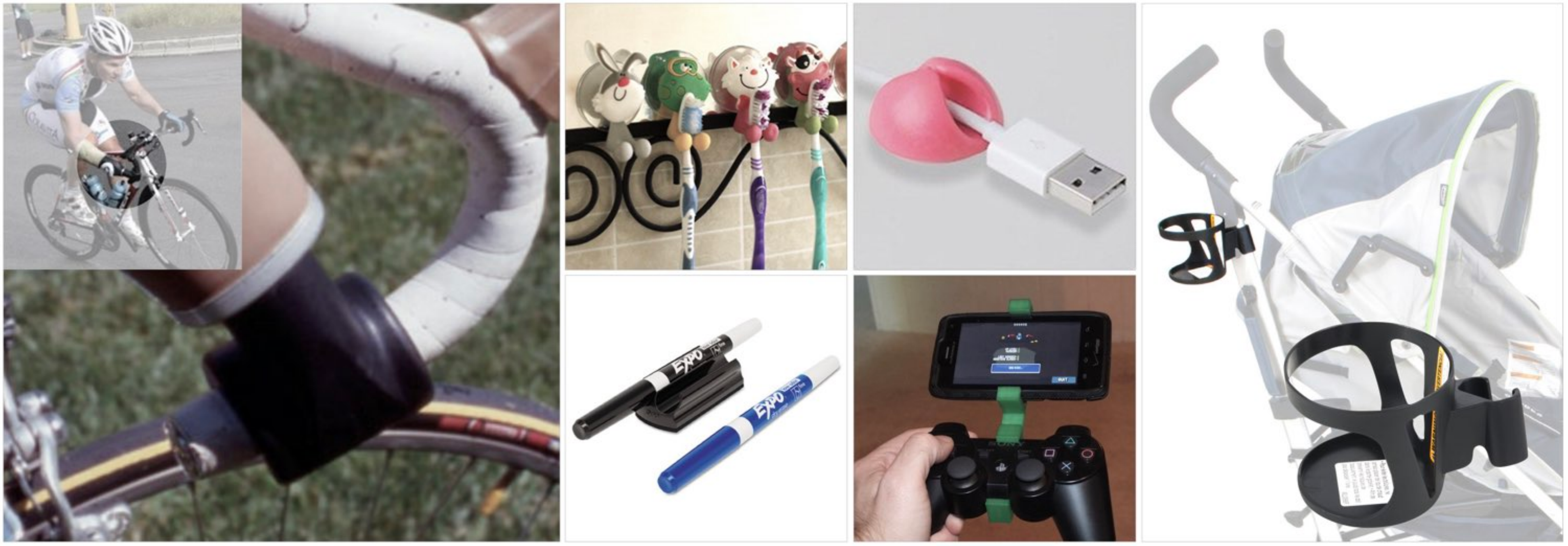}
  \caption{Example compliant couplings as part of our daily lives. Image courtesy of TRS Prosthetics, Expo, Chicco, ScribbleJ@Thingiverse.}
  \label{fig:realWorldAttachments}
\end{figure}

\begin{figure}
  \centering  
  \includegraphics[trim = 0in 0in 0in 0in, clip, width=4.5in]{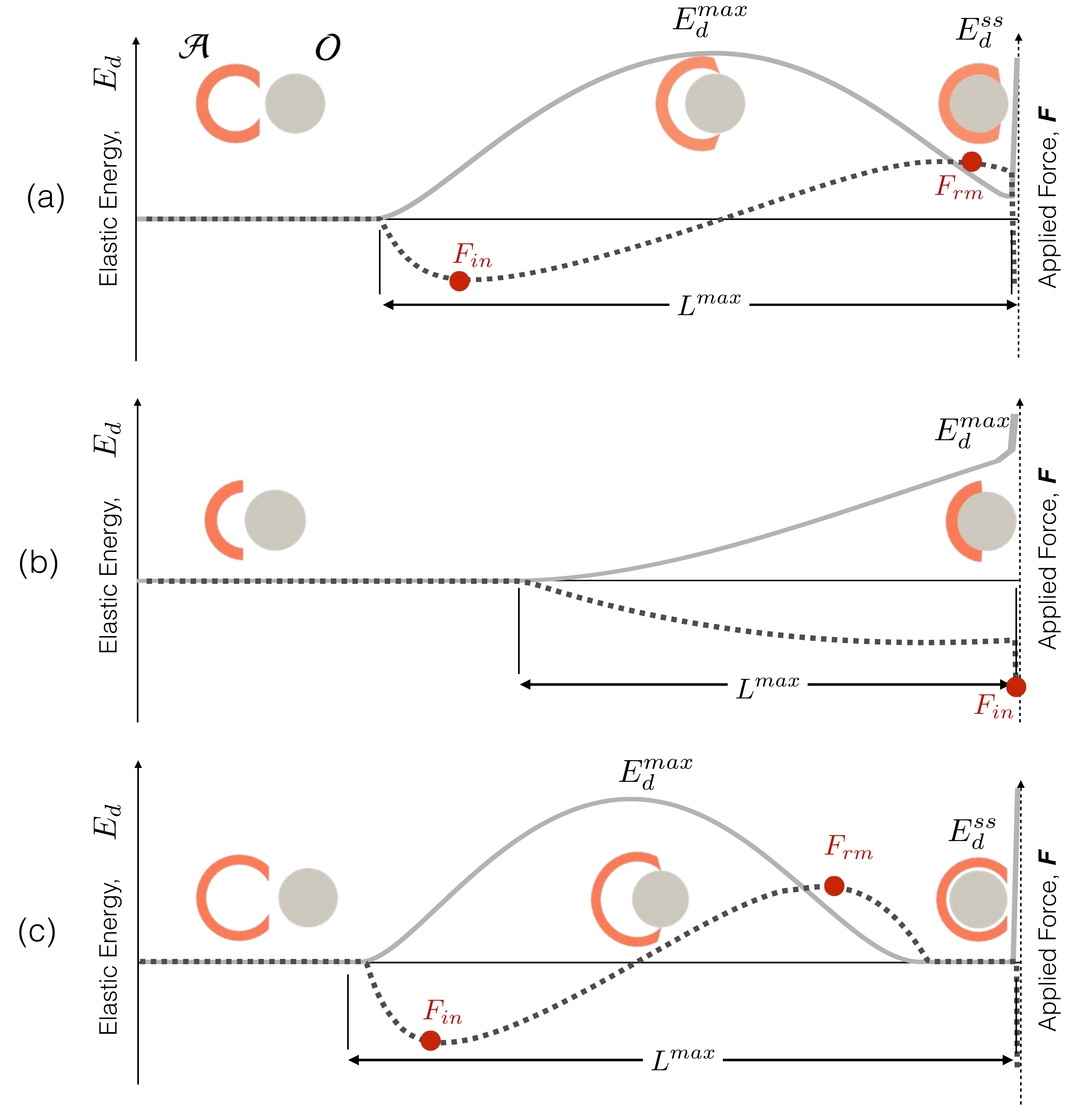}
  \caption{Deformation profiles for 3 different attachment categories~(a-c). Relative positions of the compliant structure and the inserted object corresponding to initial, maximum $E_d^{max}$, and steady state  $E_d^{ss}$ energies are shown. Solid and dashed curves represent the elastic energy and the applied force, respectively.}
  \label{fig:defEnergy}
\end{figure}

Given a compliant structure $\mathcal{A}$ and a rigid object $\mathcal{O}$ to be inserted, we optimize the rest shape of $\mathcal{A}$ (Figure~\ref{fig:overview}). We use deformation profiles to understand and characterize the coupling between $\mathcal{A}$ and $\mathcal{O}$  for the current hypothesis of $\mathcal{A}$.  

\subsection{Deformation Profiles} Figure~\ref{fig:realWorldAttachments} shows example compliant couplings ubiquitous in consumer facing products. In all such couplings, the fundamental considerations are (i) how easy it is to engage and disengage the objects, and (ii) how tightly the objects are locked together in the fully inserted states (as depicted in Figure~\ref{fig:realWorldAttachments}), with the obvious constraint that the compliant object should never break during its deformations.

As shown Figure~\ref{fig:defEnergy}, we use deformation profiles to capture both the elastic energy stored in the compliant object (solid line) as well as the resistive force it applies to the rigid object (dashed line) as a function of insertion distance.  Qualitatively, a distinct steady state minimum in the elastic energy  $E_d^{ss}$ that is attained after passing through a maximum energy state $E_d^{max}$ signals the presence of a valid coupling where no external force is necessary to keep the objects coupled (Figure~\ref{fig:defEnergy}(a),(c)). Furthermore, a non-zero $E_d^{ss}$ implies a grip force locking the objects together (Figure~\ref{fig:defEnergy}(a)), while a zero $E_d^{ss}$ implies a loose couple (Figure~\ref{fig:defEnergy}(c)). On the other hand, if a dip in the elastic energy is not present, this signals an engagement that would simply dissolve when the externally applied insertion forces are removed, \ie $\mathcal{A}$ pushes out $\mathcal{O}$ (Figure~\ref{fig:defEnergy}(b)).

\subsection{Assumptions} We assume $\mathcal{O}$ is rigid (does not deform), the engagement and disengagement are quasi-static, and the interactions are frictionless. As a result, all interaction forces develop exclusively due to the energy stored in $\mathcal{A}$. We discuss the impact of friction in the results section. We assume the mass center of $\mathcal{O}$ cannot be pushed beyond  $\mathcal{A}$'s farthest point (leftmost side of $\mathcal{A}$ in Figure~\ref{fig:defEnergy}), hence capping the maximum theoretical deformation $\mathcal{A}$ can undergo. This results in a maximum insertion length $L^{max}$. Users may provide insertion lengths that are shorter than this theoretical maximum. 

\subsection{Determining $E_d^{ss}$} We perform insertion simulations until $L^{max}$. Backtracking from the final state, we identify $E_d^{ss}$ as the first local minimum in the elastic energy profile. 

\subsection{Force Characterization} 

\paragraph{Insertion and Removal Forces} With the assumption of quasi-static coupling, the insertion force is equivalent to the resistive
force applied by $\mathcal{A}$ to $\mathcal{O}$, and can be computed as the sum of forces at the fixed boundary nodes of
$\mathcal{A}$. In Figure~\ref{fig:defEnergy}, we consider forces that resist the insertion of $\mathcal{A}$ to be negative for notational
convenience\footnote{Hence positive forces imply $\mathcal{A}$ drawing in $\mathcal{O}$.}. When disengaging the couple, the traversal is reversed.
As a result, from the user's perspective, the negative forces (\eg the maximum negative) quantify the difficulty one experiences during 
insertion, while the positive forces quantify the difficulty during removal.

\paragraph{Grip Forces} Insertion and removal forces are different than the grip force that one usually attributes to how strongly the two objects
are interlocked at $E_d^{ss}$. We define grip forces as a function of \emph{all} the forces that  $\mathcal{O}$ experiences due to its contact with $\mathcal{A}$ at $E_d^{ss}$ (Section~\ref{Optimization}). In effect, the grip force characterizes the degree of squeeze $\mathcal{A}$ imparts on $\mathcal{O}$.

\section{Insertion Simulation}

We represent the compliant attachment~$\mathcal{A}$ using a triangle mesh or a tetrahedral mesh~$\mathcal{M}$ in 2D and 3D cases. At each insertion step, $\mathcal{O}$ is displaced by a prescribed step $h_{in}$ and the corresponding deformed state of $\mathcal{A}$ is computed. $\mathcal{A}$ deforms to attain a minimum energy state while all  penetrations into $\mathcal{O}$ are precluded: 

\begin{equation}
\label{eq:optimization}
\begin{aligned}
& \underset{x}{\text{minimize}}
&&E_d(x, X) \\
& \text{subject to}
&& \psi(x_j) < 0 ,\\
\end{aligned}
\end{equation}

\noindent where $x$ and $X$ denote the deformed and rest states of $\mathcal{A}$, as column vectors with concatenated vertex positions. $E_d(x)$ and $\psi(x_j)$ are elastic energy and penetration functions, respectively. $\psi(x_j)$ is computed per vertex, $x_j$, denoting the position vector of a single vertex. 

\begin{figure}
  \centering  
  \includegraphics[trim = 0in 0in 0in 0in, clip, width=3.3in]{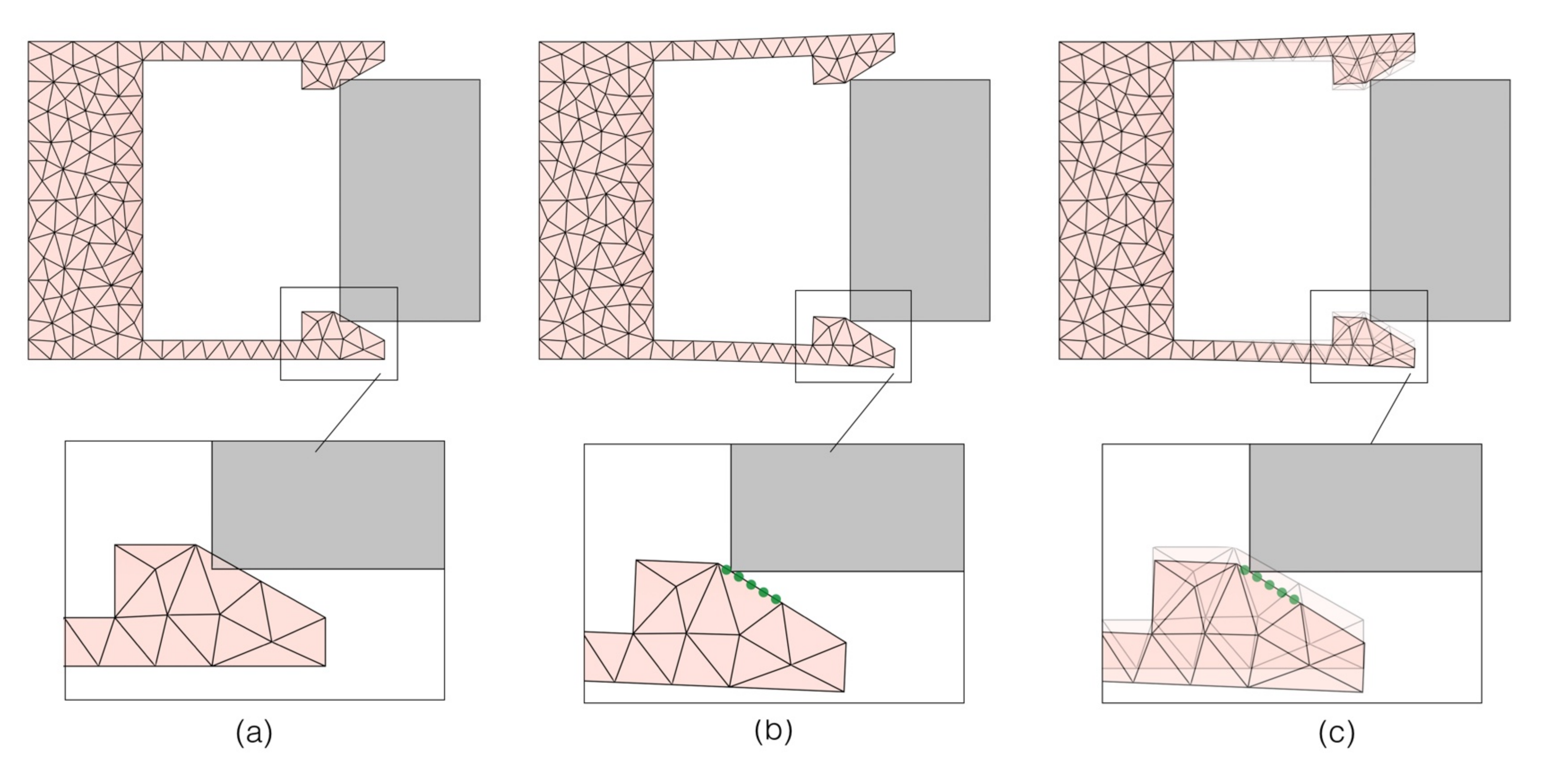}
  \caption{Effect of refinement. All vertices of $\mathcal{A}$ are out of $\mathcal{O}$ and the collision is not detected (a). With  refinement, we  interpolate new points on the colliding edges and check for collisions  (b). Solutions with and without refinement are overlaid in (c).}
  \label{fig:refinement}
\end{figure}

\subsection{Finite Element Model}
\label{sec:fem}

We find the elastic energy of $\mathcal{A}$ at the deformed state using finite element analysis. We use the rotation invariant Neo-Hookean material model to accommodate large deformations. While applicable to higher order element types, we use linear shape functions for simplicity. Using the Neo-Hookean material model and linear shape functions, the elastic energy is:

\begin{equation}
\label{eq:deformation_energy}
\begin{aligned}
E_d(x) = &V [ \frac{\mu}{2}(\|F\|_{\mathsf{F}}^{2} -\kappa_d)  - \mu log(det(F))\\
 &+ \frac{\lambda}{2} log^2(det(F))],
\end{aligned}
\end{equation}

\noindent where $\mu$ and $\lambda$ are Lam\`{e} parameters describing the material dependent  stress and strain relationship. $V$ and $\kappa_d$ are volume and dimension constants which is 2 for 2D triangular elements and 3 for 3D tetrahedral elements. $F = F(x,X)$ denotes the deformation gradient (\textit{i.e.~} $F = dx/dX$) as a function of the current state, $x$, and the rest state, $X$, of $\mathcal{A}$. Further details can be found in \cite{Sifakis:2012}.

The gradient of $E_d(x,X)$ is computed using the first Piola-Kirchhoff stress tensor. We  compute the reaction forces at the nodes of $\mathcal{A}$ using the gradient as $f = -\partial E_d(x,X)/ \partial x$. This way, we can compute the insertion forces from the reaction forces based on static equilibrium conditions. 

We use the von Mises failure criterion to determine if $\mathcal{A}$ fails ($\sigma_{vm}< \sigma_{yield}$). We compute $\sigma_{vm}$ from the Cauchy stress, $\sigma_{cauchy}$, by utilizing the already computed first Piola-Kirchoff stress tensor, $P$ with the following relation

\begin{equation}
\label{eq:cauchy}
\sigma_{cauchy} = (det(F))^{-1} P F^{T}.
\end{equation}

\begin{figure}
  \centering  
  \includegraphics[trim = 0in 0in 0in 0in, clip, width=\columnwidth]{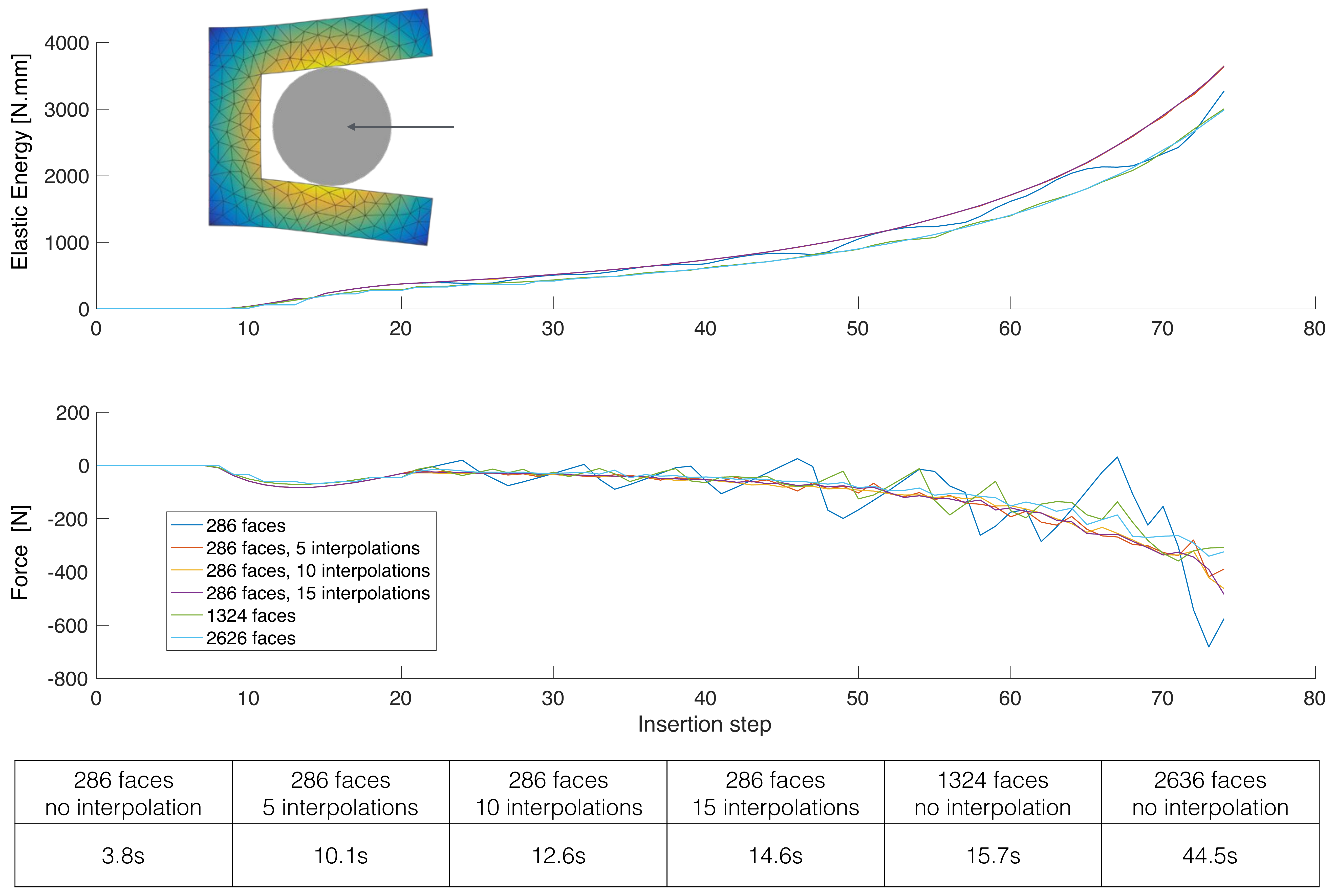}
  \caption{The effect of refinement, where smooth insertion forces are expected. The color on the attachment corresponds to the signed distance field of the circle. Note that  refinement results in smoother force curves as well as requiring less computation compared to higher mesh resolutions.}
  \label{fig:refinementPlot}
\end{figure}

\subsection{Avoiding Penetration}

In order to expel the penetrating vertices of $\mathcal{A}$ out of $\mathcal{O}$, we compute the shortest distance between all vertices of $\mathcal{A}$ and $\mathcal{O}$. This way, we quantify how much $\mathcal{A}$ has penetrated into $\mathcal{O}$. For this, we use the implicit moving least squares method (IMLS) to define a signed distance field on $\mathcal{O}$. This field is algebraically differentiable, thereby making it suitable for optimization. Using Kolluri's~\cite{Kolluri:2005} implicit surface representation, our penetration function becomes

\begin{equation}
\label{eq:IMLS}
\psi(x_j)= -\frac{\sum\nolimits_{i} n_{i}^{T}(x_{j}-v_{i})\phi_{i}(x_j)}{\sum\nolimits_{i}\phi_{i}(x_j)},
\end{equation}

\noindent where $\phi_{i}(x_j)= e^{- \| x_{j}-v_{i} \| ^{2} / \sigma^{2} }$ denotes the Gaussian kernel. $x_j$ is the position vector of a vertex on $\mathcal{A}$, $v_{i}$ and $n_{i}$ are the position and normal vectors of the $i^{th}$ vertex on $\mathcal{O}$.  $\sigma$ is set empirically per $\mathcal{O}$.

\subsection{Finding Deformed States}

\begin{figure}
  \centering  
  \includegraphics[trim = 0in 0in 0in 0in, clip, width=\columnwidth]{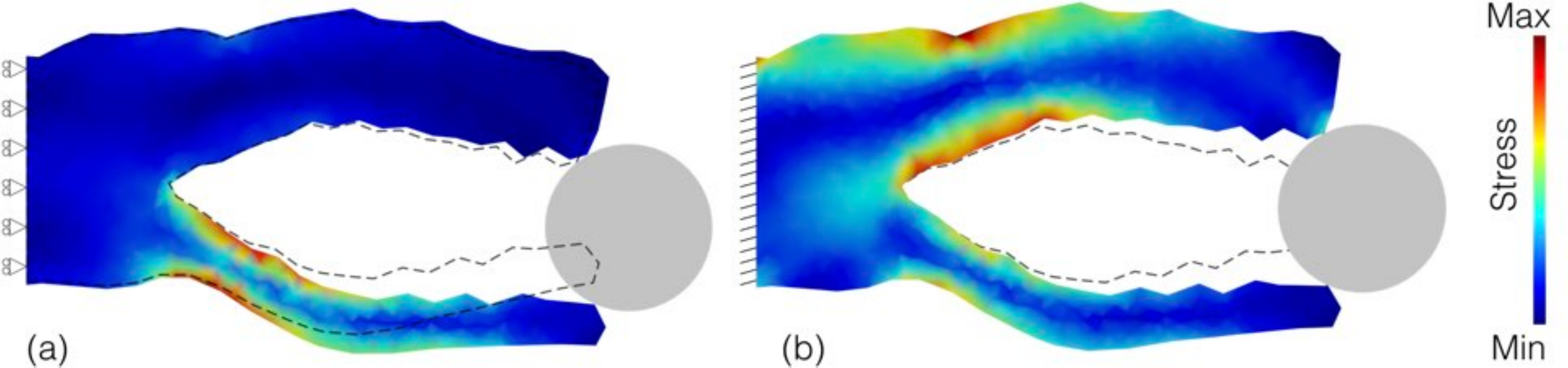}
  \caption{Effect of boundary conditions on asymmetrical attachments. The alligator is fixed on the left side. When only the insertion axis is fixed and the object is left free along the vertical axis, the results are more realistic stressing the thinner lower jaw (a). When all axes are fixed, the object pushes the thick upper jaw requiring larger insertion forces and causing higher energy states.}
  \label{fig:boundary}
\end{figure}

The deformation of a compliant attachments is governed by Eq.~\eqref{eq:optimization}, which is nonlinear in both the objective function and the constraints. At each insertion step, we employ Sequential Quadratic Programming (SQP)~\cite{Nocedal:2006} to solve it. We use MOSEK~\cite{mosek2015} to solve arising sub quadratic problems.

\begin{algorithm}
 \SetAlgoLined
 Solve \eqref{eq:optimization} with the vertices of $\mathcal{A}$\;
 \For{each edge on the boundary of $\mathcal{A}$}{
  \If{edge is in contact region}{
   \If{edge is in collision with $\mathcal{O}$}{
    add the edge into the contact edge list, $C_{L}$\;
   }
  }
 }
 \If{$C_L$ not empty}{
  \For{each edge in $C_L$}{
    interpolate new points and add to constraint list $v_c$\;
  }
  set initial conditions as the result of initial optimization\;
  solve \eqref{eq:optimization} with added constraints, $v_c$\;
 }
 \caption{Energy minimization with  refinement }
 \label{alg:refinement}
\end{algorithm}

\paragraph{Refinement} When $\mathcal{A}$ is represented with a coarse mesh,  collision checks involving only the vertices of $\mathcal{A}$ may not be sufficient to prevent penetration (Figure~\ref{fig:refinement}). Moreover, these penetrations may result in oscillatory force profiles with poor accuracy as shown in Figure~\ref{fig:refinementPlot}. As a remedy, we use the refinement described in Algorithm~\ref{alg:refinement}, where we check for collisions across the edges of $\mathcal{A}$ close to the contact regions and sample the new vertices across those edges if intersections are found. The newly sampled vertices are added as additional penetration constraints. The same approach extends to 3D by sampling across the faces of  $\mathcal{A}$. To determine the edges/faces to refine, we use the distance field already computed during the initial optimization step. We also use the IMLS surface field to check for collisions. Figure~\ref{fig:refinementPlot} illustrates the effect of refinement for various mesh and interpolation settings. In our examples, we use 10 interpolation points. 

Oscillatory or non-smooth contact forces that arise due to discrete penetration detection, contact boundary smoothness and finite element discretization are an open problem in finite element analysis. To overcome this challenge, a stabilization scheme for small deformations~\cite{Deuflhard:2008}, a continuous penalty force approach~\cite{Tang:2012}, contact based remeshing, and smoothing contact boundaries through Bezier patch approximations~\cite{Kloosterman:2002:contact} have been proposed. Unfortunately, these approaches severely impact the computational performance especially as our insertion simulations are run for each shape hypothesis of $\mathcal{A}$ being optimized. We thus limit our improvements to edge/face refinements, which only reduce contact force oscillations through a better approximation of continuous contact.  However, arbitrary shapes with non-smooth boundaries and finite element discretization still produce non-smooth force profiles (Figure~\ref{fig:refinementPlot}).

\paragraph{Boundary and Initial Conditions} During contact simulations, $\mathcal{O}$ is constrained to move along a prescribed insertion path (\eg horizontally left in (Figure~\ref{fig:boundary}). $\mathcal{A}$ is displacement-constrained only along the insertion axis while leaving the remaining two orthogonal axes unconstrained (vertical and out of page). This allows non-symmetrical rigid objects or compliant structures to be coupled in ways that minimize the total elastic energy without introducing artificial barriers to natural accommodative movements. As shown in Figure~\ref{fig:boundary}, this enables a more realistic identification of the true stress state of $\mathcal{A}$.

Users may define custom insertion paths for $\mathcal{O}$ involving both translations and rotations when desired.  However, unless specified otherwise, we assume the objects are inserted purely translationally without any rotations.

Due to the non-linear nature of ~\eqref{eq:optimization}, seeding it with a suitable initial condition is crucial for achieving robustness. We use the deformed state of the previous step as the initial condition for the current step.

\section{Rest Shape Optimization}

The insertion simulation allows its results to inform shape optimization on $\mathcal{A}$ to produce a desired coupling behavior. 

\subsection{Shape Modification}

To alter $X$, the rest shape of $\mathcal{A}$, we use linear blend skinning~(LBS) with bounded bi-harmonic weights~\cite{Jacobson:2011:BBW}. Using bounded bi-harmonic weights, we achieve smooth modifications in a localized and shape-aware manner.  We use translation and scaling  edits to facilitate a fine control over the structural dimensions of  $\mathcal{A}$. For a mesh $\mathcal{M}$ with $n$ vertices in $ \mathbb{R}^d (d = 2~ or~ 3)$ and $m$ modification handles

\begin{equation}
\label{lbs_mat}
X' = M_{LBS} T ,
\end{equation}

\noindent where $X' \in \mathbb{R}^{n \times d}$ is the matrix with the modified vertex positions in the rows, $M_{LBS} \in \mathbb{R}^{n \times ((d+1)m)}$ is the linear blending skinning matrix computed once for the original mesh $\mathcal{M}$, and $T \in \mathbb{R}^{((d+1)m \times d)}$ is a stack of handle transformation matrices that includes translations and scaling for each handle. For brevity, we refer to the translation and scale of the transformation handles as $h$.

The LBS formulation can be equivalently expressed in the following form

\begin{equation}
\boldsymbol{x}' = \sum\nolimits_{j} w_{j}(\boldsymbol{x}) T_j \boldsymbol{x}
\end{equation}

\noindent where $T_j$ represents the affine transformations of a modification handle $h_j,$ $j=1,...,m$. When transposed  $T_j$  matrices are stacked together, they form matrix $T$ in Eq.~\eqref{lbs_mat}. All vertex positions, $\boldsymbol{x}$, of $\mathcal{M}$ are modified as $\boldsymbol{x}'$. $w_{j}$ is the weight functions associated with the handle $h_j$ and we compute them using bounded bi-harmonic weights~\cite{Jacobson:2011:BBW}. These weights sum up to 1 at each vertex. Each handle has the maximum effect around its immediate neighborhood ($w_j=1$ at the handle) and its influence disappears at distant regions. $M_{LBS}$ matrix is computed combining vertex positions $\boldsymbol{x}$ with vertex weights $w_{j}(\boldsymbol{x})$.

\begin{figure}
  \centering  
  \includegraphics[trim = 0in 0in 0in 0in, clip, width=0.8\columnwidth]{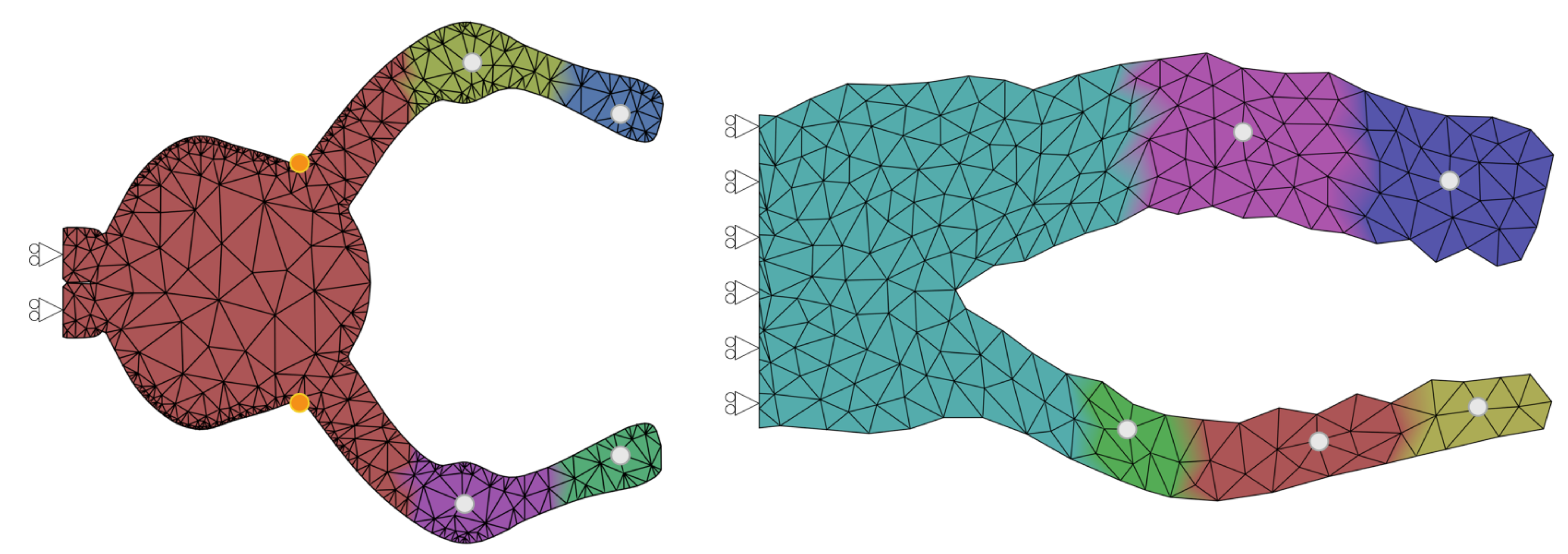}
  \caption{Handle placement through displacement based clustering. The left most edges of the objects are fixed as boundary conditions and we do not place handles on the corresponding clusters.}
  \label{HandlePlace}
\end{figure}

\begin{figure}
  \centering  
  \includegraphics[trim = 0in 0in 0in 0in, clip, width=\columnwidth]{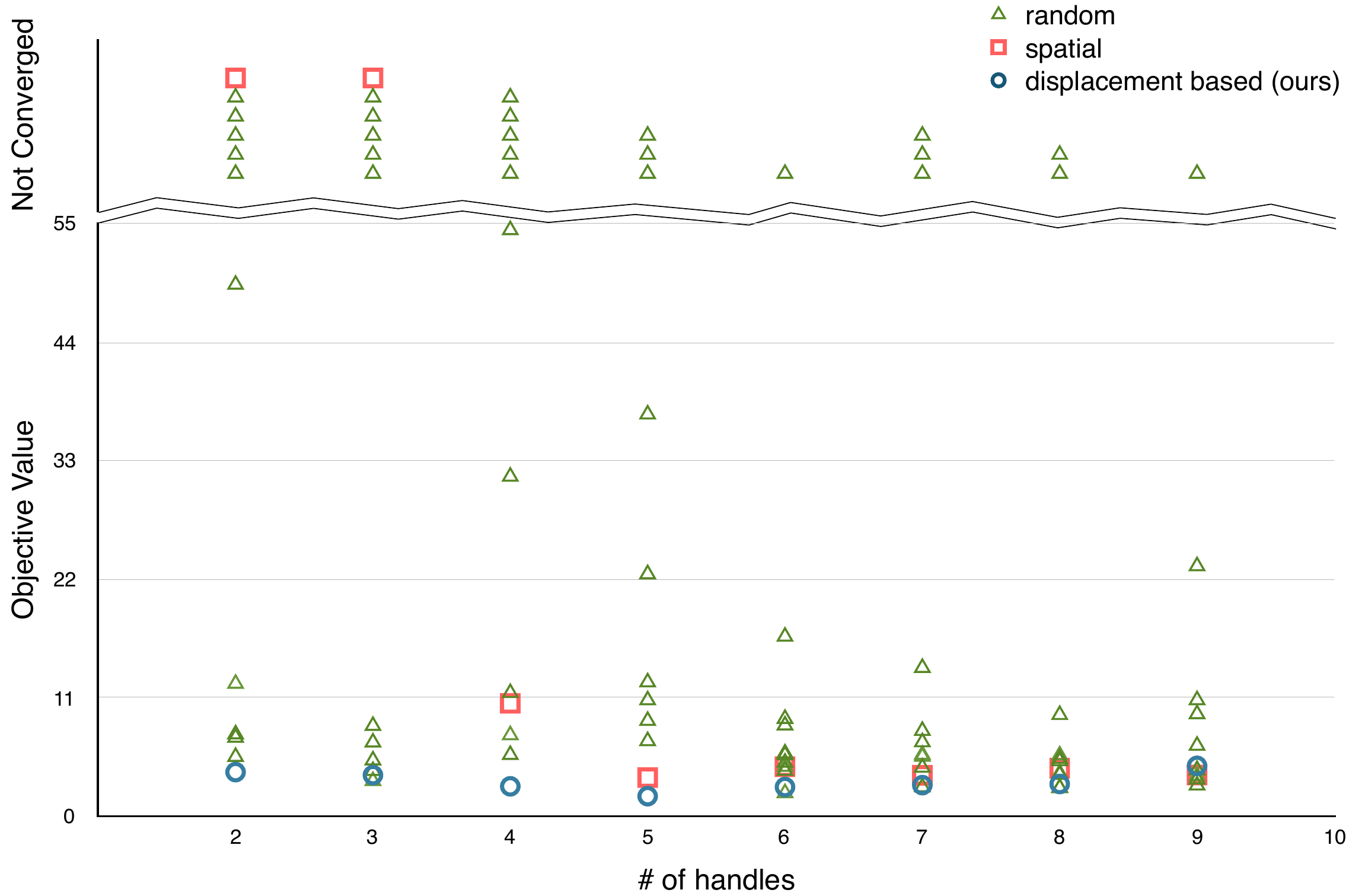}
  \caption{Displacement based, spatial, and random handle placements. Above the broken line, we report the handle placements that do not converge.}
  \label{HandlePlaceStudy}
\end{figure}

\paragraph{Displacement Based Handle Placement} To place the deformation handles $h$ within  $\mathcal{A}$, we perform an initial insertion simulation with the original $\mathcal{A}$. During the insertion process we record the maximum displacement each vertex undergoes. Then, we cluster the nodes of $\mathcal{A}$  using $\mathcal{M}$ as the connectivity graph, where the similarity between neighboring nodes is calculated based on their maximum displacements. In our implementation, we use the affinity propagation clustering method~\cite{Frey:2007}. We take the vertex closest to the centroid of a cluster as the deformation handle for that cluster~(Figure~\ref{HandlePlace}). We omit the clusters that include the displacement boundary conditions as they represent the clusters with minimal deformations. Nevertheless, we add displacement-constrained vertices as inactive handles to keep those regions unmodified during shape modifications.  Note that this only affects shape updates, and not the nature of the boundary conditions during insertion simulations. Additional inactive handles~(highlighted in orange) may be added to keep certain parts of the shape unchanged (refer to Section 6.1: Grip Control).

Figure~\ref{HandlePlaceStudy} shows the impact of various choices for $h$ for the same initial structure, objective function, and constraints (defined in Section~\ref{Optimization}). For each number of handles, we positioned them using our approach,  using spatial clustering where vertex coordinates are used for similarity, and using 10-fold random pick of the handles. The results  suggest that the handles do not have to be placed precisely. In particular, as the number of handles increases, up to 90\% of the results expectedly converge even with random handle placement. However, our approach converges to a better solution that creates a minimal deviations  in the shape. Spatial clustering also provides good handle placement and can be used if the initial insertion simulation cannot be performed.

\subsection{Optimization}
\label{Optimization}
\begin{figure}
  \centering  
  \includegraphics[trim = 0in 0in 0in 0in, clip, width=4.5in]{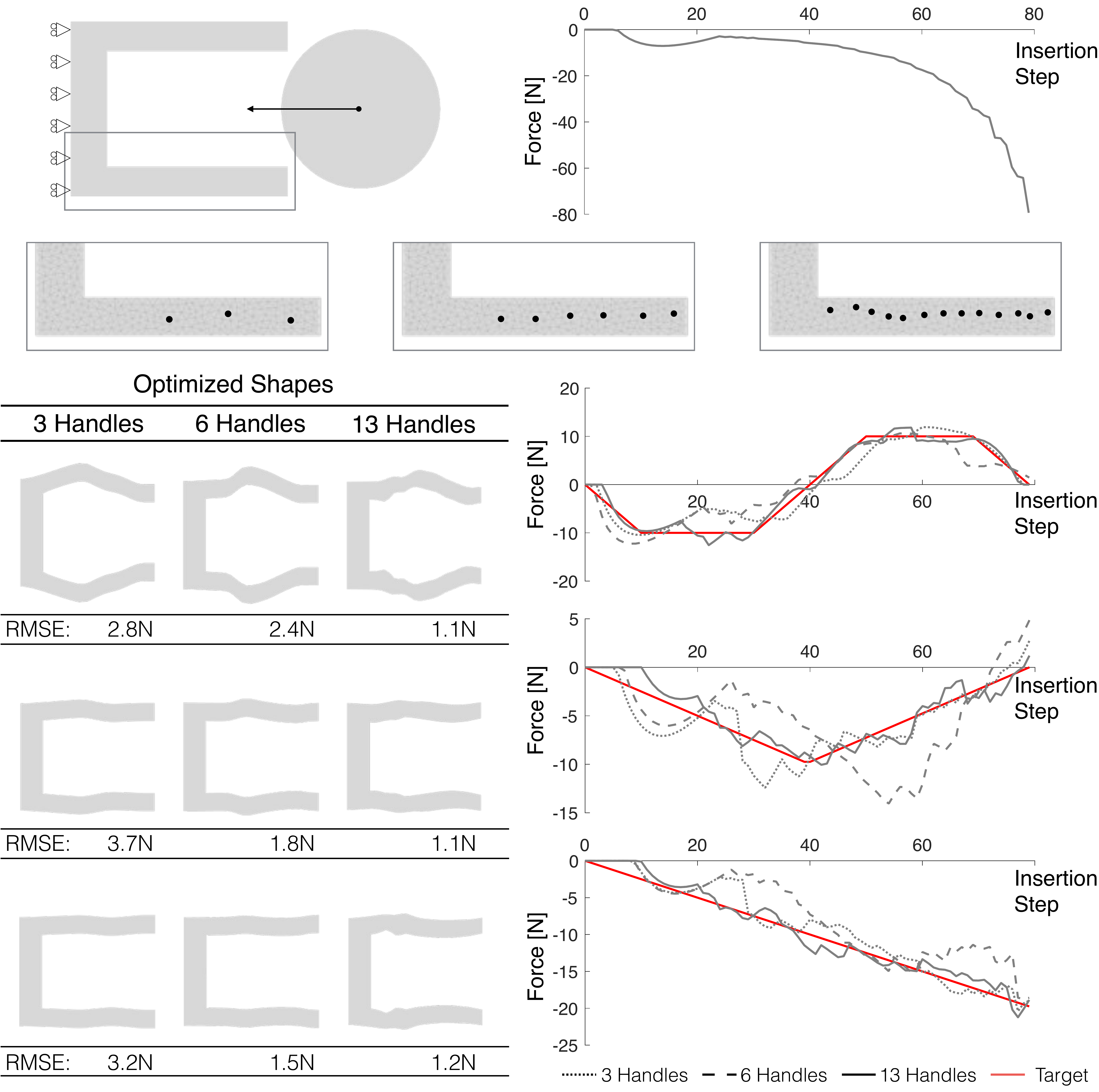}
  \caption{The compliant structure and its original force profile (top) is optimized to match prescribed force profiles. For different number of modification handles (middle),  the target and the optimized force profiles are shown together with root-mean-square errors (bottom). Resulting force profiles show that our approach can alter the structure to match the target force profiles. This image is best viewed digitally.}
  \label{curveMatch}
\end{figure}

For the current shape hypothesis of  $\mathcal{A}$, insertion simulations are performed producing the energy and force profiles. The relevant quantities are extracted from these profiles to compute the prescribed objective values. The deformation handles $h$ are then used to deform $\mathcal{A}$ in a way that improves  the objective function. Figure~\ref{curveMatch} shows a scenario where the objective is to match the prescribed force profiles as close as possible by minimizing the squared error between the target and the resulting force profiles. As shown, the target profiles can be matched arbitrarily closely with an increased number of handles, essentially allowing our approach to design custom stiffness profiles for arbitrary objects. 
 
 To facilitate high-level end-user specification of coupling behavior, we formulate our objective function as minimizing the change in the rest shape of $\mathcal{A}$ as measured by the \textit{modification energy}, using $h$ as the design variables, subject to the functional constraints of: (i) \textit{coupling ratio}, (ii) \textit{grip force}, (iii) \textit{insertion and removal forces}, and (iv) \textit{material failure}.

\paragraph{Coupling Ratio} For a pair of objects to remain stably coupled, the elastic energy profile must exhibit a peak $E_{d}^{max}$, before settling in a lower steady state energy $E_{d}^{ss}$ (Figure~\ref{fig:defEnergy}). If the value of $E_{d}^{ss}$ is too close to the value of $E_{d}^{max}$, small perturbations could cause the couple to disengage abruptly by causing $\mathcal{A}$ to eject $\mathcal{O}$. We thus define a stable coupling ratio constraint as follows: 

\begin{equation}
\label{eq:stayIn}
E_{CR} \equiv (E_{d}^{ss} / E_{d}^{max})*100 - \kappa_{cr} < 0,
\end{equation}

\noindent where $\kappa_{cr}$ is a percentile introduced to keep the inserted object inside the compliant structure. $\kappa_{cr}$  can take on two different values: (I) $\kappa_{cr} = 95\%$~if the user desires a non-zero grip, or (II) $\kappa_{cr} = 10^{-3}\%$~if the user desires a loose couple. While the actual values are set empirically, the key here is to enable a binary decision between tight versus loose grips.  If the former is elected, $E_{CR}$ simply ensures that there is a stable configuration that allows a valid coupling. The precise strength of the grip is determined by the grip forces as described next. For energy profiles that do not initially attain a steady state value~(Figure~\ref{fig:defEnergy}(b)), we use the maximum energy as the steady state. This way, $E_{CR}$ will be violated for the current $\mathcal{A}$, hence penalized during optimization, thereby helping to impart shape changes in $\mathcal{A}$ that lead to stable coupled steady states $E_{d}^{ss}$.

\paragraph{Grip Force} Qualitatively, the grip force is a measure of how firmly $\mathcal{A}$ squeezes $\mathcal{O}$ when there are no other external forces and both parts are in static equilibrium. To quantify the grip, we decompose it into three components with the $x$ component defined as: $\boldsymbol{F}_g^x = 0.5\sum_j{|f_j^x|} \quad
\forall j \in \mathcal{M}$, where  $f_j^x$ is the $x$ component of the external force on vertex $j$ imparted by $\mathcal{O}$.  Note that only a subset of the boundary vertices of $\mathcal{A}$ would contribute to $\boldsymbol{F}_g^x$. $\boldsymbol{F}_g^y$, and $\boldsymbol{F}_g^z$ are defined analogously. We finally define $F_g$ to be the $L2$ norm of the resultant force vector $[\boldsymbol{F}_g^x$~$\boldsymbol{F}_g^y$~$\boldsymbol{F}_g^z]$ and define the grip force constraint as:

\begin{equation}
\label{eq:forcegrip}
\begin{aligned}
&E_{GT_1} \equiv  F_g^{lb} - F_g  < 0, \\
&E_{GT_2} \equiv  F_g  - F_g^{ub}  < 0, 
\end{aligned}
\end{equation}

\noindent where $F_g^{lb}$, $F_g^{ub}$ are the user specified lower/upper bounds on $F_g$.

\paragraph{Insertion and Removal Forces} We define the insertion and removal forces $F_{in}$, $F_{rm}$ to be the maximum of the forces experienced by $\mathcal{A}$ extracted from the deformation profiles (Figure~\ref{fig:defEnergy}) and constrain their values within user specified upper and lower bounds as follows: 

\begin{equation}
\label{eq:forceConst}
\begin{aligned}
&E_{F_1} \equiv F_{in} - F_{in}^{ub} < 0, \\
&E_{F_2} \equiv F_{in}^{lb} - F_{in} < 0, \\
&E_{F_3} \equiv F_{rm} - F_{rm}^{ub} < 0, \\
&E_{F_4} \equiv F_{rm}^{lb} - F_{rm}< 0,
\end{aligned}
\end{equation}

\paragraph{Material Failure} We consider yielding as the failure mode and use the von Mises stress criterion to constrain the stress developed in the compliant object during insertion. We enforce the yield criterion as a single constraint where we ask the maximum von Misses stress observed in the deformable object to be less than the yield stress.

\begin{equation}
\label{eq:materialFailure}
E_{MF} \equiv max(\sigma_{vm}) - \sigma_{yield} < 0
\end{equation}

\paragraph{Modification Energy} We use the Laplacian editing energy~\cite{Sorkine:2004} of the surface mesh to quantify the total modification energy, also used by~\cite{Prevost:2013,Bacher:2014}.

\begin{equation}
\label{eq:LaplaceEditing}
minimize~~~ E_{\mathcal{L}} \equiv {v_s}^{T}  M_{\mathcal{L}} v_s,
\end{equation}

\noindent where $M_{\mathcal{L}} = \mathcal{L}^T \mathcal{L}$ is a positive semidefinite matrix constructed using the surface Laplacian and calculated once for the original boundary mesh of $\mathcal{A}$. $v_s$ denotes the modified boundary vertices.

\begin{figure}
  \centering  
  \includegraphics[trim = 0in 0in 0in 0in, clip, width=\columnwidth]{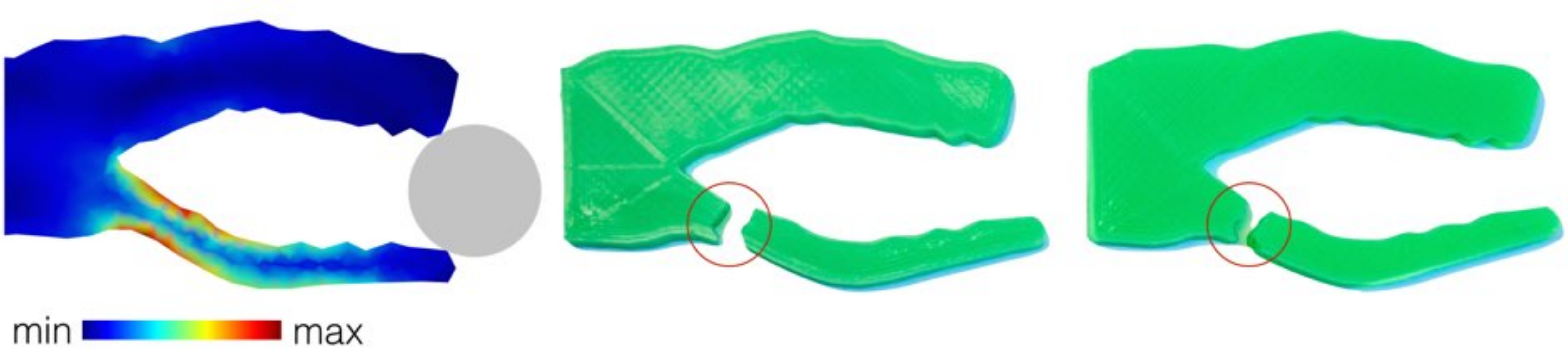}
  \caption{As our insertion simulations predict, the initial shape breaks at the high stress region.  Color plot shows the computed stresses ~(left). Two 3D prints of the initial design~(middle, right) consistently break at the same location.}
  \label{fracture}
\end{figure}

The optimization seeks to minimize $E_{\mathcal{L}}$ by modifying $\mathcal{A}$  through the design vector $h$, while satisfying the constraints $E_{CR}$, $E_{GT}$, $E_{F}$, and $E_{MF}$. We transform this into an unconstrained problem by augmenting the objective with constraints as penalties. For any constraint $c(h)$, we transform it into a penalty $c(h) \leftarrow max(0, c(h)) ^2$ and add its contribution using a large penalty constant.

While we aim to establish a congruent scale for all constraint terms (percentages:0-100, stress: $\sigma_{yield}$ around 60MPa for materials of interest, forces:1-100N range based on ergonomic limits~\cite{astin1999finger}), this scheme would need adjustments if the scales of interest vary significantly.

We use simulated annealing for optimization and implemented it as described in~\cite{Kirkpatrick:1983}. A pseudo-code of our rest shape optimization is given in Algorithm~\ref{alg:restShapeOpt}. In each iteration of the optimization, we generate new states by sampling a new design vector around the current vector drawn from a Gaussian distribution, with an adaptive decaying variance. For cooling schedules, we experimented with linear, logarithmic, exponential cooling and observed similar convergence for our problem. In all of our examples, we are using exponential cooling.

\begin{algorithm}
 \SetAlgoLined
$E_{curr} = E_{initial}$,  $h_{curr} = h_{initial}$\;
$E_{opt} = E_{initial}$, $h_{opt} = h_{initial}$\;
 \For{$i=1$ to $iterations_{max}$}{
  $h_i \leftarrow$ generate new neighbor state\;
  $T_{curr} \leftarrow$ temperature cooling\;
  $\mathcal{A}(h_i) \leftarrow$ update current shape hypothesis\;
  perform an insertion simulation with $\mathcal{A}(h_i)$\;
  $ E_{d}^{ss}, E_{d}^{max}, F_{g}, F_{in}, F_{rm} \leftarrow$ process deformation profiles\;
  $ E_{i} \leftarrow$ augment objective $E_{\mathcal{L}}$ with penalty constraints $ E_{CR}$, $E_{GT}$, $E_{F}$, $E_{MF}$\;
  
  \uIf{ $E_{i} < E_{curr}$ }{
    $E_{curr} = E_{i}$,  $h_{curr} = h_{i}$\;
   \If{ $E_{i} < E_{opt}$ }{
     $E_{opt} = E_{i}$, $h_{opt} = h_{i}$\;
   }
   }
   \ElseIf{$exp( (E_{curr}-E_{i} )/T_{curr} ) > rand(0,1) $}{
    $E_{curr} = E_{i}$,  $h_{curr} = h_{i}$\;
   }
 }
\Return $\mathcal{A}(h_{opt})$ 
 
 \caption{Rest Shape Optimization}
 \label{alg:restShapeOpt}
\end{algorithm}

\section{Results}
\label{sec:coupling:results}

We apply our shape optimization through deformation profiles on a variety of compliant structures with different behaviors.  We present further details in the accompanying video.

\subsection{Validation}

\paragraph{Insertion Simulations}

\begin{figure}
  \centering  
  \includegraphics[trim = 0in 0in 0in 0in, clip, width=0.9\columnwidth]{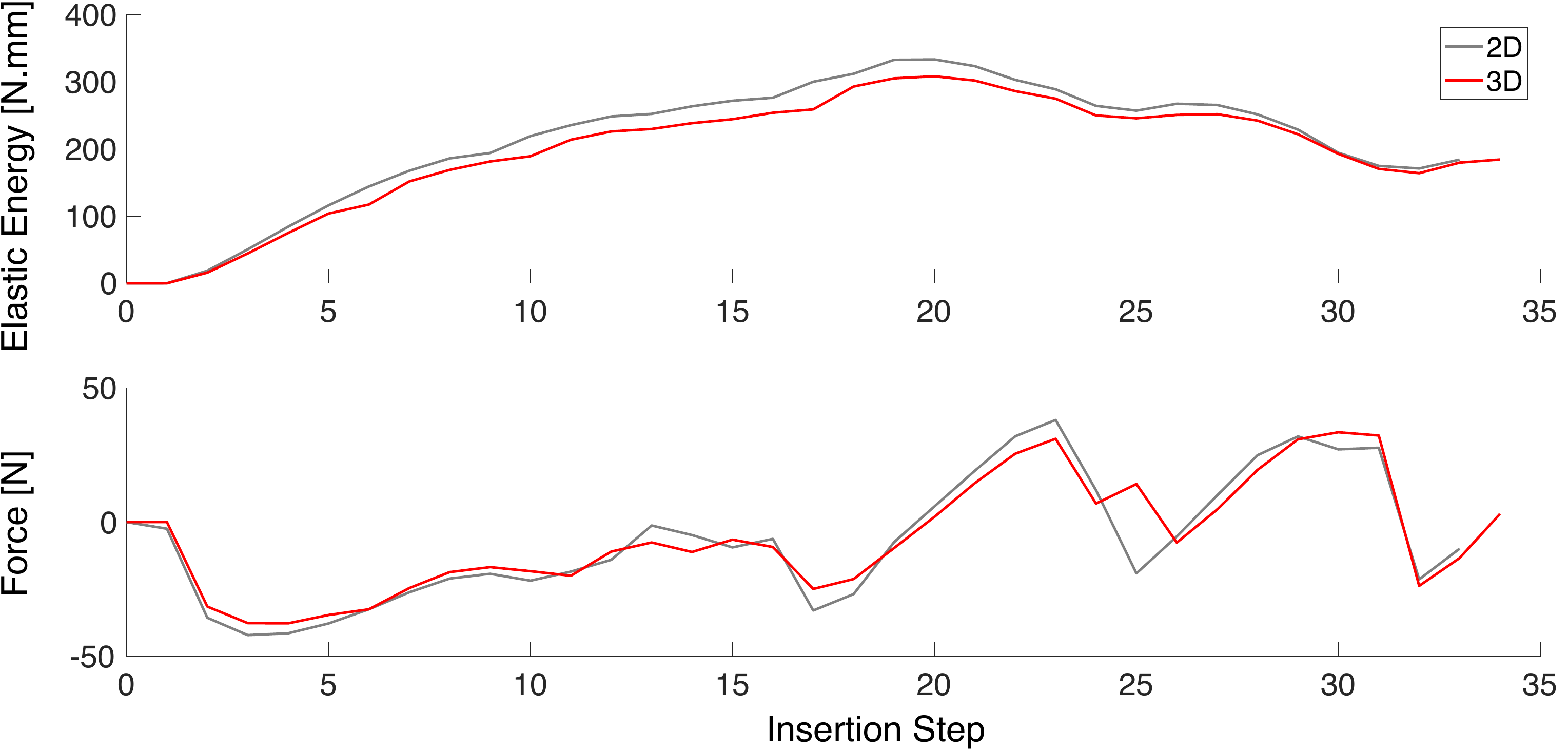}
  \caption{Comparison of insertion simulations for alligator model using 2D and 3D analysis. Plots are shown for the initial shapes.}
  \label{alligator2d3d}
\end{figure}

\begin{figure}
  \centering  
  \includegraphics[trim = 0in 0in 0in 0in, clip, width=\columnwidth]{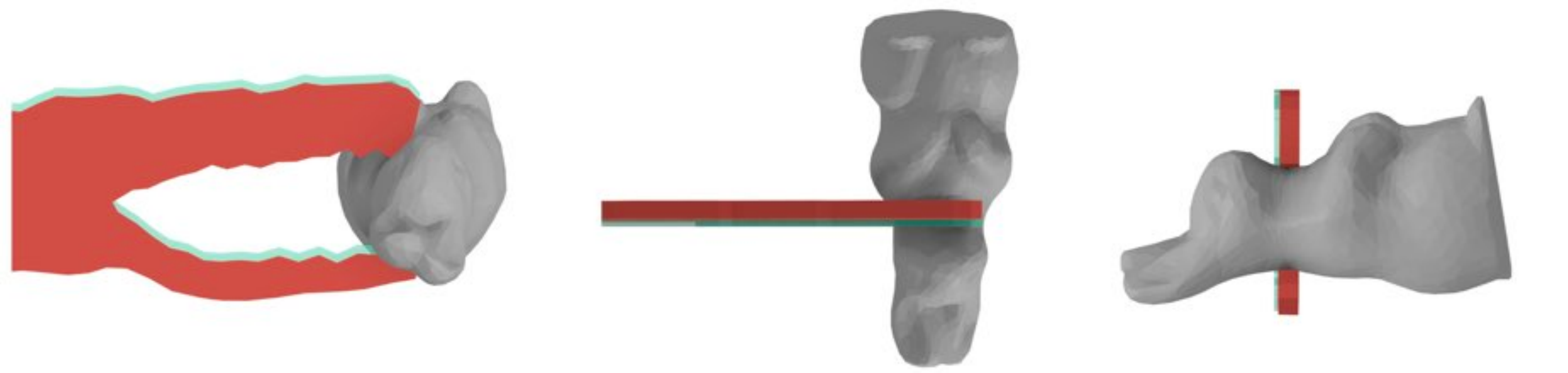}
  \caption{The shift in the alligator attachment during the insertion process from front, top and side views. The transparent green illustrates the original position.}
  \label{alligatorRabbitMove}
\end{figure}

For the alligator presented in Figure~\ref{fig:overview}, our simulations suggest that the initial unoptimized shape breaks during engagement with the stress distribution given in Figure~\ref{fracture}. Printed physical samples break at the maximum stress region predicted by our simulations, supporting the boundary condition specifications~(see the discussions of Figure~\ref{fig:boundary}). For this model, we also investigate the performance gain of a 2D analysis over a full 3D analysis. We obtain similar deformation profiles as shown in Figure~\ref{alligator2d3d}. With 2D analysis, a considerable reduction in computation time is obtained due to the decrease in dimensionality as well as a reduction in the total number of vertices and elements. See Table~\ref{tab:hresult} for a comparison of the computation time and the number of elements. 

Figure~\ref{alligatorRabbitMove} shows the bunny-in-alligator coupling. A vertical and out of page shift can be observed that correctly accommodates the tapered surface of the bunny. 

\paragraph{Force Control}

\begin{figure}
  \centering  
  \includegraphics[trim = 0in 0in 0in 0in, clip, width=\columnwidth]{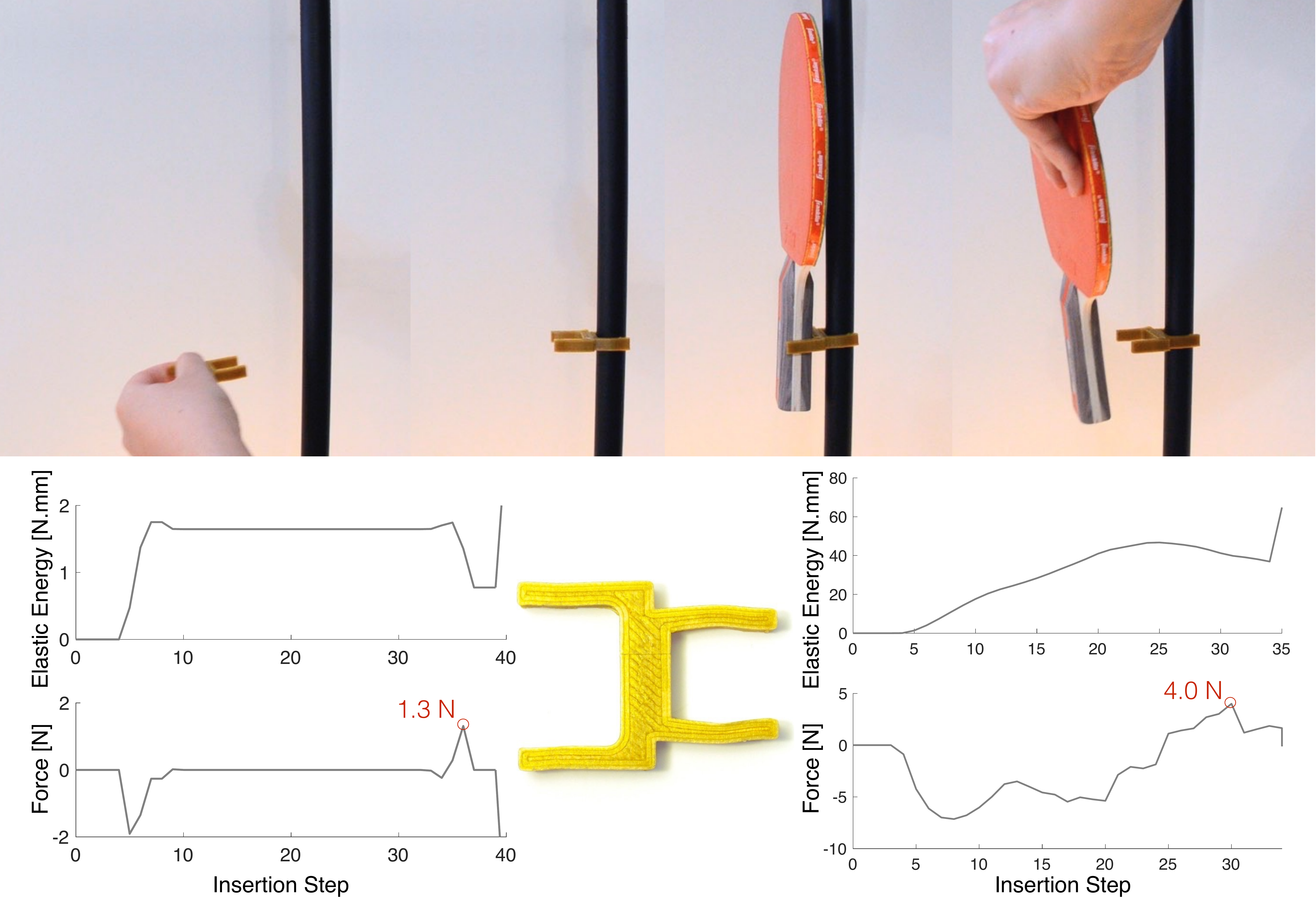}
  \caption{The paddle side is designed for a removal force less than the removal force on the stand side. The physical pull out of the paddle validates the design. Deformation profiles for both sides of the connector are also shown. Both sides have tight grips, but the paddle side requires a smaller removal force.}
  \label{paddle}
\end{figure}

In Figure~\ref{paddle}, we design a connector to attach a table tennis paddle to a cylindrical stand, with the paddle to be removed easily while the stand side remaining firmly attached. The resulting deformation profiles show this behavior with the stand side requiring a larger removal force. 

\paragraph{Grip Control}

\begin{figure}
  \centering  
  \includegraphics[trim = 0in 0in 0in 0in, clip, width=5in]{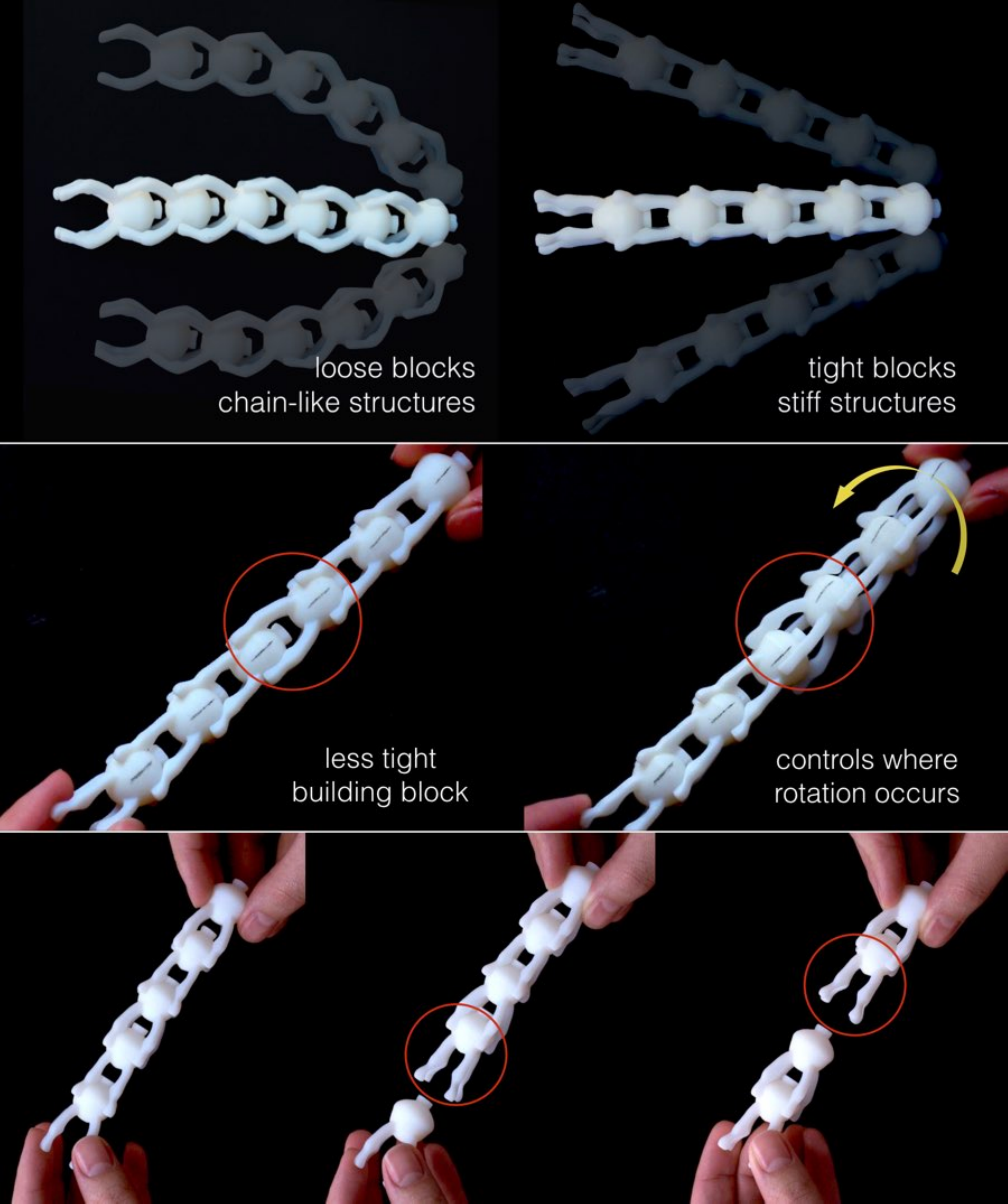}
  \caption{Starting from an initial geometry that barely stays together, we design building blocks with loose, tight and tighter grips. Using building blocks with different behaviors, we can build chain like structures, stiff structures~(top) or combine them to control the motion transfer~(middle). When we arrange the building blocks in an alternating order of tight and tighter grip ones, it separates from a block with tighter grip every time we pull from both ends~(bottom). Notice tighter grips do not necessarily correspond to higher insertion/removal forces.}
  \label{linklingPull}
\end{figure}

\begin{figure*}
  \centering  
  \includegraphics[trim = 0in 0in 0in 0in, clip, width=\textwidth]{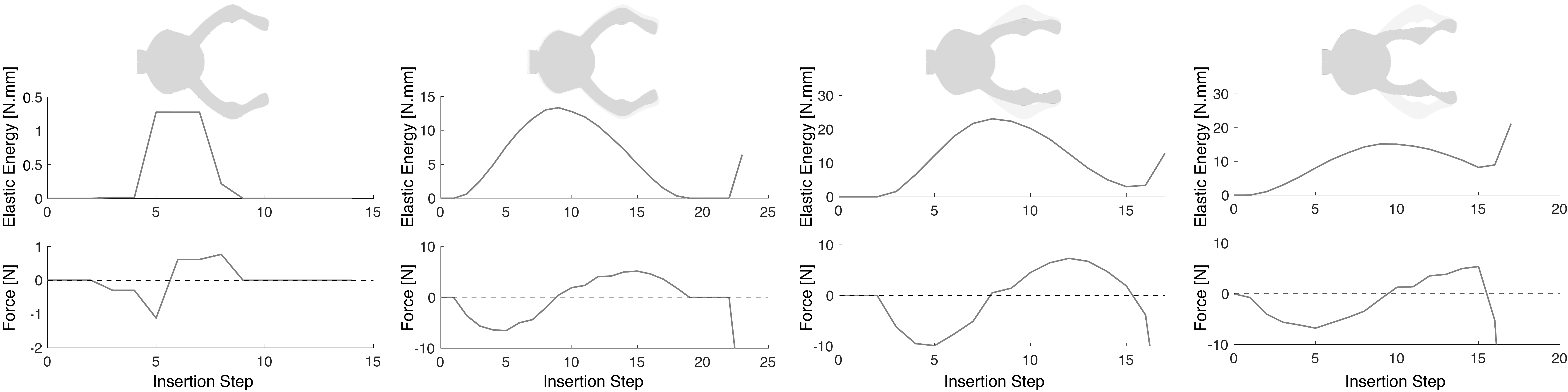}
  \caption{The initial shape is optimized through geometric changes to achieve loose, tight and  tighter grip~(from left to right). Deformation profiles are shown at the bottom with corresponding optimized shapes. Underlying silhouettes represent the initial geometries. Note that the loose design has zero steady state energy and the tight design has a non-zero steady state energy that is lower than the tighter design. }
  \label{linklingPlot}
\end{figure*}

We design a set of building blocks with fine-tuned physical grips as shown in Figure~\ref{linklingPull}. Our approach decouples the grip force from the insertion and removal forces, thereby creating coupling behaviors that are not easily achievable through  intuition alone. The deformation profiles capture the different grips as well as the tighter block interestingly requiring less insertion and removal forces~(Figure~\ref{linklingPlot}).  For this chain structure, to keep the rigid, inserted part  identical in all designs, we add fixed handles at the base of the arms so that the spherical body remains unchanged. We optimized the building blocks using a simplified 2D analysis and synthesized the 3D versions by revolving the base and replicating the optimized arms.

\FloatBarrier

\begin{figure}
  \centering  
  \includegraphics[trim = 0in 0in 0in 0in, clip, width=\columnwidth]{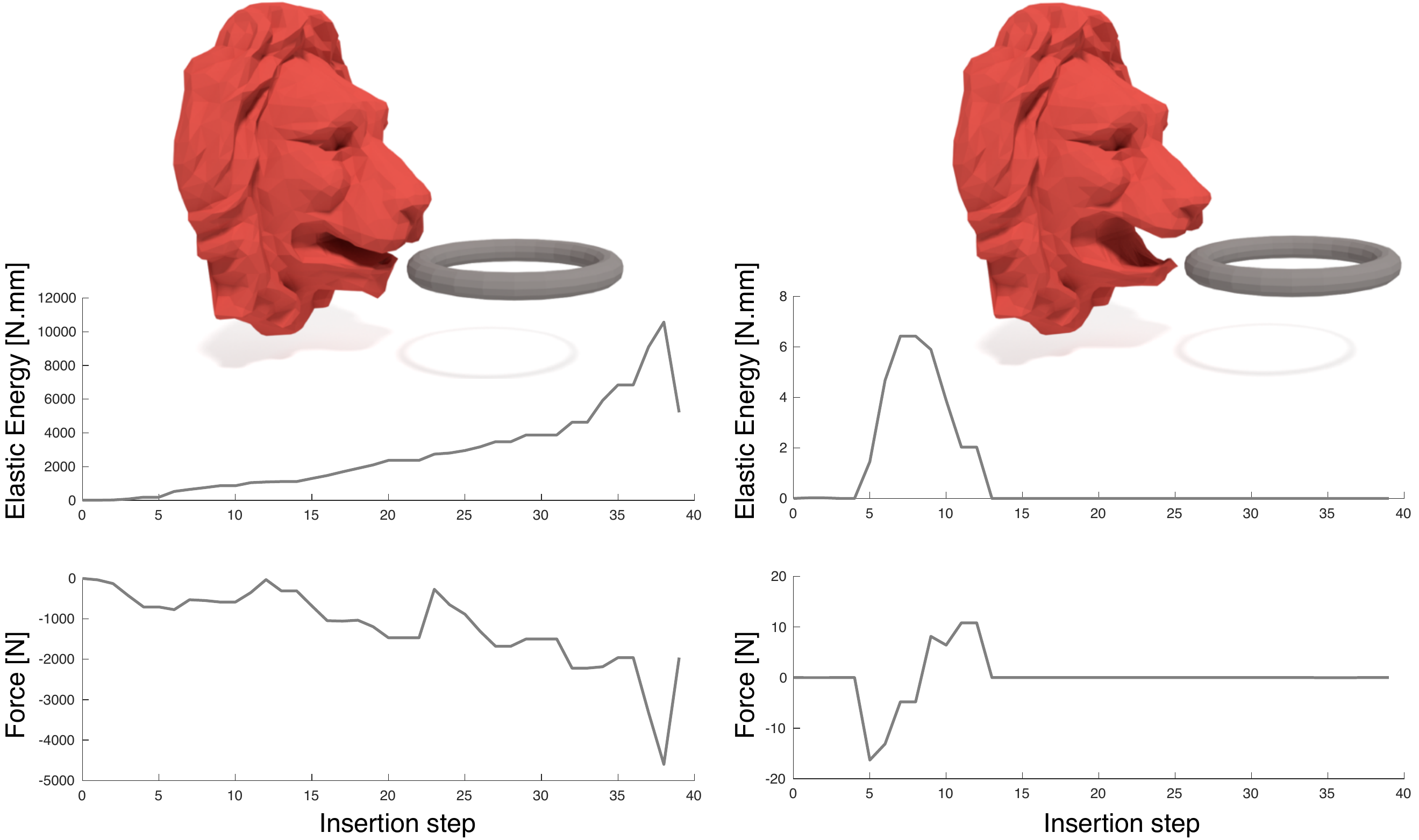}
  \caption{Initial~(left) and optimized~(right) lion head models with their corresponding insertion simulations.}
  \label{lionPlot}
\end{figure}

Figure~\ref{lionPlot} shows a door knocker that exhibits a loose grip. To simulate the coupling process, we use a rotating trajectory where the ring is initially inserted horizontally and rotated afterwards following the natural knocking motion. The initial shape not only requires excessive insertion forces on the order 4000~N, but it also readily yields with stresses up to 1425~MPa. When optimized, the insertion force is reduced down to 20~N, with a von Mises stress of 41~MPa (less than 60~MPa of yield stress).

\begin{figure}
  \centering  
  \includegraphics[trim = 0in 0in 0in 0in, clip, width=\columnwidth]{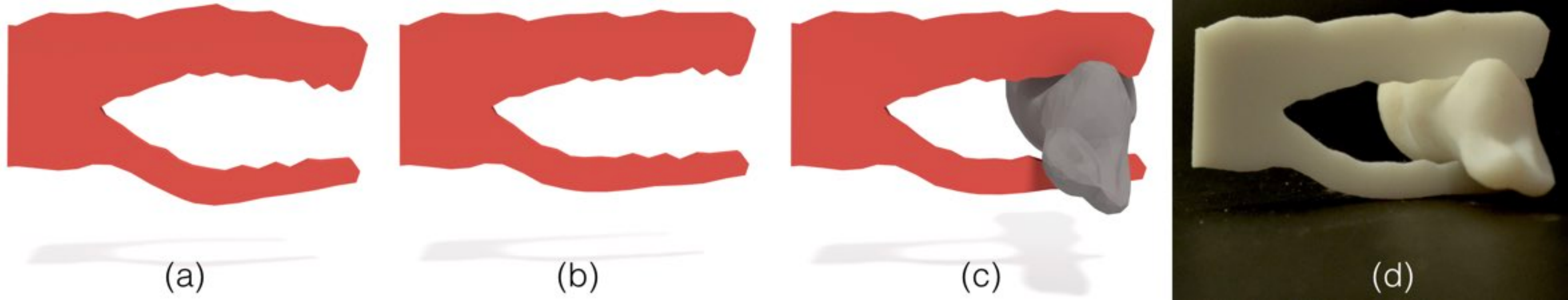}
  \caption{Initial alligator ~(a) is optimized~(b) to hold the bunny ~(c). Printed and attached objects (d) exhibit the simulated/optimized behavior.}
  \label{alligatorRabbit}
\end{figure}

\begin{figure}
  \centering  
  \includegraphics[trim = 0in 0in 0in 0in, clip, width=\columnwidth]{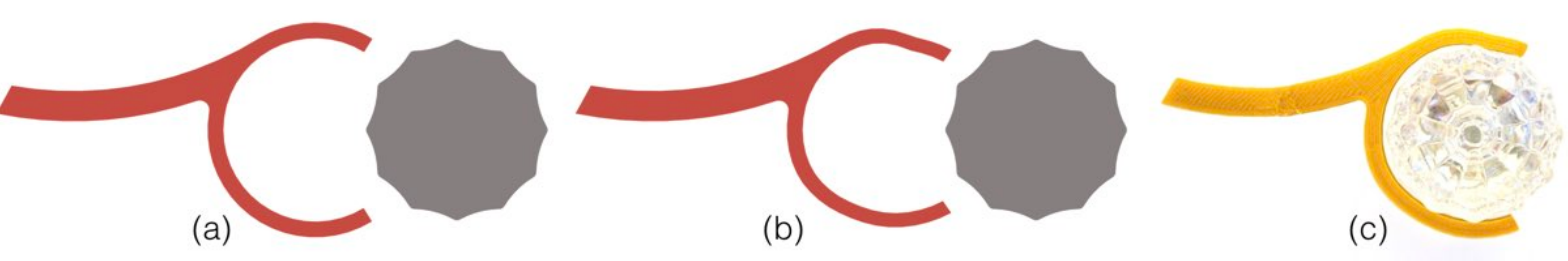}
  \caption{Assistive door handle (a) is optimized (b) to be attached on the door knob. The attachment (c) remains coupled while in use.}
  \label{doorHandlePhoto}
\end{figure}

\begin{figure}
  \centering  
  \includegraphics[trim = 0in 0in 0in 0in, clip, width=\columnwidth]{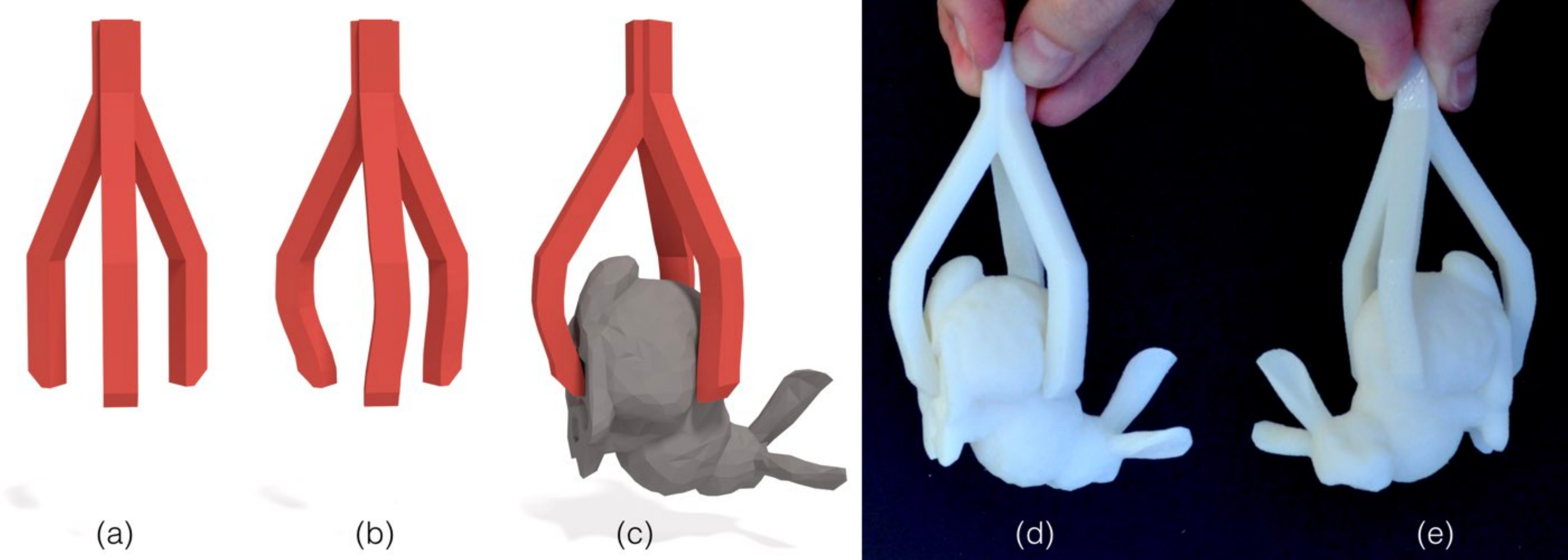}
  \caption{The initial mechanical claw~(a) is optimized~(b) to hold the bunny~(c). Printed and attached optimized result is shown in (d) and (e)}
  \label{bunnyCatcher}
\end{figure}

Ergonomic grippers for handling common objects are popular items in 3D printing repositories. We demonstrate the design of an assistive door handle attachment with a tight grip for those with difficulty gripping door knobs in Figure~\ref{doorHandlePhoto}. We also optimized the alligator to hold the bunny  as shown in Figure~\ref{alligatorRabbit}. While the initial pairs cannot be coupled, the optimized alligator attaches to the bunny with a firm grip. In Figure~\ref{bunnyCatcher}, we optimize an initial mechanical claw that is unable to grab onto the bunny so that the resulting shape can latch onto it.

\paragraph{Comparison}

\begin{figure}
  \centering  
  \includegraphics[trim = 0in 0in 0in 0in, clip, width=\columnwidth]{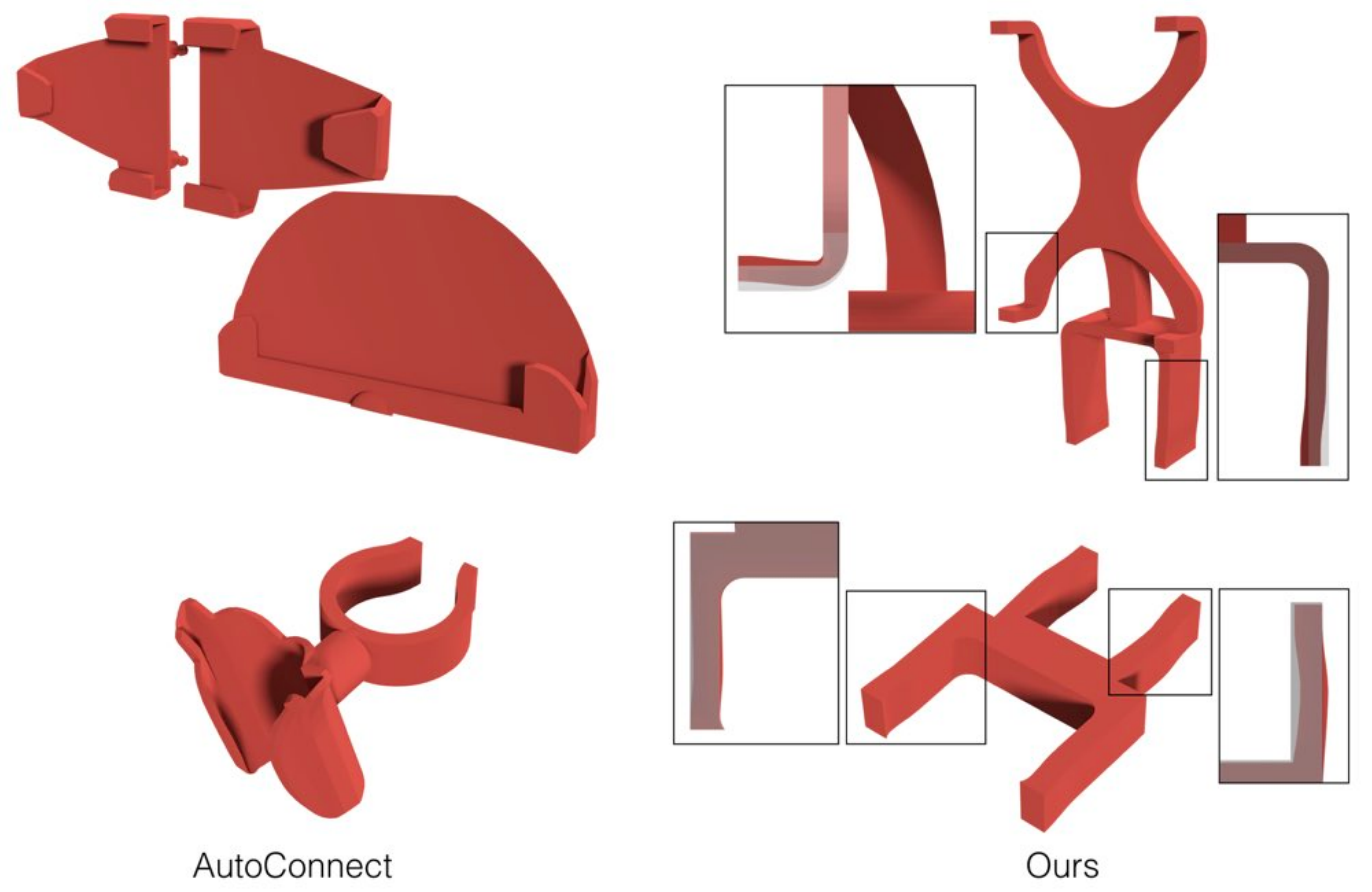}
  \caption{A comparison between AutoConnect~\cite{Koyama:2015} and our method. Our approach can create compliant attachments for phone (top) and paddle (bottom), while AutoConnect generates rigid holders when the objects are not cylinders or rectangular prisms.}
  \label{comparison}
\end{figure}

In Figure~\ref{comparison}, we compare our approach with  AutoConnect~\cite{Koyama:2015} for two attachments. We optimize a connector that mounts the phone on its charger and a paddle attachment that connects the paddle to a cylindrical stand. For non-cylindrical and non-rectangular objects, Autoconnect generates rigid structures that the object is either slid in without deformations, or requires partitioned rigid attachments that are post-joined. Unlike AutoConnect, our method creates compliant structures for both the phone side and the paddle side of the connector. It modifies a starting geometry enabling easy insertion and removal, while producing a tight grip to hold the objects in place. For the stand side of the paddle attachment, our approach converges to a solution similar to AutoConnect's standard C-clamp, while it begins with a square geometry. Similarly, we initialize the paddle side with a perfect match to the paddle's cross section when attached  (hence no deformation or grip) and converge to a firmly gripping structure. 

\FloatBarrier

\subsection{Physical Tests and Performance}

\paragraph{Fabrication} We use a low cost FDM  Printrbot as well as a Stratasys Objet Connex for fabrication, printing PLA and VeroWhite, respectively. 

\begin{figure}
  \centering  
  \includegraphics[trim = 0in 0in 0in 0in, clip, width=\columnwidth]{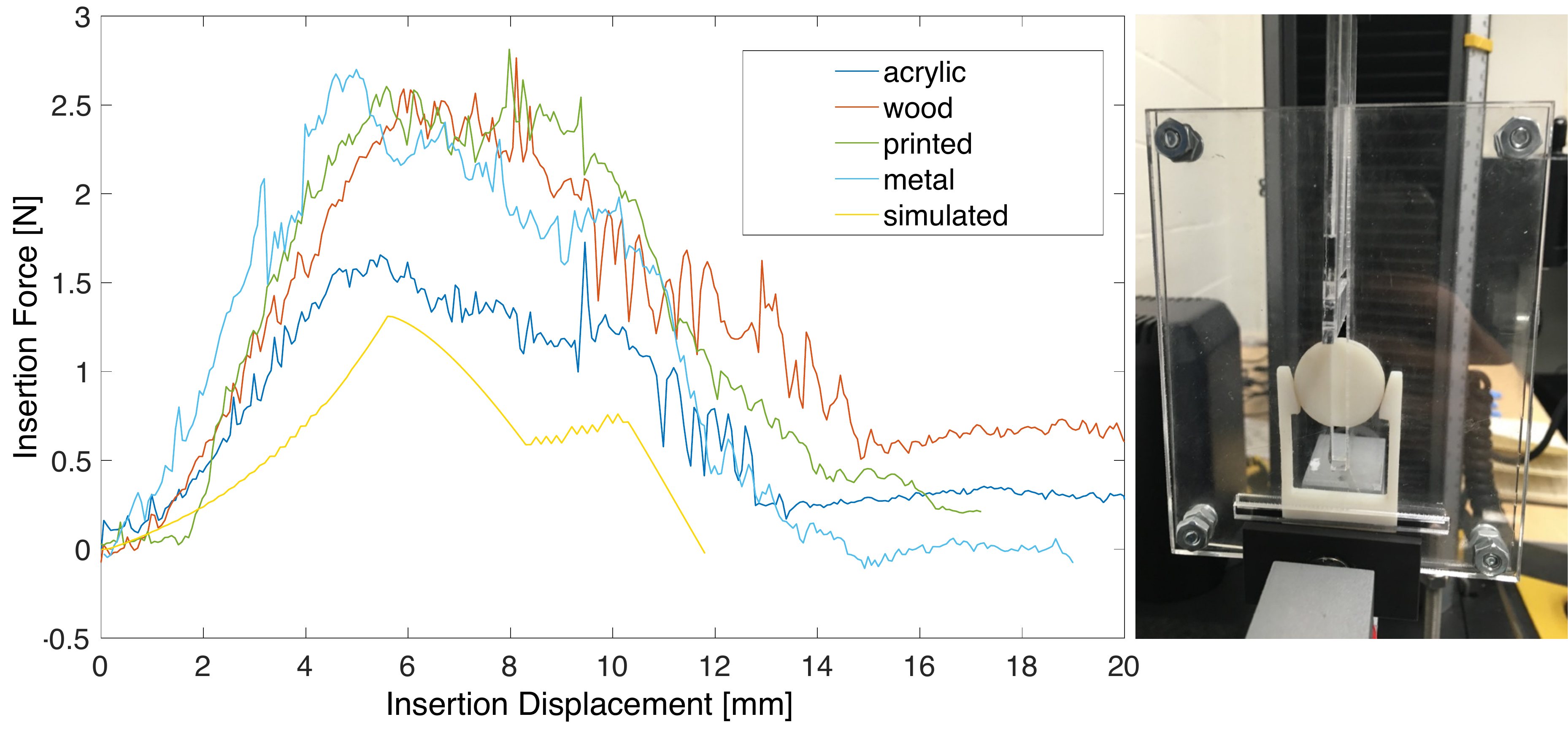}
  \caption{Effect of friction on insertion forces. The plots show insertion forces measured during the insertion of the same geometry manufactured with four different materials into one 3D printed attachment. Our  simulations and the measurements exhibit similar profiles.}
  \label{friction}
\end{figure}

\paragraph{Friction tests}To study the impact of friction, we measured the insertion force of the same geometry with four different materials as 3D printed VeroWhite, acrylic, wood and metal using an Instron tensile testing machine~(Figure~\ref{friction}). As expected, friction increases the required insertion force. Nonetheless, the physical measurements and our simulations exhibit similar trends in the profiles, differing primarily by a shift. 

 \begin{table}
\small
\caption{Statistics of the models used in our tests. For each model, we present \# of simplex elements for attachment, \# of vertices on the inserted object, \# of modification handles, \# of insertions steps, computation cost per iteration~(one full insertion simulation).} 
\centering 
\begin{tabular}{l ccccc} 
\hline\hline 
\multirow{2}{*}{Model} & \multirow{2}{*}{Elements} & \multirow{2}{*}{Vertices} & \multirow{2}{*}{Handles}  & Insertion & \multirow{2}{*}{Time[s]} \\ 
& & & & Steps & \\ [0.5ex]
\hline 
Stand Side           & 450  & 50    & 4 & 35& 4.8\\ 
Paddle Side          & 1024 & 168   & 4 & 41& 5.7\\ 
Building B.-L        & 896  & 142   & 4 & 25& 4.5\\
Building B.-T1       & 896  & 142   & 4 & 25& 4.5\\
Building B.-T2       & 896  & 142   & 4 & 25& 4.5\\
Door Handle          & 1222 & 264   & 5 & 36& 13.9\\
Alligator 2D         & 422  & 50    & 5 & 35& 4.9\\ 
Alligator 3D         & 1040 & 1154  & 5 & 35& 16.7\\
Alligator-Bunny      & 1040 & 1687  & 5 & 35& 48.3\\
Door Knocker         & 5080 & 288   & 5 & 40& 121.0\\
Charger Side         & 716  & 182   & 6 & 30& 5.2\\
Phone Side           & 4110 & 20378 & 8 & 22& 172.3\\
Claw                 & 12063& 11655 & 9 & 25& 877.2\\[1ex] 
\hline 
\end{tabular}
\label{tab:hresult}
\end{table}

\paragraph{Performance} Shape updates and modification energy computations are expectedly fast. The insertion simulation is the computationally most demanding step in our approach. Table~\ref{tab:hresult} shows the performance of our method using a 3.2GHz Intel Core i5 computer with 8GB memory. Because our rest shape optimization involves stochastic optimization, convergence speed may vary across different runs of the same problem setting. For the results in this work, convergence is achieved under 300 iterations.

\subsection{Limitations and Future Work}
We do not remesh the compliant object during optimization; our estimations will likely become increasingly inaccurate due to degenerate elements if severe deviations from the original rest shape is required to achieve the desired coupling behavior. In this work, we meshed our initial geometries as uniform as possible to mitigate mesh dependency issues. We use Triangle~\cite{triangleSoftware} for 2D Delaunay meshing and TetGen~\cite{TetGen} for tetrahedral meshing.

We do not model friction. Although friction based attachments are prone to wear and tear, there are use cases where a friction based attachment is required due to the object's geometry and limiting surroundings. It is an interesting future direction to study friction-dominated coupling behaviors. Likewise, our method does not model the true surface contact area between the compliant structure and the inserted object. In some cases, increasing the contact area might be desirable to further increase grip tightness. For example, our door handle and the knob contact each other at a limited set of discrete points whereas a larger contact area is likely more desirable to increase grip. However, this issue also highlights the difference between our definition of grip and that one experiences in real life. We quantify grip solely by the normal forces acting on the rigid object (squeeze force), while daily experience would also incorporate friction as part of this quantification (\eg how easy is it to rotate the objects relative to one another?).  

Even though our algorithm reduces the stress and guarantees structural soundness during insertion, stress might still be concentrated around certain regions of the object which may limit fatigue performance. While we do not consider fatigue in this work, our work can be extended to account for this failure mode in the future.

While our method handles insertion paths that involve prescribed curved paths as well as prescribed rotations to the rigid object during engagement, a future extension would be to relax the rotary degrees of freedom during insertion. This would more closely mimic humans' natural, accommodatory motions when coupling two objects.

\section{Conclusions}
\label{sec:coupling:conclusions}

Given a pair of arbitrary objects, we present a method to design targeted coupling behavior through shape modifications on the compliant object.  Our approach  does not rely on the knowledge of known contact points or deformed states. With our approach, non-intuitive coupling behaviors, such as firm grips requiring weak insertion/removal forces, can be designed. Our results demonstrate that free-form geometries can be geometrically altered to produce compliant attachments that expand the basic geometries of traditional attachments. This offers an opportunity for assistive design technologies where unique personal needs are present. 

\chapter{Buckling Behavior Design For 3D Surface Deployment From Flat Patterns}
\label{chap:inserts}

3D printing thin surface objects may turn into a very costly process due to support material requirements and production time. 2D manufacturing techniques usually are not viable due to surface developability limits and mechanical stability issues after the deployment to a 3D shape. Hence, facilitating a simpler and more convenient fabrication of complex 3D surfaces is an important challenge. We present 2D laser cut patterns that deploy into 3D surfaces after the addition of rigid inserts. The work exploits the idea of creating protrusions/buckling on a 2D sheet by adding rigid inserts that create internal constraint. Specific challenge addressed here stem from  understanding the highly non-linear underlying mechanism in the deployment of flat patterns into 3D surfaces. 

\begin{figure}
  \centering  
  \includegraphics[trim = 0in 0in 0in 0in, clip, width=\textwidth]{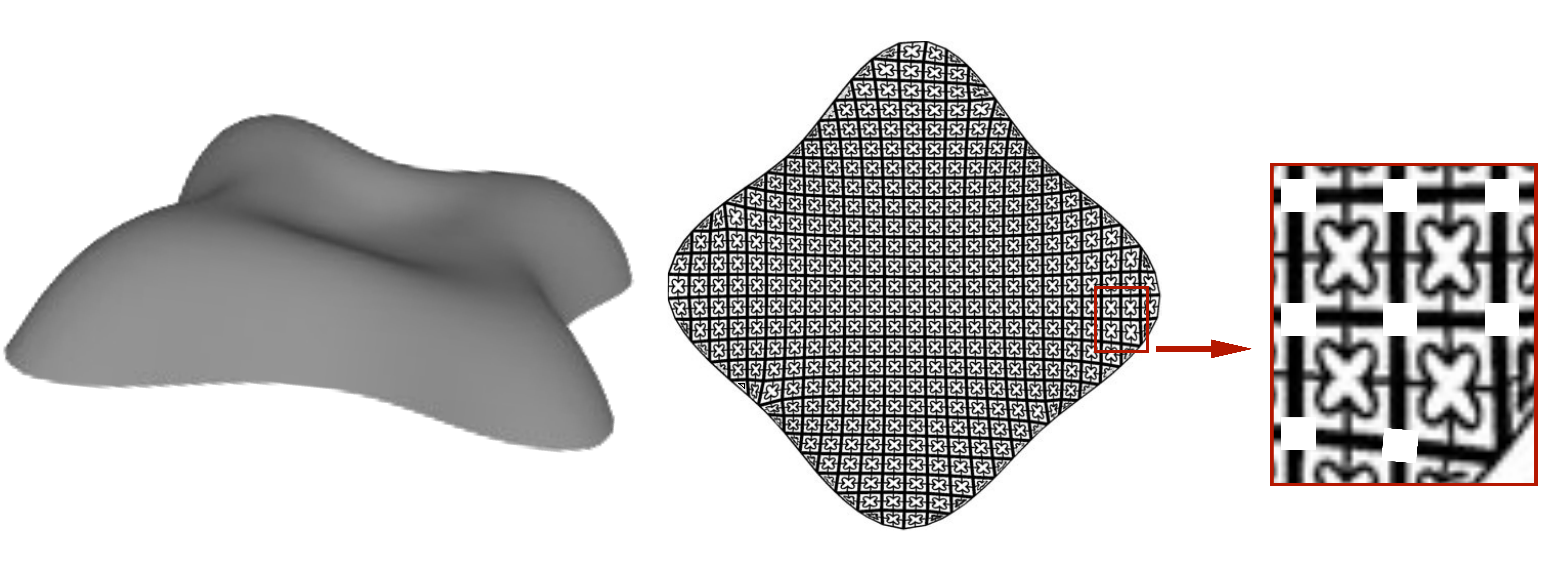}
  \caption{A 3D complex surface and corresponding 2D pattern layout.}
  \label{inserts:lilium}
\end{figure}

\section{Introduction}

Additive manufacturing methods offer high flexibility in shapes they can produce but result in slow production times very quickly as the shapes get larger. Laser cutting provides a fast alternative, yet it works in 2D space. Achieving 3D surfaces from 2D laser cut shapes is a challenge due to developability constraints. Additionally, ensuring mechanical stability of 3D surfaces presents another obstacle in the realization of 3D surfaces from flat pieces.

\begin{figure}
  \centering  
  \includegraphics[trim = 0in 0in 0in 0in, clip, width=0.7\textwidth]{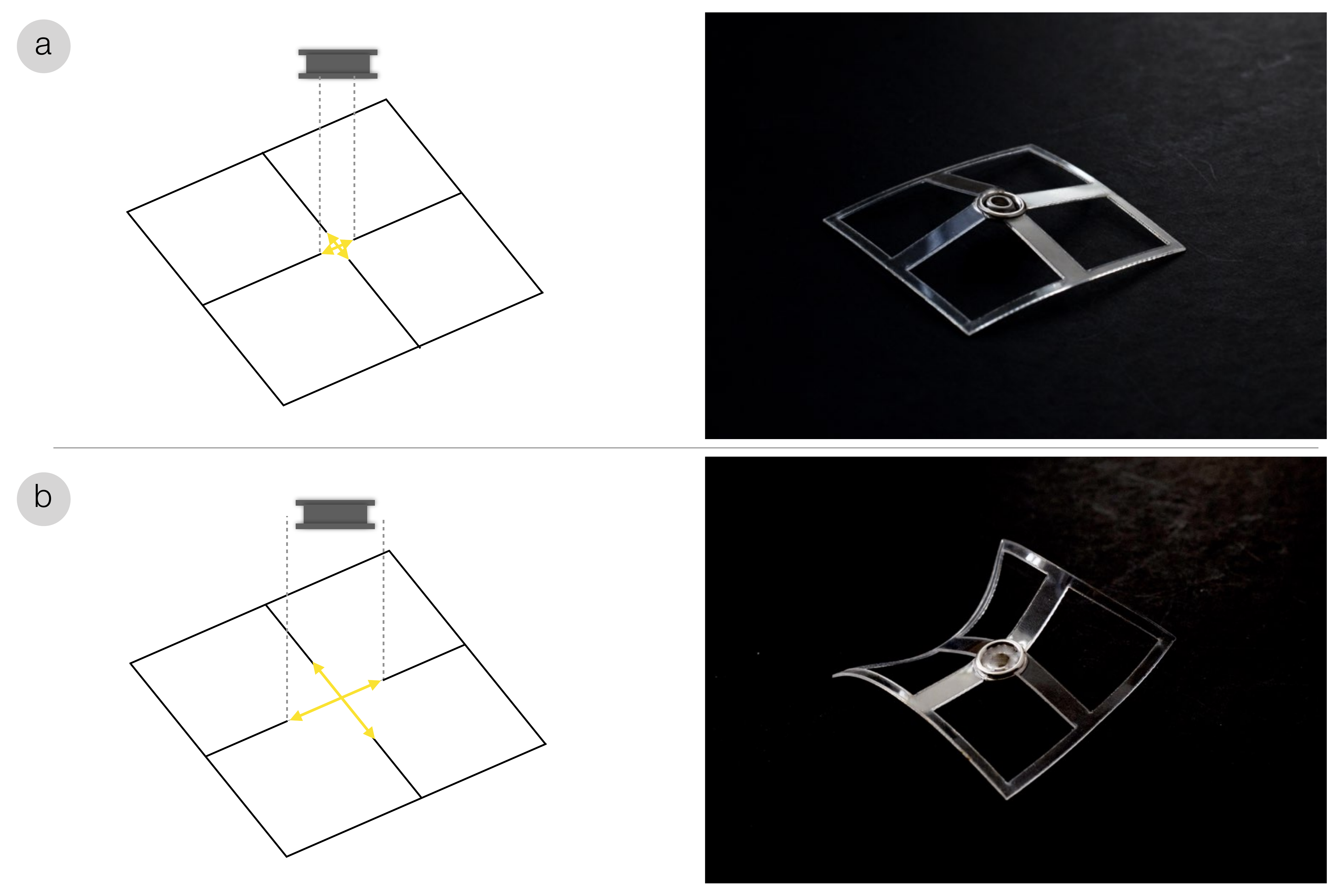}
  \caption{Buckling via rigid insert concept. }
  \label{inserts:insert_concept}
\end{figure}

We propose a shape parameterization method to map 3D complex surfaces into 2D flat patterns so that the 3D surfaces can be realized by installing rigid inserts (Figure~\ref{inserts:lilium}). Our method is based on fundamental buckling concept that is observed when a rigid object is inserted into a gap that is different than its size as depicted in ~Figure~\ref{inserts:insert_concept}. This way, the whole system settles into a new minimum energy state that creates a protrusion. Here, we exploit this mechanism to produce complex structures. Our method takes input as a target surface and produces a pattern layout in 2D. The 2D pattern then can be laser cut and realized into 3D surface by installing rigid inserts. 

The idea of obtaining 3D surfaces from flat pieces has been recently explored. Achieving developable surfaces through auxetic cuts has been studied in~ \cite{konakovic2016beyond} but this work requires meticulous forming of the flat surface onto some sort of template mold for deployment. In~\cite{guseinov2017curveups,perez2017computational}, developability and deployability issues have been investigated through 3D printing on pre-stretched elastic materials. One attribute that makes our approach attractive is the simplicity of just using a regular laser cutting process rather than using 3D printers with special methods and materials.

Harnett et al.~\cite{harnett2013digital} presents geometrically frustrated tiles which are regular grids of structures that can attain an energetically favorable state  due to internal constraints. While the idea of geometrically frustrated tiles is demonstrated, the problem of understanding their folding pathways and determining the folded configurations in order to match a given shape remains a challenge. Assessing stability of the entire configuration provides an even more challenging problem. Drawing inspiration from the geometrically frustrated tiles, we undertake the challenge of matching target shapes with mechanical stability.

\section{Deployment Mechanism}

\subsection{Buckling Beams and Rigid Inserts}

Figure~\ref{inserts:insert_concept} illustrates the proposed deployment mechanism that is based on buckling of beams through internal constraints. If the gap is smaller than the size of the rigid insert, the resulting surface is a dome shape with positive Gaussian curvature~(Figure~\ref{inserts:insert_concept}.a). On the other hand, we obtain saddle shape surfaces with negative Gaussian curvature when the gap between the beams is larger than the size of the rigid beam~(Figure~\ref{inserts:insert_concept}.b). Section~\ref{inserts:sec:flattening} gives geometric insights into behavior of the deployment mechanism.

\begin{figure}
  \centering  
  \includegraphics[trim = 0in 0in 0in 0in, clip, width=\textwidth]{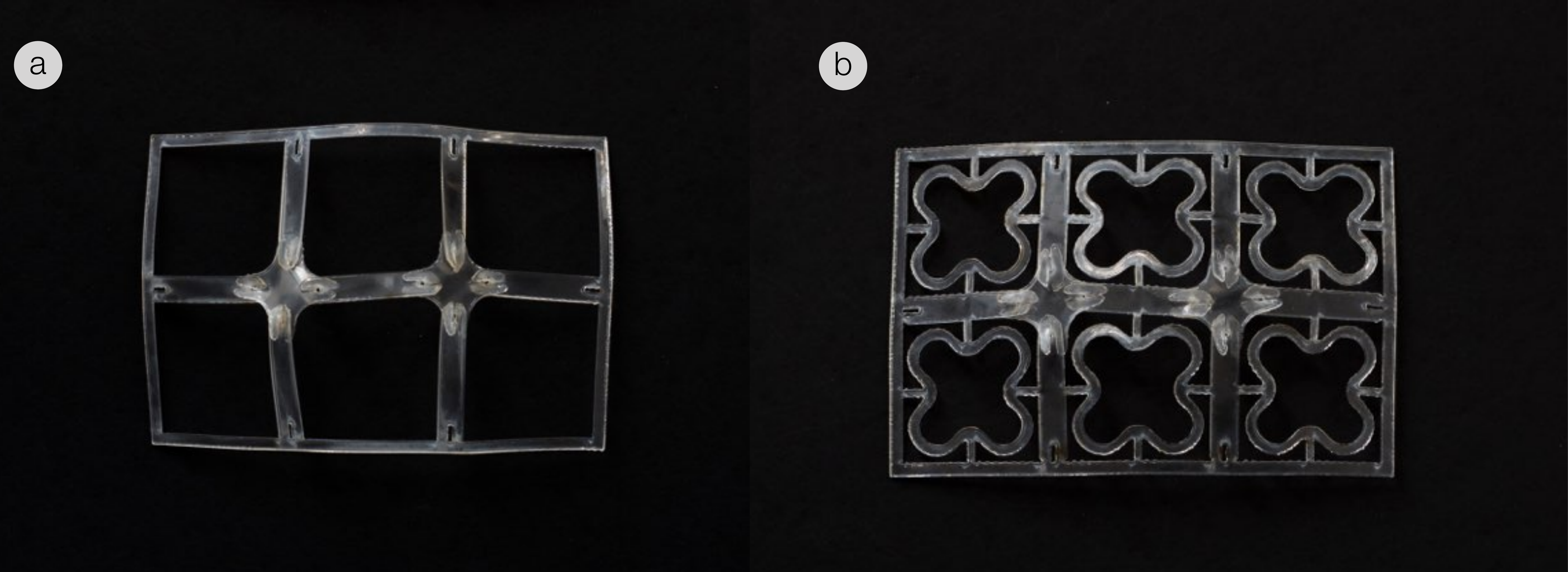}
  \caption{Design Choices: Torsional springs keep the whole structures together before the inserts are assembled. }
  \label{inserts:torsional_spring}
\end{figure}

\subsection{Design Choices}

We use the same size inserts and only alter the gap size through beam lengths. This way, we can create different amounts of protrusions using the same inserts thereby making the assembly process simpler and more practical. In addition, we place torsional springs~(the curvy beams) to keep the structure together before the rigid inserts are installed (Figure~\ref{inserts:torsional_spring}). We also observe that they serve as stiffening components that enhance structural stability of the pattern after deployment.

\section{Surface Parameterization}

\subsection{Flattening}
\label{inserts:sec:flattening}

\begin{figure}
  \centering  
  \includegraphics[trim = 0in 0in 0in 0in, clip, width=\textwidth]{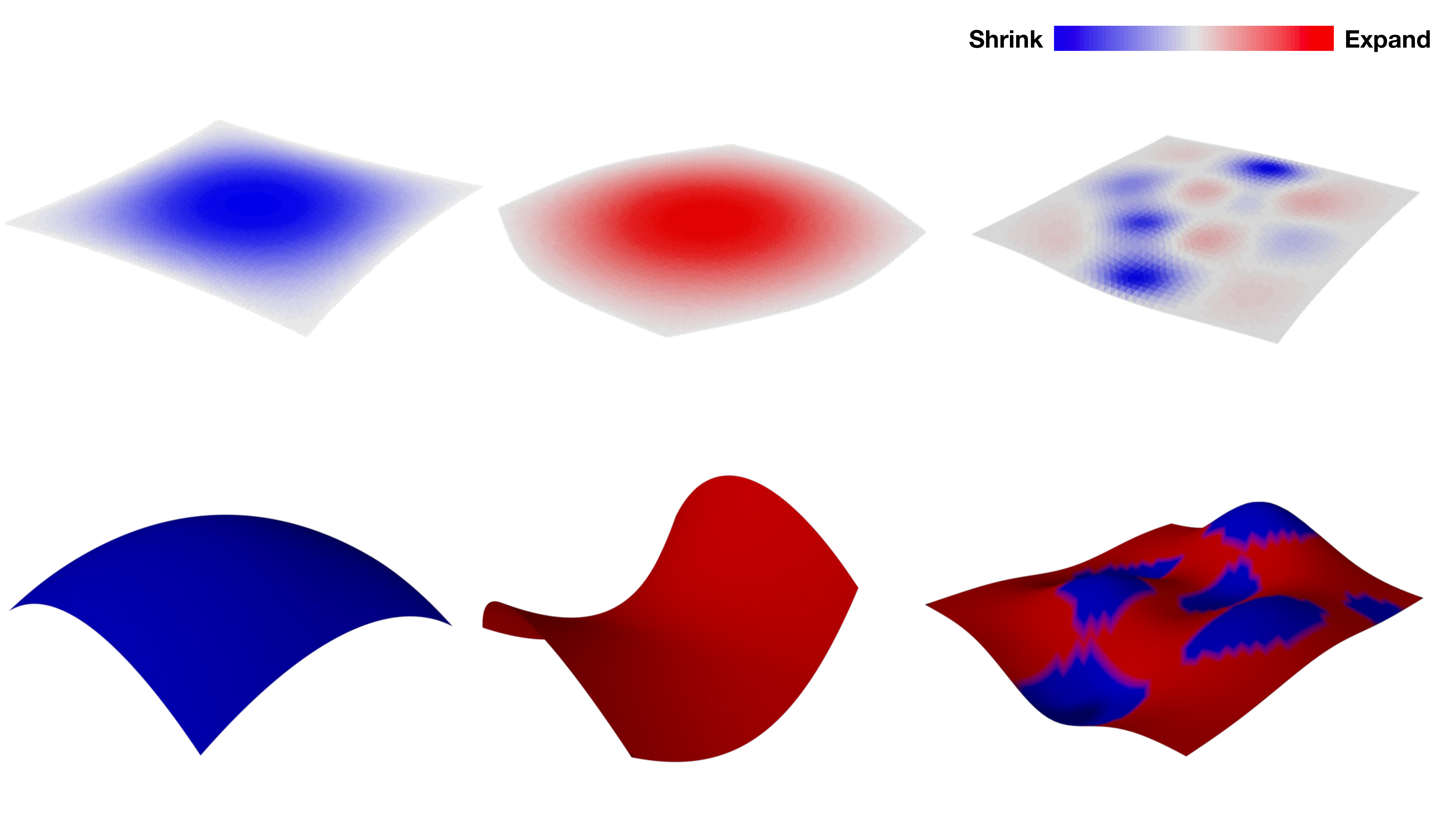}
  \caption{Top: Flattened mesh with colors representing area distortions during flattening. Bottom: Original surfaces with blue and red colors representing positive and negative Gaussian curvatures, respectively.}
  \label{inserts:flat_tests}
\end{figure}

To obtain the layout of the 3D surface in 2D domain, we use conformal mapping.  Main challenge in this context is to understand the deployment process and use an appropriate conformal mapping technique. For our buckling patterns, one important characteristic is that the edge lengths on the boundary do not change during deployment. Therefore, one property to look for in the conformal mapping algorithm is to be able to prescribe zero scale factors along the boundary. Conformal mapping techniques that allow such prescriptions exist and indeed it turns out that such prescriptions result in the flattening with minimal area distortions~\cite{Springborn:2008:CETM}. Note that achieving the minimal area distortions in the flattening is a very desirable property. To achieve this flattening, CETM~\cite{Springborn:2008:CETM} or BFF~\cite{Sawhney:2017:BFF} methods can be used. These two methods produce virtually indistinguishable results. Due to its lower computational cost, we use BFF in our algorithm.

Equipped with the flattening algorithm that characterizes our deployment mechanism, we conducted a few simple experiments that give insights into how different types of surfaces can be realized using our mechanisms. Figure~\ref{inserts:flat_tests} illustrates three surfaces with their corresponding flattened versions. We observe that a spherical surface (positive Gaussian curvature) only shrinks while flattening. Thus, by using rigid inserts that are larger than the gap sizes, we can expand the shape during the deployment and create positive Gaussian curvature surfaces. On the other hand, a saddle surface (negative Gaussian curvature) only expands while flattening. Therefore, the rigid internal constraints must shrink the shape during deployment to 3D shape for a negative curvature region. We observe the same behavior on a complex surface that includes both positive and negative Gaussian curvature regions.

\begin{figure}
  \centering  
  \includegraphics[trim = 0in 0in 0in 0in, clip, width=\textwidth]{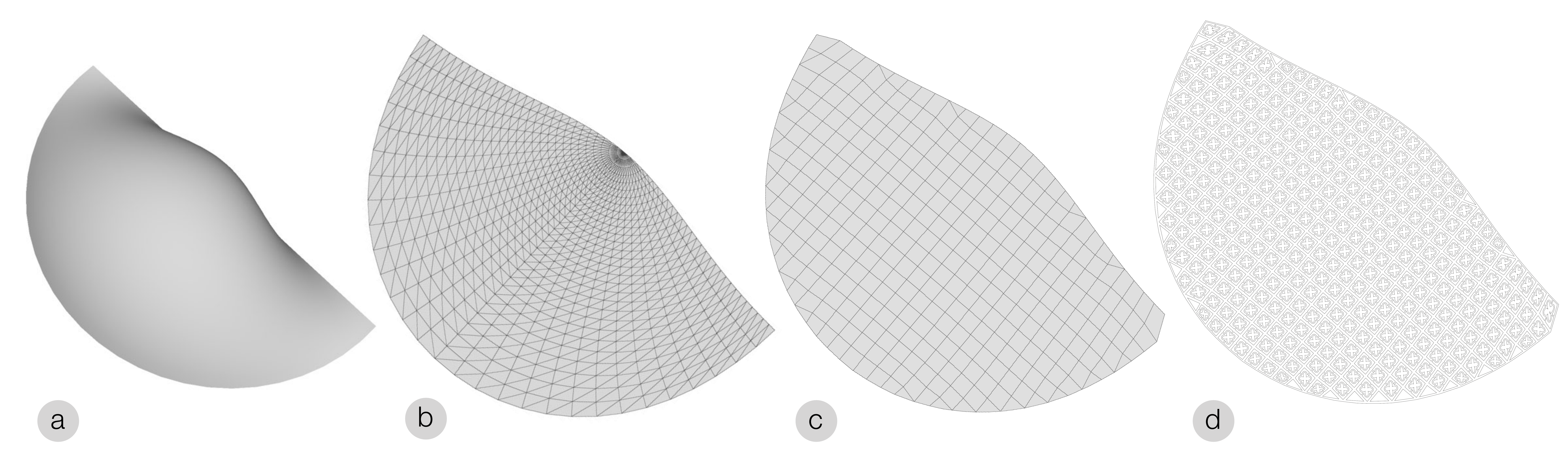}
  \caption{Pattern generation: (a) Input surface, (b) Flattened triangulation, (c) Quad Mesh, (d) Our patterns.}
  \label{inserts:snail_patterns}
\end{figure}

\subsection{Pattern Generation}

Figure~\ref{inserts:snail_patterns} depicts the pattern generation process. After the flat surface is obtained, we generate a quad mesh on the flat triangulation. Then, the whole pattern is generated on the quad mesh. Each quad edge is offseted with beam thickness and and the torsional spring pattern is placed inside each quad. We represent torsional spring patterns using Bezier curves. Some quads on the boundary may be degenerate, i.e., two neighboring edges are almost colinear. In those degenerate quad instances, we do not place torsional springs. Since these instances only occur on the boundary, not having torsional springs does not result in disconnected components since the beams are already connected to the boundary. After, the pattern is generated, we initiate gaps between the beams depending on the area distortions of the flattening and the rigid inserts size. Note that the rigid insertions allow only a limited amount of extension or shrinkage in the area distortions. If the area distortions exceed the achievable limits, cuts may be introducec during flattening.

\section{Physical Simulations}

Our patterns can be modeled using beam elements. To model the large deflections and post buckling behavior of beam elements, so far, we have investigated Elastica~\cite{shoup1971use} and discrete elastic rods~\cite{bergou2008discrete}. We observe that Elastica gives a simpler formulation yet discrete elastic rods cover more generalized deformation scenarios such as twists. For the physical analysis, we use discrete elastic rods. We use ODE~\cite{ode2008} library to simulate the rigid inserts. We enforce, rigid body coupling using fast manifold projection as presented in ~\cite{bergou2008discrete}.

\section{Applications and Future Work}
\begin{figure}
  \centering  
  \includegraphics[trim = 0in 0in 0in 0in, clip, width=\textwidth]{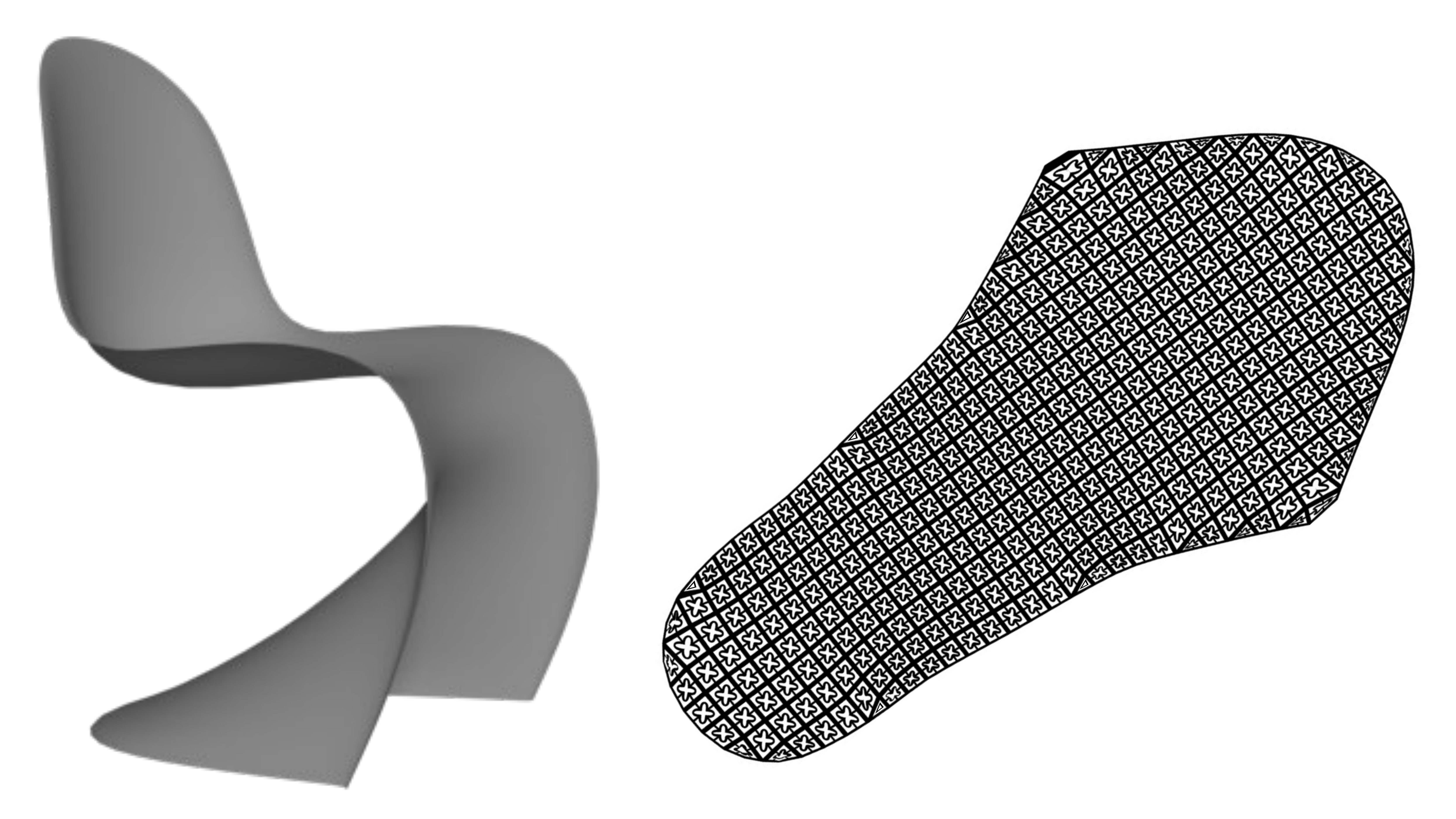}
  \caption{A concept chair design that can be realized with our approach and can be further optimized (through possible future research) for structural stability under in-use forces}
  \label{inserts:vitra_panton}
\end{figure}

In this work, the focus is the realization of complex surfaces through 2D fabrication techniques. We believe this work will fuel further research efforts into functional use of deployed 3D surfaces. Figure~\ref{inserts:vitra_panton} shows a conceptual chair design that can be realized with our approach and can be further optimized for structural stability under in-use forces. 

In addition to the advantages in fabrication, our method can enable flat-packaging and quick assembly of rigid inserts to achieve a 3D end product similar to IKEA concept. To enable easy assembly, the rigid inserts could be placed on a stencil that facilitates an easy transfer to the 2D pattern. Robotic assembly of the rigid inserts is another possible direction for the realization of the proposed method in real life applications.

The units depicted in~Figure~\ref{inserts:insert_concept} also have the bistable property that may open up a world of new possibilities in terms of future applications.

\section{Conclusions}

In this work, we present a deployment mechanism  that is based on buckling of beams through inserting rigid pieces into gaps that are different in size. To exploit this deployment mechanism to generate arbitrary 3D surfaces, we have developed a shape parameterization method that maps 3D surfaces into 2D flat patterns such that the 3D surfaces can be realized by installing the rigid inserts. In addition, we have developed a routine for the physical simulation of the buckling phenomena. While the shape parameterization gives a flat layout of the 3D shape, the resulting layout only considers geometry. Thus, using the physical simulations is important to ensure mechanical stability and improve the accuracy of the resulting 3D surface. Therefore, an immediate future work is to use the solution of the shape parameterization as initial condition of an optimization routine where approximation to the input 3D surface is improved using the physical simulations.

\chapter[Wisdom of Micro-Crowds]{Wisdom of Micro-Crowds in Evaluating Solutions to Esoteric Engineering Problems}
\label{chap:crowd}
\blindfootnote{This chapter is based on Ulu et al., 2018 \cite{ulu2018crowdsourcing}.}

A multitude of studies in economics, psychology, politics, and social sciences have demonstrated the wisdom of crowds (WoC) phenomenon,  where the collective estimate of a group can be more accurate than estimates of individuals. While WoC is observable in such domains where the participating individuals have experience or familiarity with the question at hand, it remains unclear how effective WoC is for domains that traditionally require deep expertise or sophisticated computational models to estimate objectively answers. This work explores how effective WoC is for engineering design problems that are esoteric in nature, that is, problems (1) whose solutions traditionally require expertise and specialized knowledge, (2) where access to experts can be costly or infeasible, and (3) in which previous WoC studies with the general population have been shown to be highly ineffective. The main hypothesis in this work is that in the absence of experts,  WoC  can be observed in groups that consist of practitioners who are defined to have a base familiarity with the problems in question but not necessarily domain experts. As a way to emulate commonly encountered engineering problem solving scenarios, this work studies WoC  with practitioners that form micro-crowds consisting of 4 to 15 individuals, thereby giving rise to the term the wisdom of micro-crowds (WoMC).  Our studies on design  evaluations show that WoMC produces  results whose mean is in the 80th percentile or better across varying crowd sizes, even for problems that are highly non-intuitive in nature.

\section{Introduction}
\label{sec:crowd:intro}

Crowdsourcing is emerging as a cost-effective, rapid approach to problem solving in a variety of disciplines where the collective estimate of a group can outperform the individuals, even in the presence of domain experts. This phenomenon is known as the wisdom of crowds (WoC)  and has been demonstrated across a range of problem domains~\cite{galton1907ballot,hooker1907mean,surowiecki2005wisdom}. Traditional crowdsourcing naturally focuses on tasks that are human easy and computer hard, such as vision problems where crowds are asked to identify and label objects in large sets of images~\cite{wah2006crowdsourcing}. In such problems, the task is typically very intuitive for humans, and thus the correct answer can be inferred from a crowd consensus. In engineering problems requiring domain expertise, however, crowdsourcing has proven to be significantly less effective, in part due to the limited number of experts in the sampled crowd~\cite{burnap2015crowdsourcing}. This suggests that extending traditional crowdsourcing to tasks requiring expertise is non-trivial, especially if experts are scarce. As an alternative to crowdsourcing, expert collaboration has been extensively studied~\cite{summers2010mechanical,yang2010consensus,gurnani2008collaborative,takai2010game,cabrerizo2014building}. However,  interactions among group members have been shown to lead to similarity of experts \cite{hong2004groups}, which may result in experts being outperformed by diverse groups \cite{lorenz2011social}. As such, it remains unclear how conventional crowdsourcing can be made truly effective for engineering design problems,  especially for tasks that require expertise. 

As one step toward addressing this gap, this work explores the effectiveness of WoC  for engineering design  problems that are esoteric in nature. Esoteric problems are defined as those (1) that traditionally require expertise and specialized knowledge, (2) where access to experts can be costly or infeasible, and (3) in which previous WoC studies with the general population have been shown to be highly ineffective~\cite{burnap2015crowdsourcing}. The main hypothesis in this work is that in the absence of experts,  WoC  can be observed in groups that consist of \emph{practitioners} who are defined to have a base familiarity with the domain and the problems in question, even though no single individual may have the expertise to correctly solve the problem. With this definition,  experts are a subset of practitioners. However in this work, in contrast to purely expert crowds,  practitioner crowds are characterized by individual responses that exhibit both significant accuracy (deviation from the ground truth) and precision errors (variation among the responses). This new definition and focus on  practitioners stands in contrast to previous studies that explore WoC in design that  rely either on the general population crowds where experts are extremely scarce and unknown~\cite{burnap2015crowdsourcing}, or on teams of experts~\cite{yang2010consensus} as the basis of crowds. Additionally, as a way to emulate commonly encountered engineering problem solving scenarios, this work studies WoC  with practitioners that form micro-crowds consisting of 4 to 15 individuals (rather than tens or hundreds of individuals), thereby giving rise to the term the \emph{wisdom of micro-crowds} (WoMC) which is central to the presented work.

As part of this study, four design assessment questions with varying levels of difficulty and intuitiveness were deployed where the participants were asked to assess the quality of the candidate design solutions.  Several data aggregation methods were developed and tested on the acquired data.  The results suggest that WoMC with practitioners can indeed be observed, where the crowd estimate  outperforms the individuals in the vast majority of instances. To facilitate benchmarking, these results have been obtained for problems in which there already exists an objectively true solution (\ie~benchmark results obtained through optimization).  As such,  it could be argued that crowdsourcing is remarkably unnecessary for such problems where  solution methods already exist. However, the most significant conclusion of the presented work is that for current or future engineering design problems where  algorithmic solutions  may currently not exist, small groups of practitioners may in fact provide very effective solutions. Note that, in this context the current lack of  solution methods implies a lack of experts, which reinforces the importance of practitioners. 

An interesting limitation of the presented work, however, is that when applied to open-ended, conceptual design problems where no objectively true solution exists, the performance of WoMC declines significantly. The results indicate that in such cases, the individuals in the crowd tend to make significant estimation errors when benchmarked against  expert ratings. Nevertheless, it remains unclear whether these estimation errors are due to the practitioners' inability to accurately assess candidate solutions, or whether there exists issues even with expert ratings of such open-ended problems.

\section{Experimental Design}

In order to study the wisdom of crowd in esoteric engineering applications, it is necessary to understand the relationship between crowds and problem types. This section explains the characteristics of the crowd participants and the design problems used in this work.

\subsection{Crowd Population}

Two key factors in the WoC are diversity of opinion and independence. Therefore, a crowd should include people with a variety of opinions rather than a group of elites or experts that may create bubbles and conform to each other's opinions~\cite{surowiecki2005wisdom}. To support independence, we collected survey results through a web-based survey  providing anonymity and independence across participants. To support diversity of opinion, we collected crowds through AMT or students specializing various topics in mechanical engineering.

This work considers two types of crowds: AMT workers and practitioners. AMT crowds consist of individuals from the population at large, with no explicit control over an individual's level of expertise. On the other hand, the practitioner group represents individuals who have familiarity and knowledge within the target domain, however are not necessarily domain \textit{experts} for the given task. For example, a practitioner would be an individual who has studied or currently practices mechanical engineering, but does not necessarily specialize in the field of a given task such as heat transfer, structural mechanics, or manufacturing. For a practitioner group, performance of individuals may have significant variation yet the base domain knowledge pushes the estimation method to accurate levels. Note that with this definition, experts are a subset of practitioners.

For the practitioner group, 15 mechanical engineering graduate students at Carnegie Mellon University were recruited to participate. Each participant was compensated monetarily for their time. The 15 practitioners were recruited from an available pool of over 300 graduate students.
It is important to note that these students have different skill levels. As later will be shown, this can be observed by large individual estimation errors and significant performance variation among the group members. For the AMT surveys, groups of 100 people were gathered through Amazon Mechanical Turk, receiving monetary compensation. In order to remain true to the notion of general public as closely as possible, no specific demographic groups were targeted. For the structural mechanics questions (discussed in detail below), the study used the data provided by Burnap \etal~\cite{burnap2015crowdsourcing}.

\subsection{Survey Design and Questions}
\label{crowd:surveydesign}

\begin{figure}
  \centering  
  \includegraphics[trim = 0in 0in 0in 0in, clip, width=2.5in]{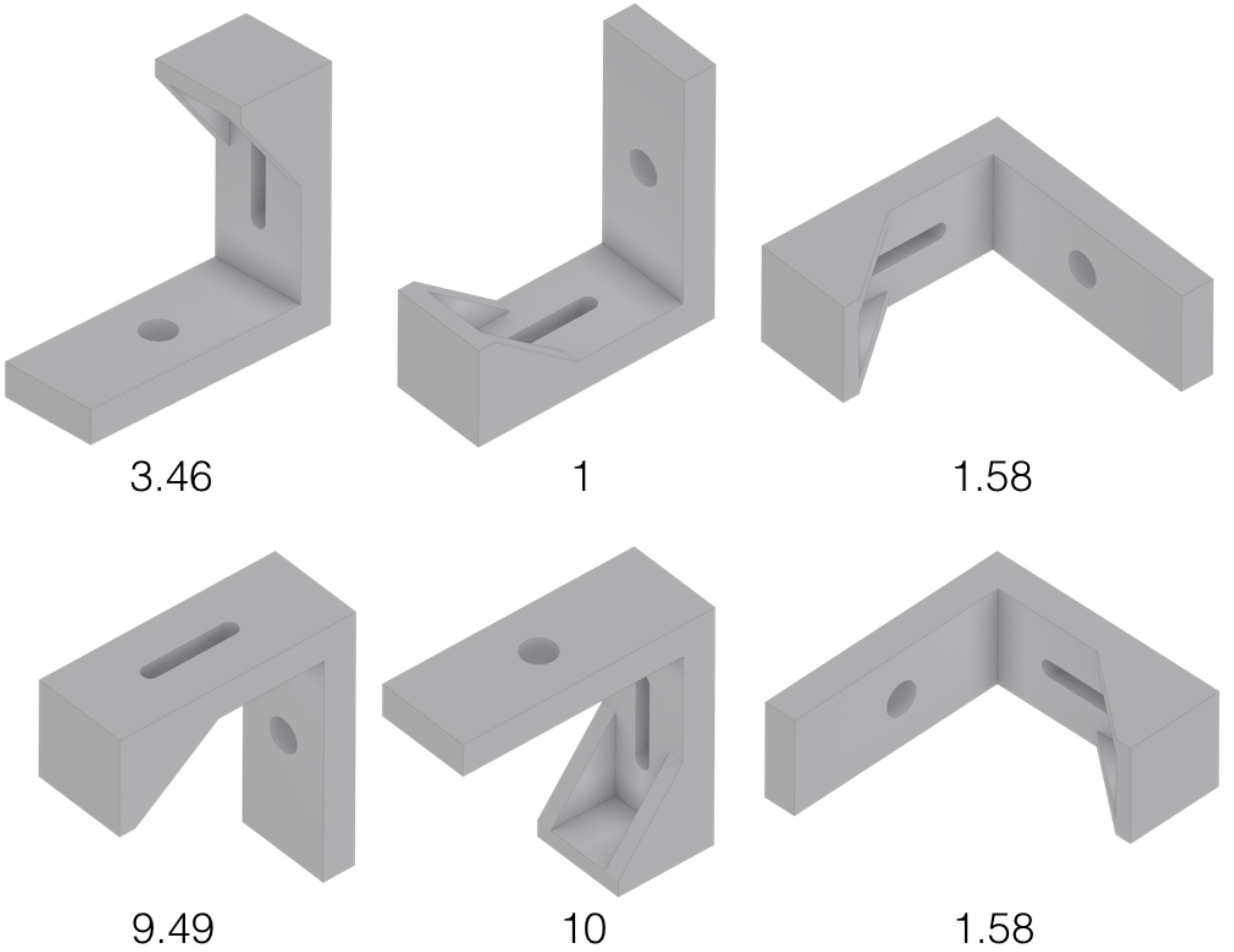}
  \caption{3D printing-1: support material question.}
  \label{crowd:3dpQ1}
\end{figure}

\begin{figure}
  \centering  
  \includegraphics[trim = 0in 0in 0in 0in, clip, width=2.5in]{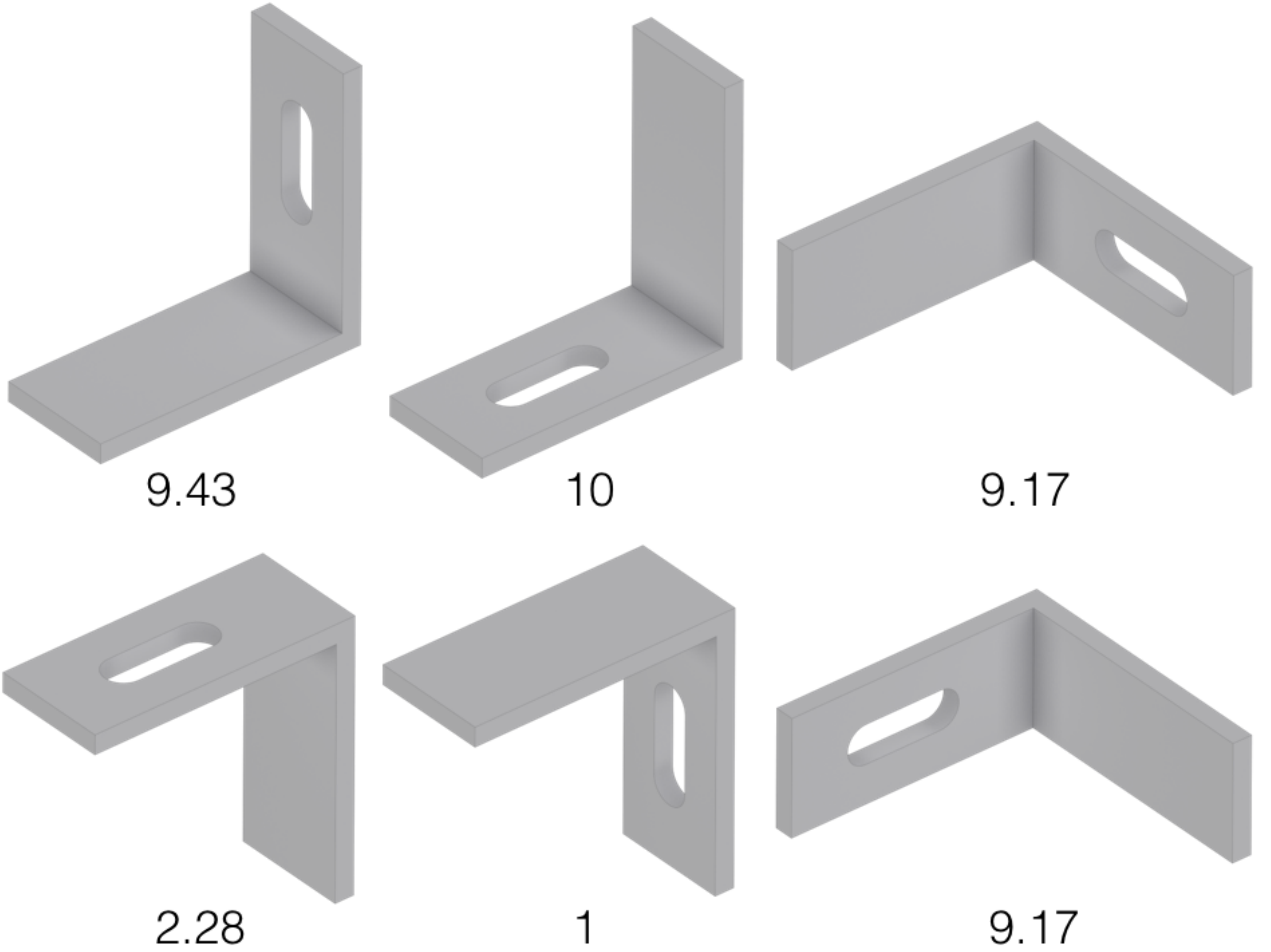}
  \caption{3D printing-2: surface finish question.}
  \label{crowd:3dpQ2}
\end{figure}

\begin{figure}
  \centering  
  \includegraphics[trim = 0in 0in 0in 0in, clip, width=2.5in]{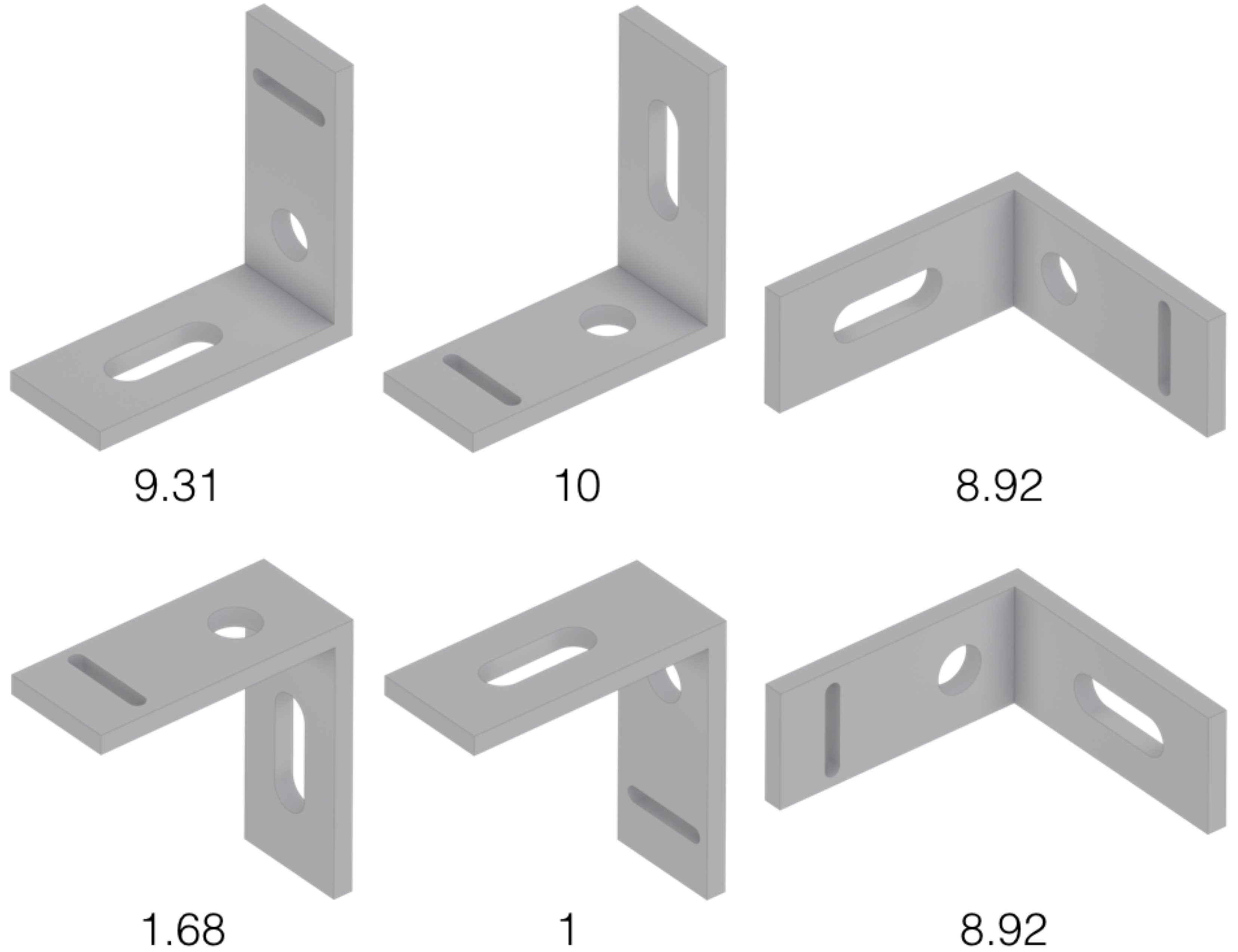}
  \caption{3D printing-3: surface finish question.}
  \label{crowd:3dpQ3}
\end{figure}

\begin{figure}
  \centering  
  \includegraphics[trim = 0in 0in 0in 0in, clip, width=0.5\columnwidth]{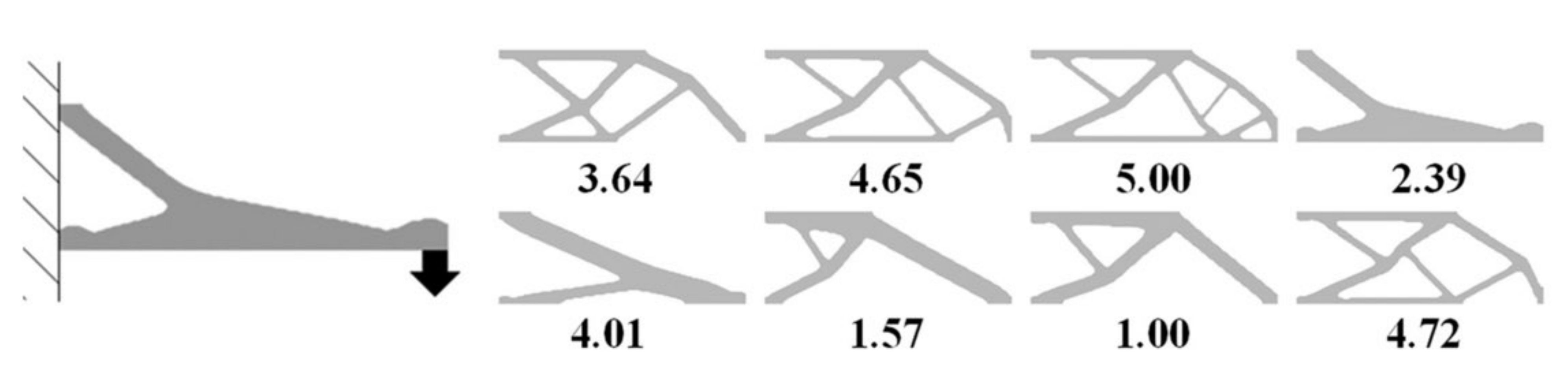}
  \caption{The structural mechanics problem \cite{burnap2015crowdsourcing}.}
  \label{crowd:topOptQ}
\end{figure}

This study investigates the WoC with four different surveys that range in the challenge they present to a human. All surveys require the respondents to be knowledgeable about the terminology used in the questions. 3D printing  questions (Fig.~\ref{crowd:3dpQ1},~\ref{crowd:3dpQ2},~\ref{crowd:3dpQ3})  aim to probe broadly intuitive perception skills involving  visual estimations of areas and volumes. However, they are designed to be increasingly more challenging. Conversely, the structural mechanics problem that involves estimating  shape deformations~(Fig.~\ref{crowd:topOptQ}) presents a much greater challenge to  humans, even for experts.

Although engineering problems are often computer easy, human hard, they are solved using expert intuition when no computational tools are available. A series of surveys for problems with known solutions such that the crowd evaluation accuracy could be determined, assessed whether such situations could benefit from WoC. The structural mechanics problem~(Fig.~\ref{crowd:topOptQ}) provides a good example, as such structural design problems had been solved primarily by experts' knowledge and intuition until  the introduction of topology optimization techniques in the 1990s~\cite{bendsoe1989optimal}. Therefore, there now exists the tools to computationally evaluate the aggregated crowd evaluations and benchmark the performance against true values. As such, practitioners' performance on such problems (which can now be objectively assessed) may provide insights into whether crowd-evaluations of design proposals may yield successful outcomes especially for  engineering challenges for which computational modeling and analysis tools may not yet exist.

A rating-assignment approach within a predefined scale is utilized. Each survey consists of multiple questions (\eg~rating the amount of support material for six different orientations) to facilitate expertise inference later in the crowd aggregation stage. In all surveys, participants are presented with the problem statement and the candidate solutions to be rated. 

Figure~\ref{crowd:3dpQ1} shows 3D printing-1 survey where participants are asked to rate the amount of support material required to print an object at various orientations using a fused-depositon printer. For each of the given orientations, participants are required to evaluate the amount of support material needed on a scale from 1 (very little) to 10 (a lot). The benchmark analysis computes the required support material as the volume that is created by the projection of overhangs to the base with zero overhang angle. Then, the scores are scaled linearly between 1 and 10 to create the benchmark values.

3D printing-2 survey is about evaluating the surface finish quality of an object in  various  orientations~(Fig~\ref{crowd:3dpQ2}). The participants are asked to rate the quality of the printed object considering the amount of surfaces in contact with support material for each presented orientation. Surface quality rating is between 1 (poor) and 10 (excellent). To compute the true surface finish, the overhang areas are computed with zero overhang angle. Then, the overhang areas are scaled inversely between 1 and 10 such  that 1 represents large support material contact with poor finish and 10 is very good finish with the least amount of support material contact. 3D printing-3 survey~(Fig.~\ref{crowd:3dpQ3}) asks the same question on an object with more features that increase the difficulty of evaluation.

In the structural design survey, participants are presented with eight different bracket designs intended to support a downward force at the end of the bracket~(Figure~\ref{crowd:topOptQ}). Then, they are asked to rate the strength of each bracket on a scale from 1 (weak) to 5 (strong), where strength is defined to be the amount of deformation under the given load~\cite{burnap2015crowdsourcing}. The main reason  we use this problem is that estimating the strength of arbitrary shapes is significantly more demanding  compared to volume/area evaluations. While humans are exposed to volume/area computations in daily life, rating the strength of an arbitrary design requires a specific experience~\cite{nobel2016improving}, which is highly unlikely to be prevalent in the general population. 

\section{Crowd Estimate Aggregation Techniques}
The choice of aggregation method affects the collective estimate of the group. For instance, previous studies show that the median or geometric mean can result in  estimates that are more accurate over the arithmetic mean~\cite{galton1907ballot,lorenz2011social}. This section explains  the different  aggregation methods used in this work. 

The following metrics are used: Arithmetic mean, geometric mean, median, majority voting and Bayesian networks. In a crowd of $n$ participants with a set of estimates $Y: y_1, ..., y_n$ where $y_i \in \mathbb{Z}: 1\leqslant y_i\leqslant 10$ for all $i$, the arithmetic mean is $y^{agg} = \frac{1}{n} \sum_{j=1}^{n}y_i$. The geometric mean is $\exp( \frac{1}{n} \sum_{j=1}^{n}\ln (y_i))$. The median is the median value in $Y$. The majority vote is the mode of $Y$. 

\begin{figure}
  \centering  
  \includegraphics[trim = 0in 0in 0in 0in, clip, width=0.5\columnwidth]{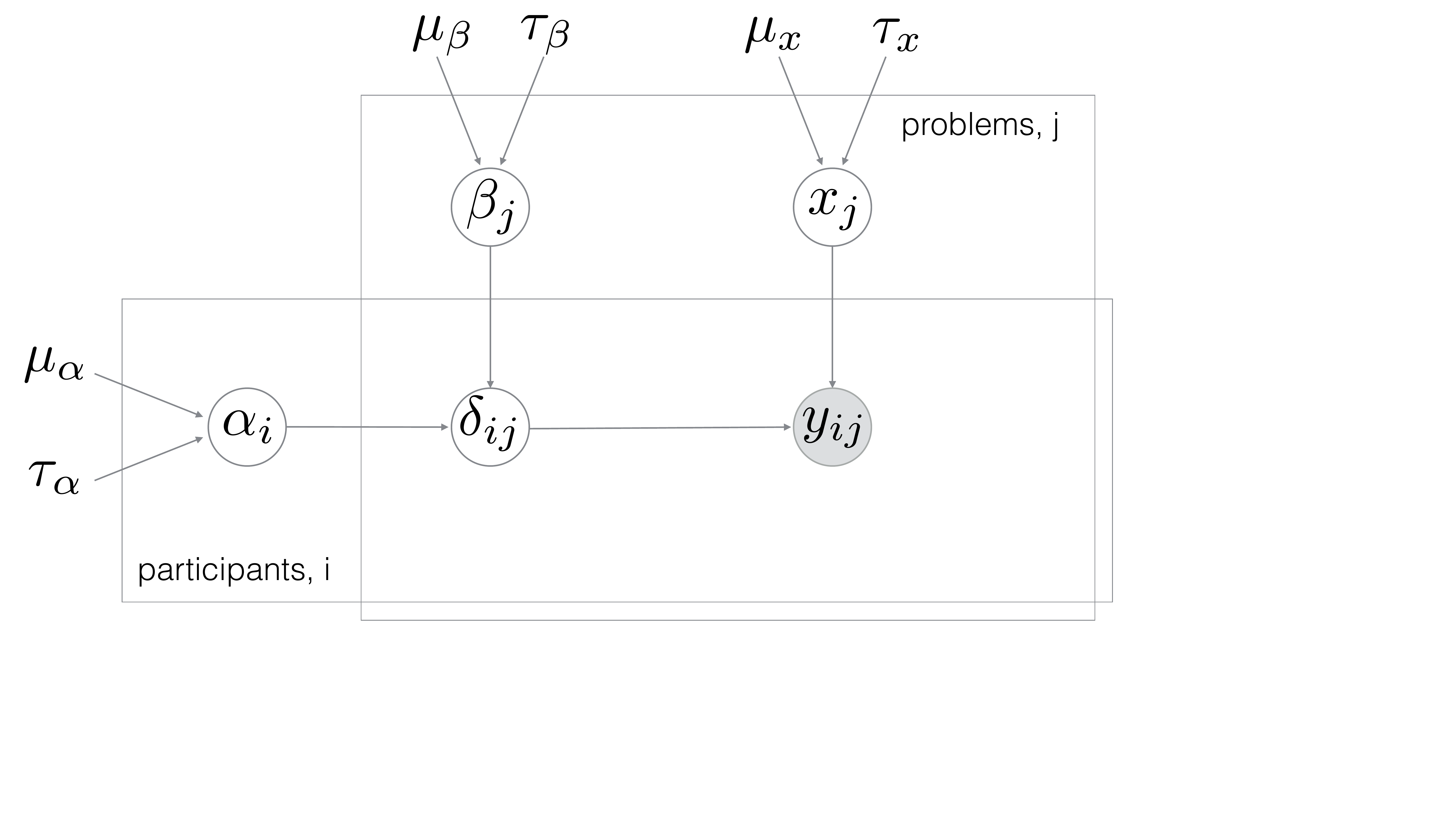}
  \caption{The Bayesian network model.}
  \label{crowd:model}
\end{figure}

Bayesian networks have been widely used in crowdsourcing to mitigate the noise from biased responses. Relevant studies model the sources of bias using models that consider problem difficulty and the competence of participants~\cite{whitehill2009whose,wah2006crowdsourcing,bachrach2012grade,welinder2010multidimensional,lakshminarayanan2013inferring,wauthier2011bayesian,burnap2015crowdsourcing}. Similar to these approaches, this work adopts a Bayesian model as shown in~Fig.~\ref{crowd:model}. The evaluation process is modeled such that for participant $i$ working on problem $j$, participant expertise $\alpha_i$, and problem difficulty $\beta_j$, result in evaluation error $\delta_{ij}$. The evaluation of participant $i$ on problem $j$, $y_{ij}$, is obtained when the true score of the problem, $ x_j$ is combined with the evaluation error, $\delta_{ij}$. Note that the Bayesian model  does not require prior knowledge of the true answers, participant expertise or problem difficulty. The only observed variable is the participant answer for each question.

The evaluation error is obtained using participant expertise and problem difficulty. This work assumes that a participant may be malicious, inexperienced or experienced. Also, a problem can be easy, difficult or unintuitive. Defining both parameters on a continuous range, the evaluation error is modeled as follows:

\begin{equation}
\delta_{ij} = \frac{exp(-\alpha_i / \beta_j)}{1+exp(-\alpha_i / \beta_j)}
\end{equation}

\noindent where the participant expertise is modeled by the parameter $\alpha_i \in (-inf,+\inf)$ and the problem difficulty is $\beta_j \in (0,+\inf)$. The resulting evaluation error becomes $\delta_{ij} \in [0,1]$. The evaluation process is modeled as a random variable with a truncated Gaussian distribution around the true score~($\mu = x_j$) with a variance as evaluation error, $\delta_{ij}$. To bring everything into the same scale, evaluations, $y_{ij}$, are scaled to $[0,1]$ from the original survey scale. The true scores are also represented as  $x_j \in [0,1]$.

\begin{figure}
  \centering  
  \includegraphics[trim = 0in 0in 0in 0in, clip, width=0.5\columnwidth]{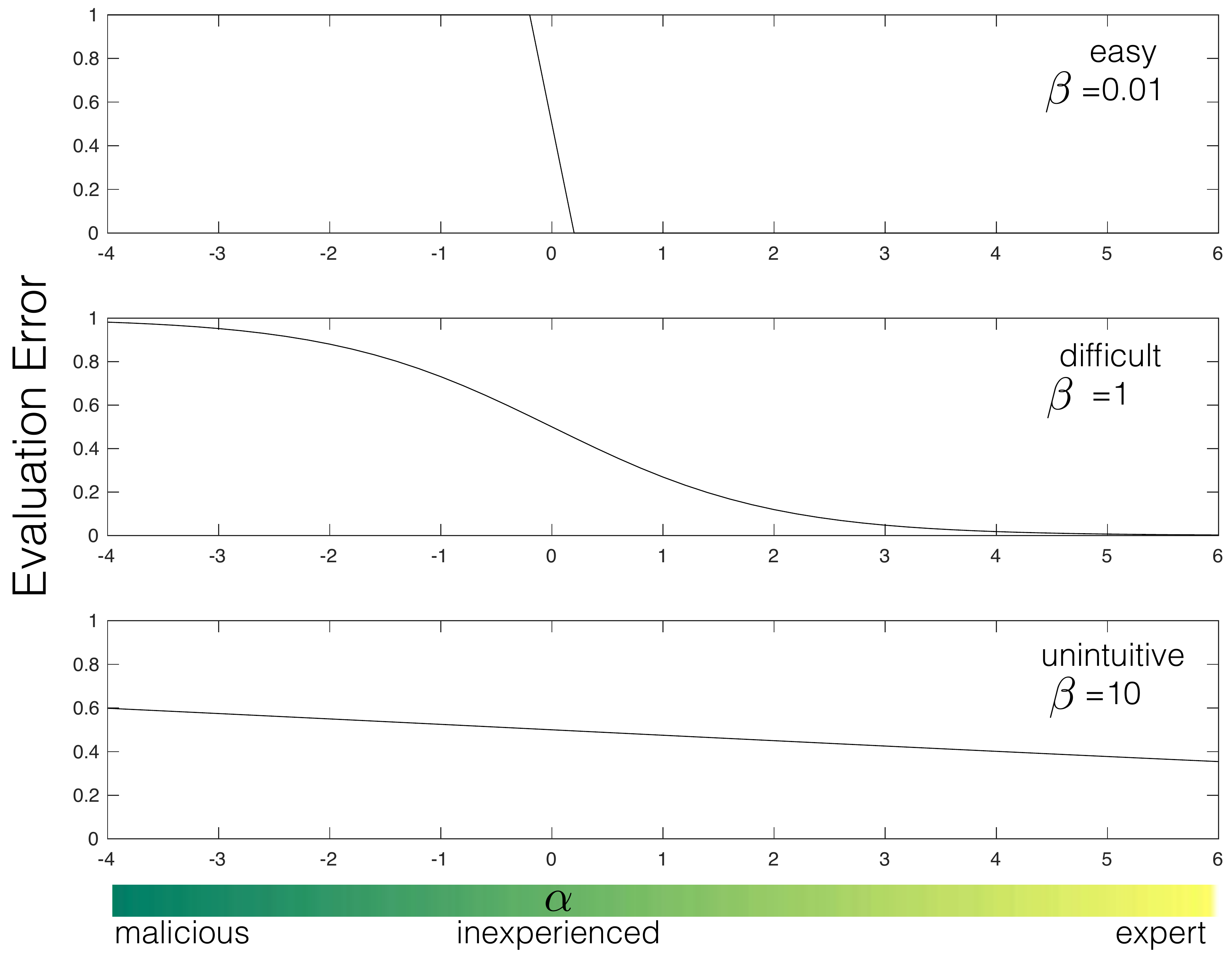}
  \caption{The estimation error variance with participant expertise shown at three different problem difficulty levels as easy, difficult and unintuitive.}
  \label{crowd:expertise_difficulty}
\end{figure}

The  relationship between the evaluation error with participant expertise and problem difficulty is further explained in Figure~\ref{crowd:expertise_difficulty}. The variance of evaluation error with respect to participant expertise is presented for three problem difficulty levels that correspond to easy, difficult and unintuitive. A similar trend can be observed in the variance of evaluation error with varying participant expertise for all problem difficulty levels. As anticipated, participants with high expertise provide accurate answers with very small errors while non-experts can give answers with large errors. Yet there is a potential for malicious participants who intentionally give the wrong answers. Since the answers are maliciously wrong, the amount of error is even more than that of a non-expert that randomly guesses the answers. On the other hand, for a very easy question, even unskilled participants can give answers with a small error and anyone malicious can make the most damage~(Figure~\ref{crowd:expertise_difficulty}-Top). As the questions get more difficult, expertise affects the accuracy of answers more~(Figure~\ref{crowd:expertise_difficulty}-Mid). Yet, an unintuitive question can not be evaluated with good accuracy by participants at any skill level and evaluated with similar error values since all participants evaluate the problem with random guesses~(Figure~\ref{crowd:expertise_difficulty}-Bottom).

The causal structure explained above leads to the graphical model shown in~Figure~\ref{crowd:model}. In the model, participant expertise, $\alpha_i$, problem difficulty, $\beta_j$, and true scores, $ x_j$, are sampled from a known prior distribution and these determine the observed evaluations,  $y_{ij}$. Given a set of observed evaluations, the task is to infer the most likely values of true scores, $ x_j$, together with the participant expertise, $\alpha_i$, and problem difficulty, $\beta_j$, parameters. Assuming a Bayesian treatment with priors on all parameters, the joint probability distribution can be written as

\begin{equation}
\begin{aligned}
p(\mathbf{y},\mathbf{x},\mathbf{\delta},\mathbf{\alpha},\mathbf{\beta})  =& \prod_i p(\alpha_i) \prod_j p(\beta_j) p(x_j)\\
&  \prod_{ij} p(y_{ij}|\delta_{ij},x_j) p(\delta_{ij}|\alpha_i,\beta_j)
\end{aligned}
\end{equation}

The model excludes hyper-parameters for brevity. In our implementation, we use Gaussian priors for $\alpha$ with mean, $\mu_{\alpha} = 1$, and precision, $\tau_{alpha} = 1$. Since the value of $\beta$ needs to be positive, the implementation imposes a truncated Gaussian prior with mean, $\mu_{\beta} = 1$, and precision, $\tau_{beta} = 1$, with a lower bound as $+\epsilon$. For the true scores, $x_j$, we use a truncated Gaussian with bounds $[0,1]$, mean $\mu_x = 0.5$ and precision $\tau_x=0.1$.

Markov Chain Monte Carlo~(MCMC) simulations are employed to infer the results utilizing a Metropolis step method. Empirically, we observe that using thinning interval of 3 and burn-in length of $10^5$ works well with $5\times10^5$ iterations.


\begin{table*}[t]
\caption{The wisdom of crowd effect exist in engineering problems with expert groups and Bayesian model gives the best estimate in most cases. While AMT groups result in high errors that suggest poor accuracy, no statistical aggregation method consistently performs better.}
\begin{center}
\label{crowd:table_all_RMS}
\begin{tabular}{l c c c c c}
& & \multicolumn{3}{c}{RMS error in crowd estimation} &  \\ 
\cline{2-6}
Question                 & Arithmetic & Geometric  & Median & Majority  & Bayesian \\
                                &  mean        &  mean         &               &  voting    &  model\\

\hline
3D printing-1, practitioner    & 0.111 & 0.091 & 0.136 & 0.079 & 0.055\\
3D printing-1, AMT       & 0.403 & 0.378 & 0.430 & 0.336 & 0.363\\
3D printing-2, practitioner    & 0.202 & 0.236 & 0.197 & 0.163 & 0.113\\
3D printing-2, AMT       & 0.438 & 0.462 & 0.473 & 0.540 & 0.600\\
3D printing-3, practitioner    & 0.196 & 0.198 & 0.136 & 0.111 & 0.116\\
3D printing-3, AMT       & 0.402 & 0.431 & 0.363 & 0.453 & 0.561\\
Structural Mech., practitioner    & 0.197 & 0.217 & 0.198 & 0.342 & 0.173 \\
Structural Mech., AMT       & 0.339 & 0.352 & 0.385 & 0.395 & 0.392\\
\hline
\end{tabular}
\end{center}
\end{table*}

\begin{figure*}
  \centering  
  \includegraphics[trim = 0in 0in 0in 0in, clip, width=\textwidth]{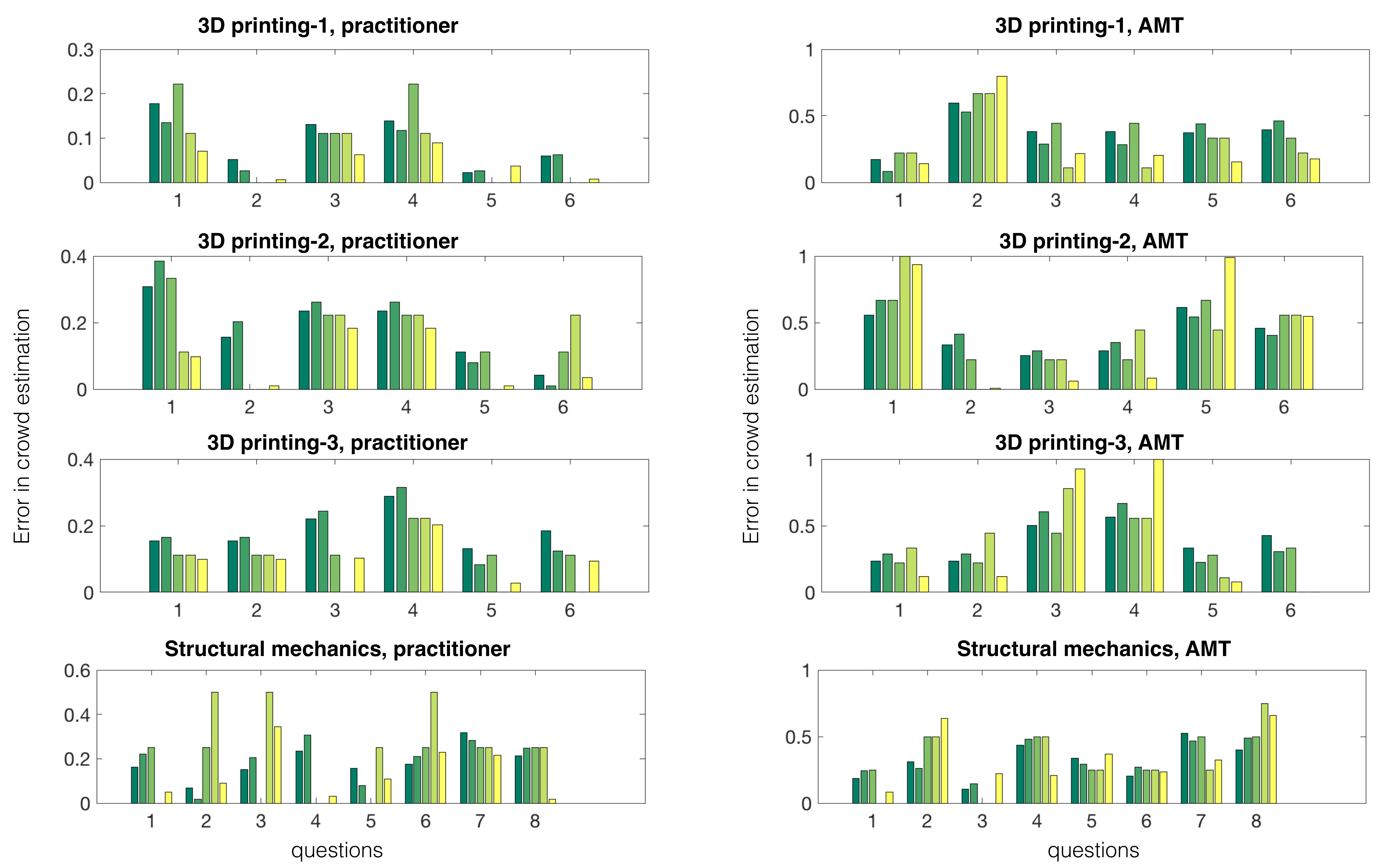}
  \caption{Error of crowd estimation for each question in the four survey groups. Each bar group represents the error of the crowd estimation aggregated through arithmetic mean, geometric mean, median, majority voting and Bayesian model, respectively.}
  \label{crowd:eachQ}
\end{figure*}

\section{Results}
\label{sec:crowd:results}

To demonstrate the WoMC in esoteric engineering problems, we  conducted four surveys on two sets of crowds (practitioners and AMT workers) having different skill levels as explained in the previous sections. This section presents the results of the surveys and compares the performance of the aggregation methods.

\textbf{Survey results.~}
The results of the surveys with different crowds and aggregation methods are summarized in Table~\ref{crowd:table_all_RMS}. All scores are scaled between 0 and 1 for direct comparison across surveys. In addition to the overall survey results, Figure~\ref{crowd:eachQ} includes estimation errors for each  question in the surveys. While the collective error can be defined as the difference between the true answer and the aggregated answer $(y^t-y^{agg})$ for a single question, this work uses root mean square (RMS) error
for multi-question surveys since it provides a performance measure in the same scale as the individual questions. For a survey containing $m$ questions, the collective error can be computed as $\sqrt{\frac{1}{m} \sum_{j=1}^{m}(y^t_j-y^{agg}_j)^2}$. Note that the participant responses  are discrete scores rather than continuous variables. While arithmetic mean, geometric mean, and Bayesian networks produce a real number  from discrete inputs, median and majority voting  remain  discrete values. For consistency, we compare continuous and discrete aggregates with true continuous answers and their rounded values, respectively.

\begin{figure*}[t]
  \centering  
  \includegraphics[trim = 0in 0in 0in 0in, clip, width=\textwidth]{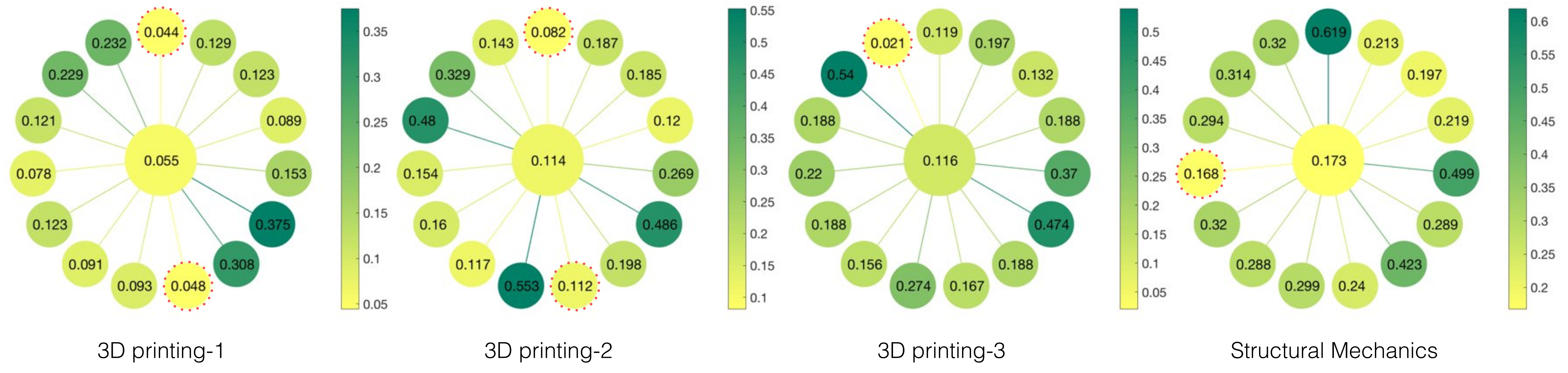}
  \caption{Estimation error significantly varies in the practitioner group. Collective estimate of the practitioner crowd is more accurate than vast majority of individual practitioners. Collective error of the crowd and errors of individual practitioners in the crowd are given in the center node and surrounding nodes, respectively. The color of the circles represents the error and individuals who perform better than the collective answer are marked with a dashed circle. The results for 3d printing-1,2,3 and structural mechanics questions are given from left to right.}
  \label{crowd:expertNetworks}
\end{figure*} 
 
 \textbf{Crowd expertise and aggregation methods.~}
As shown in Table~\ref{crowd:table_all_RMS}, with the  AMT  groups, there is no accurate estimations with any of the aggregation methods, with RMS errors around 40\% and as high as 60\%. Moreover, the Bayesian network method is outperformed by the other methods in all of the AMT studies. This outcome is consistent with previous findings that argue crowdsourcing  AMT populations for engineering design evaluations may produce unreliable results~\cite{burnap2015crowdsourcing}. On the other hand, the results of the practitioner  studies suggest that crowdsourcing can indeed be useful for the same kinds of problems, where consistently more accurate estimations are obtained relative to the AMT groups.

\begin{table}
\caption{Percentile rank of crowd estimation in individual estimations for the practitioner crowd.}
\begin{center}
\label{crowd:table_Percentile}
\begin{tabular}{l c c}
& \multicolumn{2}{c}{Percentile rank of crowd estimation}\\ 
\cline{2-3}
Question & Continuous & Discrete \\
\hline
3D printing-1    &87\% &100\%\\
3D printing-2    &87\% & 93\%\\
3D printing-3    &93\% & 93\%\\
Structural Mech. &93\% &100\%\\

\hline
\end{tabular}
\end{center}
\end{table}

When the aggregation methods are compared,  no single method appears to be best in the AMT studies. On the other hand, for the practitioner groups,  the results indicate that the Bayesian network  consistently produces accurate crowd estimations.  Of note, for both the practitioner and the AMT groups, the geometric mean method never emerges as the best approach. This can be explained by the fact that the responses are constrained within particular upper and lower bounds (1-10 for the 3D printing and  1-5 for the structural design problems) where the range spans only one order of magnitude, whereas the geometric mean is most useful when input data varies in orders of magnitude~\cite{lorenz2011social}.

\textbf{WoMC and individuals.~}
To analyze the WoC effect, the performance of the aggregated crowd estimation is compared against the  individuals~(Fig.~\ref{crowd:expertNetworks}). Only practitioner crowds are included in this analysis as we do not observe a reasonable accuracy in AMT surveys. The collective answers aggregated with Bayesian networks are employed as they consistently perform well in practitioner group studies.

Figure~\ref{crowd:expertNetworks} shows that the collective estimation of the crowd is more accurate than most of the individuals\footnote{WoC is not expected to outperform all individuals. Rather, its effectiveness is proportional to the fraction of individuals  it is able to outperform. In actual use, which individuals have the best answer is unknown.}. Note that the practitioner group is composed of individuals with different skill levels and estimation errors significantly vary in the group. This confirms that Bayesian networks can produce an accurate measure of the WoC for the problems that are of esoteric nature. This can be explained by the participant expertise and problem difficulty based inference that considers all answers of an individual to multiple questions collectively rather than a single one. Moreover, these results suggest that the Bayesian networks approach does not undermine the WoC effect by erroneously honing in on only an elite group of experts in the group, and instead allows  diverse perspectives to be incorporated. This can be explained by the fact that the level of expertise is not prescribed but rather inferred as a latent variable in the Markov Chain Monte Carlo simulations.


Table~\ref{crowd:table_Percentile}  further quantifies the WoC effect by revealing the fraction of people that are outperformed by the collective answer. A higher percentile  suggests that a higher fraction of individuals are outperformed, hence a stronger WoC effect is achieved. The percentile rank  of the crowd is computed  using two error metrics as continuous and discrete: the continuous percentile rank computed as the distance between the true answers and participant ratings; the discrete measure rounding the true answers to the nearest integer while computing the individual estimate errors. Note that the discrete measure can be significantly affected  by these round off errors. The difference between continuous and discrete percentile ranks can be explained by this fact. Of note is the distinction between the percentile rank and the accuracy of the collective estimate. The percentile rank reveals the relative performance of the collective estimate compared to the individual estimates, while the accuracy refers to the RMS error between the estimate and ground truth benchmark.

\begin{figure}
  \centering  
  \includegraphics[trim = 0in 0in 0in 0in, clip, width=0.5\columnwidth]{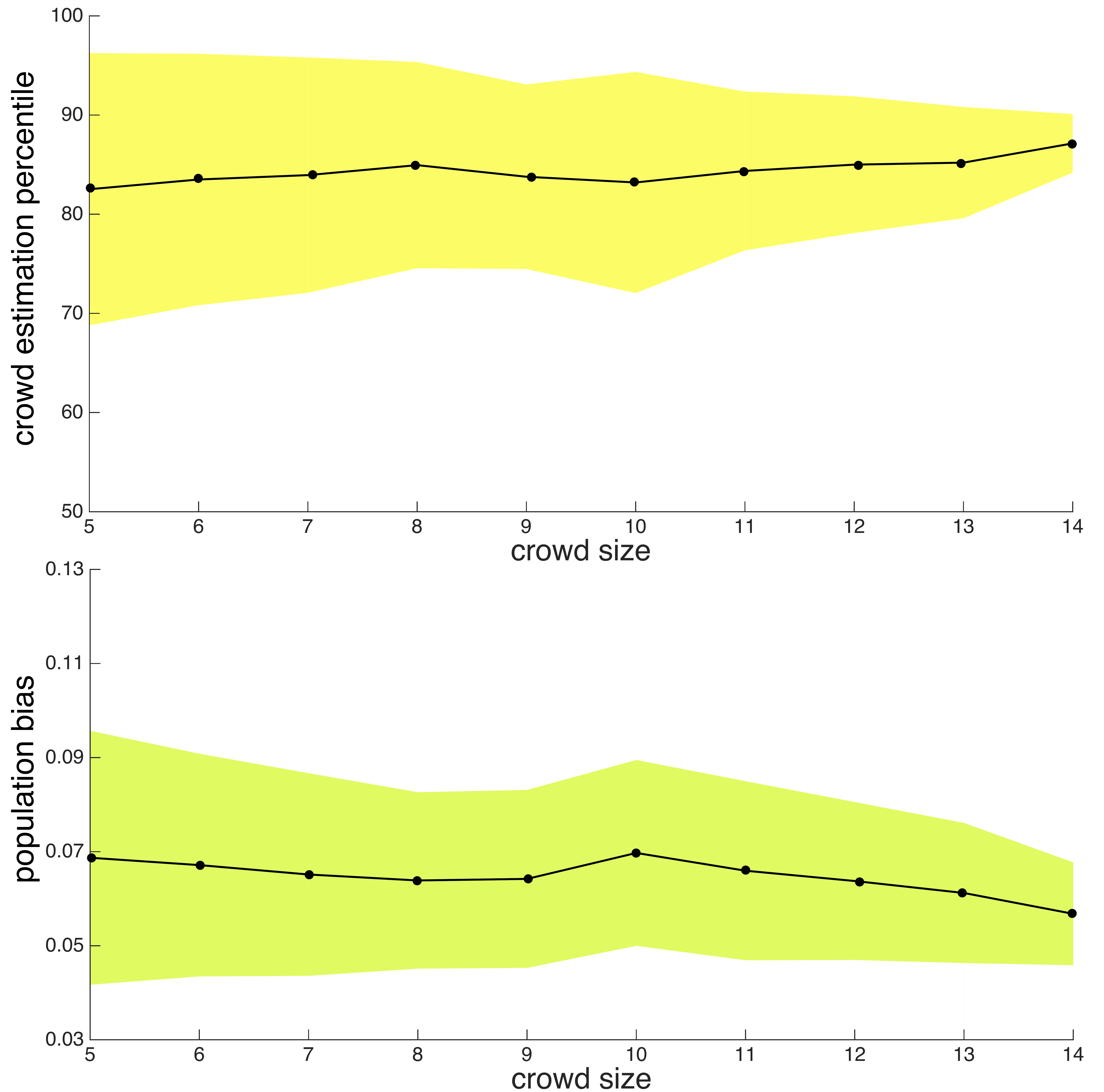}
  \caption{ Effect of crowd size on the success of crowd answer success as percentile and population bias. A slightly increasing trend in the percentiles and a significant decrease in the standard deviation (yellow shaded) as the crowd size increases suggests that higher percentile ranks can be achieved with higher probability in larger crowds.}
  \label{crowd:crowdSize}
\end{figure}

\textbf{Effect of crowd size.~}
Platforms such as AMT enable access to large and diverse groups. However, in most practical problem-solving settings,  only a limited number of practitioners are likely to be accessible for the solution of the engineering challenge. To gain insight into the impact of small-sized practitioner groups, we analyze the WoC effect across even smaller group sizes, leading to the term micro-crowds (WoMC). 

Figure~\ref{crowd:crowdSize} shows that  WoMC can still be observed in smaller groups. The crowd size is analyzed with the 3D printing-1 survey and crowd estimation computed using Bayesian networks~(Table\ref{crowd:table_all_RMS}). Initially, practitioner studies are conducted with 15 participants. To simulate micro-crowds with smaller number of participants, a subset of 500 randomly generated  combinations of 5 to 14 individuals were generated from the original 15 participant set. The results suggest that the WoC effect can still be observed in diminishing group sizes. The probability of obtaining crowd estimations with higher success (percentile) increases with larger crowds. An approximately 6\% increase in percentile rank with 10\% decrease in standard deviation is observed as the crowd size is increased from 5 to 14.  Figure~\ref{crowd:crowdSize} also shows the effect of crowd size on population bias, defined as the error of aggregated estimate across the crowd~\cite{vul2008measuring}.  Both the mean and standard deviation slightly decrease with the increasing crowd size.

\textbf{Conceptual design evaluations.~}
As an extension of the methods presented in this work, the feasibility of using a practitioner-sourced Bayesian network model within the context of conceptual designs was explored. To accomplish this, a practitioner evaluation study was run in which each individual practitioner evaluated a pre-existing set of conceptual design solutions that had also previously been evaluated by two trained experts. Fifteen practitioners were recruited from Carnegie Mellon University, each specializing in Mechanical Engineering (Design focus), or Product Development. Participants were allowed a maximum of 120 minutes to complete the ratings, and were monetarily compensated for their time. 

\begin{figure}
  \centering  
  \includegraphics[trim = 0in 0in 0in 0in, clip, width=0.6\columnwidth]{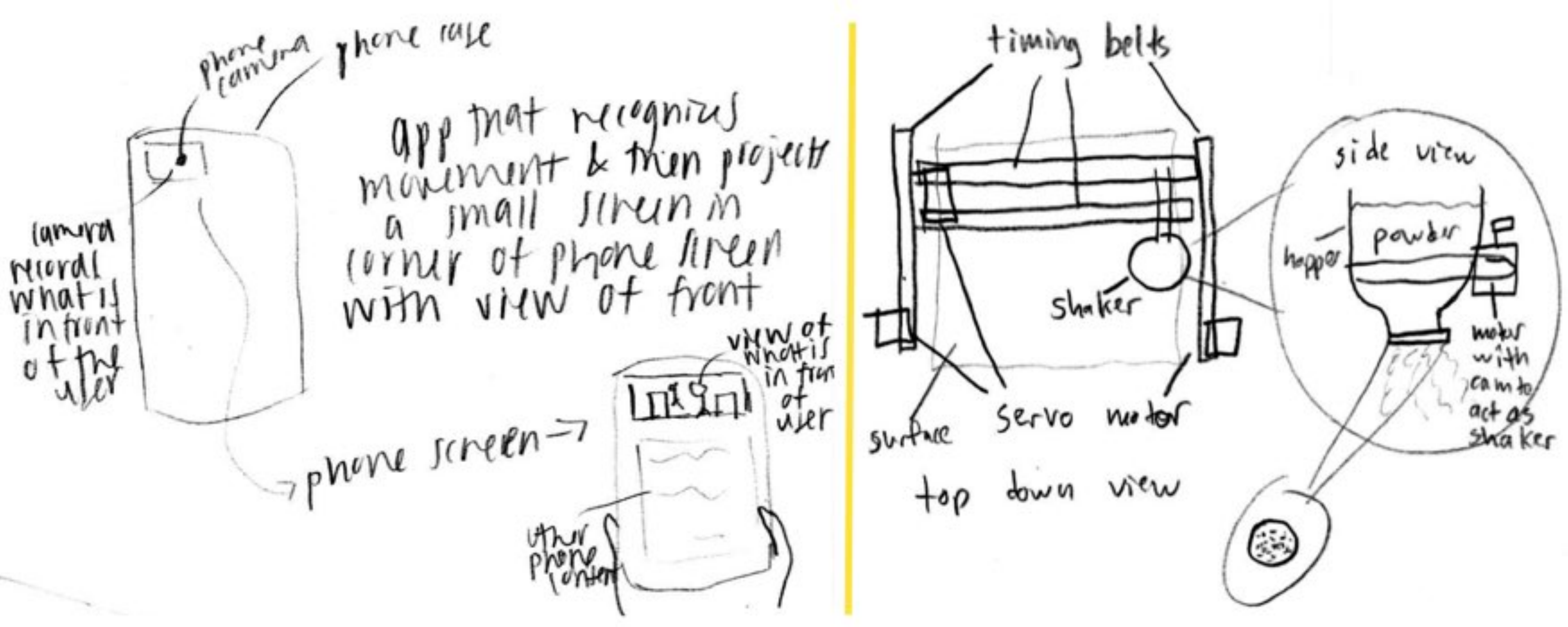}
  \caption{Example conceptual designs.}
  \label{crowd:conceptualQue}
\end{figure}

\begin{table}
\caption{RMS error in crowd estimation for the conceptual design evaluations.}
\begin{center}
\label{crowd:table_Conceptual}
\begin{tabular}{l | c}
\hline
Aggregation method & RMS error \\
\hline
Arithmetic mean     &0.2388\\
Geometric mean     &0.6028\\
Median                    &0.3256\\
Majority voting       &0.3652\\
Bayesian model      &0.3268\\
\hline
\end{tabular}
\end{center}
\end{table}

\begin{figure}
  \centering  
  \includegraphics[trim = 0in 0in 0in 0in, clip, width=0.7\columnwidth]{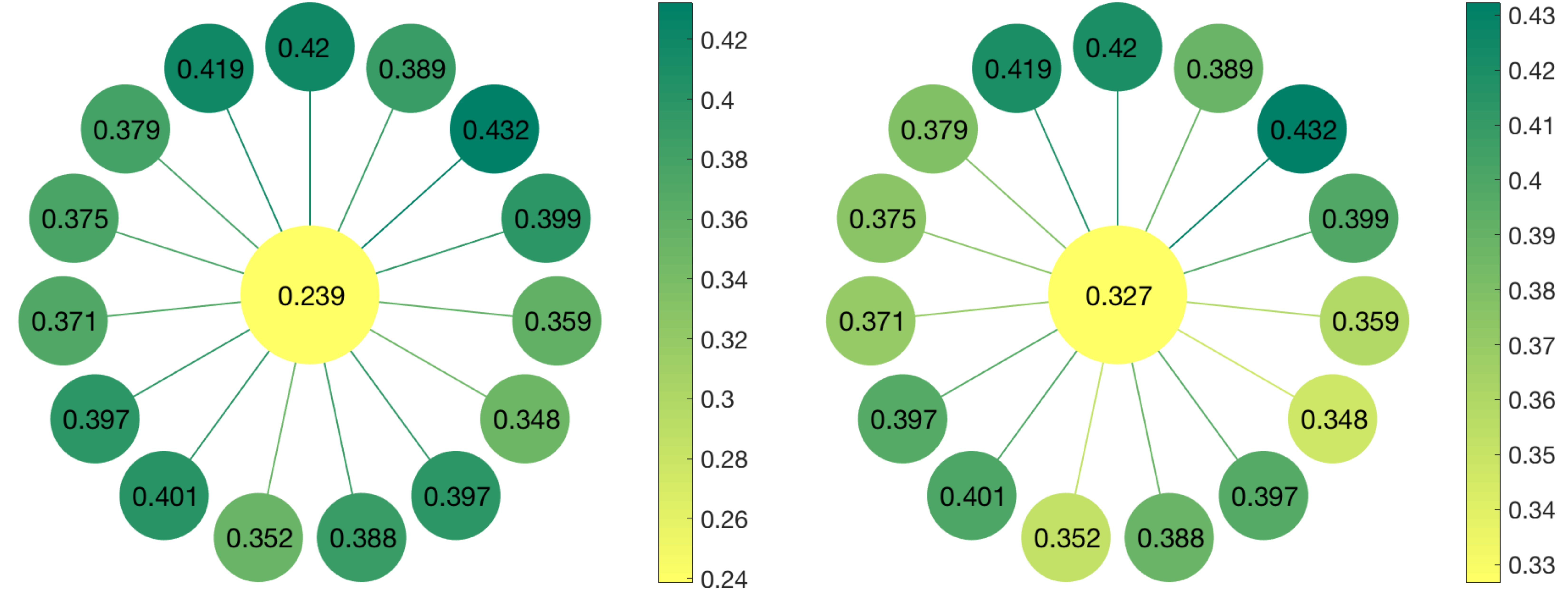}
  \caption{The conceptual design survey illustrates significant estimation errors for each individual practitioner. Individual estimation errors of practitioners are given at the surrounding nodes and the collective estimation error is the center node. Left: arithmetic mean, Right: Bayesian model.}
  \label{crowd:conceptualNet}
\end{figure}

Each practitioner evaluated 114 conceptual designs, corresponding to one of four design problems. These problems are as follows: a device that disperses a light coating of a powdered substance over a surface~\cite{Linsey2008}, a way to minimize accidents from people walking and texting on a cell phone~\cite{Miller2014}, a device to immobilize a human joint~\cite{wilson2010} and a device to remove the shell from a peanut in areas with no electricity~\cite{viswanathan2013design}. This set of conceptual design solutions was taken from a solution set collected for prior work by Goucher-Lambert and Cagan~\cite{goucher2017using}. In that study, inter-rater reliability was assessed using the 114 solution concepts included here. Each design was evaluated across four metrics: usefulness, feasibility, novelty, and quality. Practitioners were provided with one-sentence criteria for each metric (including scoring), and did not see any example solutions prior to rating designs. Example concepts for two of the problems are shown in Figure~\ref{crowd:conceptualQue}. The goal here is to determine the accuracy of the Bayesian network model for a class of problems with extremely low structural and functional similarity.

Table~\ref{crowd:table_Conceptual} summarizes the collective estimation errors aggregated with different methods. Here, the Bayesian model does not perform well and is outperformed by arithmetic mean. In addition to the large collective estimation errors, Figure~\ref{crowd:conceptualNet} illustrates that individual estimation errors of practitioners are also significantly large.

\section{Discussion}
\label{sec:crowd:discussion}

The analyses conducted identified some key insights on how WoMC can be achieved in esoteric engineering problems, highlighted as follows.

\textbf{Problem intuitiveness and difficulty.~}
All of the surveys require specific knowledge about the engineering problem at hand but they range in intuitiveness and difficulty levels. 3D printing questions are based on qualitative area/volume estimations in 3D scenes, which humans are expected to be relatively comfortable with. On the other hand, the structural design problem is significantly more demanding since  estimating the strength of complex geometries  requires a deeper familiarity and experience within the domain~\cite{nobel2016improving}. While individuals are able to make more accurate estimations in the 3D printing questions than they can in the structural design question, an interesting observation is that no significant difference in the wisdom of crowds~(\ie~ percentile rank) is observed implying that crowdsourcing works equally effective in both cases. Moreover, no significant difference between the results of 3D printing-2 and 3D printing-3 surveys occur, even though the latter is more demanding with a larger number of geometrical features. These results suggest that even for problems that are demanding, the WoC is attainable at levels comparable to those attained in less demanding problems. 

\textbf{Level of expertise.~}
Populations of ordinary people (\eg~AMT crowds) perform poorly on esoteric engineering problems. Results indicate that the wisdom of crowds can be achieved in practitioner micro-crowds of the domain of such problems. This suggests that people who are still gaining experience in the domain may prove to be a valuable asset as problem solvers. This is especially important as practitioner crowds may be more accessible than experts.

\textbf{Aggregation methods.~}
In the context of practitioner populations, the most effective aggregation method found in this work is the Bayesian network. For practitioner groups, the exposure to the domain of the esoteric problem builds \emph{true} consistency in the data and allows the Bayesian network to mitigate the mistakes made by individual practitioners. In the AMT groups, however, we observe consistently wrong answers due to lack of expertise. For that reason, Bayesian network method performs worse than arithmetic mean here for AMT populations as also discussed in~\cite{burnap2015crowdsourcing}. This work indicates that Bayesian network method is more effective given a minimum level of expertise in the group. 

\textbf{Crowd size.~}
As shown in Figure~\ref{crowd:crowdSize},  as the crowd size increases, the mean percentile performance increases (albeit modestly) while the standard deviation of the percentile rank of the group estimates decreases over sets of different micro-crowds. This indicates  larger practitioner crowds will likely lead to better and more consistent outcomes. On the performance of WoMC on an absolute scale, our results indicate group estimates in the 90$^{th}$ percentile can be achieved with as few as 5 to14 practitioners. This suggests that in cases where computational tools are not readily available, high quality assessments on engineering problems can be gleaned from small groups of practitioners. 

\textbf{Conceptual design evaluations.~}
When assessing solutions to a set of open-ended, conceptual problems, practitioner crowds struggle to give answers at a level that experts do. For these problems, estimation error in crowd estimation aggregated with the Bayesian model is significant and it is outperformed by arithmetic mean. Looking into individual estimation errors gives an insight into why the Bayesian model is not performing well for these conceptual designs that lack the structural and functional similarity. Figure~\ref{crowd:conceptualNet} demonstrates that every individual in the practitioner group makes a significant estimation error. Even though the estimation aggregated through the Bayesian model is better than all individuals, it is still very high due to large estimation errors of each practitioner.  In contrast to the previous esoteric engineering problems, conceptual design problems have no \textit{true} solution. We believe the open ended nature of conceptual design problems creates a challenge for consistent evaluation in crowd sourced environments and requires further exploration.

\textbf{Crowdsourcing scenarios in esoteric domains.~}
\mycolor{In engineering, crowdsourcing is used in the form of grand challenges to gather candidate solutions but the crowdsourced solutions are assessed by experts juries. However, use of experts to assess these solutions may not be ideal since these grand challenges are created for very complex problems that can not be solved by experts. The idea of wisdom of micro-crowds with practitioners that is presented here can be an alternative to experts juries. This way, both the generation and assessment of the candidate solutions can be crowdsourced through practitioner crowds that are exposed to the esoteric domain. Another esoteric crowdsourcing scenario includes online communities in advancing fields. Thingiverse, an online community for 3D printing designs or GrabCAD, an online community for sharing CAD designs can be examples of communities of advancing esoteric fields. These communities already include large groups of people that are familiar with their respective domains, practitioners. Such communities can benefit from the utility of our work to asses candidate designs that may in fact produce successful outcomes, which would be critically important in cases where no appropriate computational evaluation techniques exist for problems in advancing fields. }

\section{Conclusions}
\label{sec:crowd:conclusions}

This work explored the ability of crowdsourced populations to estimate accurate values for a variety of esoteric problems within the domain of engineering design. Results demonstrate that the wisdom of crowd is most effective in practitioner groups, or groups of individuals who possess some level of domain knowledge, but are not necessarily experts. Aggregated crowd results of practitioners achieve high accuracy across a range of problems. By simulating small groupings of 5 to15 practitioners, called micro-crowds, it is found that crowd estimates perform more accurately than individual estimates across the majority of the studies. These results suggest that the WoMC can provide a powerful tool for answering difficult problems in which computational methods have not been established. In addition, these results argue for the establishment of online communities of practitioners, which could facilitate the solution of future engineering challenges. However, the results also suggest that the practitioner crowds struggle to evaluate open-ended conceptual design problems at a level that experts do. An open research questions is thus the utility of crowdsourcing  for problems involving open-ended synthesis.

\chapter{Conclusions}
\label{chap:conclusions}

Digital fabrication includes all steps from conceptual ideation to manufacturing physical prototypes. Hence, enhancing utilization of digital fabrication is a holistic system problem across the process, from design to fabrication and not just about using computer controlled equipment. Despite the increasing availability of fabrication equipment and services, true power of digital fabrication is hindered by the lack of computational tools that can support the engagement of the general population in content creation. The overarching objective of this thesis has been to develop computational methods in order to enhance overall digital fabrication experience and catalyze its widespread dissemination. Toward this goal, this thesis has explored two key aspects of content creation, (1) function-driven design (2) design assessment.

To overcome challenges in design aspect, first, a generative shape modeling framework that facilitates easy geometry specification and modification for novice users is introduced. The results of this study indicates that this approach works well for a variety of design problems  with the presented actual 3D printed results alongside their digital models. In addition, the user study conducted supports the practical usage of the presented framework. A critical observation that arises from this study is that the most geometrical specifications are dictated by functional requirements. 

This observation lends itself naturally to the immediate insight of prescribing function instead of geometry. To support design by high-level functional specifications, a physics based shape optimization method for compliant coupling behavior design has been developed. In this part, we presented a method for computationally designing the mechanical coupling behavior between a rigid object and a compliant enclosure based on high-level functional specifications such as the ease of engagement and grip

In line with the idea of function-driven design, producing complex 3D surfaces from flat 2D sheets by exploiting the concept of buckling beams has also been explored. This work facilitates a simpler and convenient fabrication of complex 3D surfaces. We believe this work will pave the way for faster manufacturing and convenient flat shipping of free-form 3D surface products.

Design evaluation, the second key aspect, presents itself crucial in problems where computational solutions may currently not exist. For these problems, this work investigates crowdsourcing as a way to empower non-experts in esoteric design domains that traditionally require expertise and specialized knowledge. Here, we explore the ability of crowdsourced populations to estimate accurate answers for a variety of esoteric problems within the domain of engineering design. With this work, we show that our Bayesian model can be used to produce crowd estimations that exceed the accuracy of individuals in the vast majority of instances indicating a wisdom of crowd effect. Our observations also show that the wisdom of crowd effect is maintained for micro-crowds of practitioners, 4-15 individuals, for less intuitive problems suggesting that the challenging engineering design problems can vastly benefit from crowdsourcing.

\appendix\pagestyle{plain}


\bibliographystyle{IEEEtran}

\bibliography{content/references}

\begin{thebibliography}{100}
\providecommand{\url}[1]{#1}
\csname url@samestyle\endcsname
\providecommand{\newblock}{\relax}
\providecommand{\bibinfo}[2]{#2}
\providecommand{\BIBentrySTDinterwordspacing}{\spaceskip=0pt\relax}
\providecommand{\BIBentryALTinterwordstretchfactor}{4}
\providecommand{\BIBentryALTinterwordspacing}{\spaceskip=\fontdimen2\font plus
\BIBentryALTinterwordstretchfactor\fontdimen3\font minus
  \fontdimen4\font\relax}
\providecommand{\BIBforeignlanguage}[2]{{%
\expandafter\ifx\csname l@#1\endcsname\relax
\typeout{** WARNING: IEEEtran.bst: No hyphenation pattern has been}%
\typeout{** loaded for the language `#1'. Using the pattern for}%
\typeout{** the default language instead.}%
\else
\language=\csname l@#1\endcsname
\fi
#2}}
\providecommand{\BIBdecl}{\relax}
\BIBdecl

\bibitem{gerald2012fa}
\BIBentryALTinterwordspacing
N.~Gershenfeld, ``How to make almost anything: The digital fabrication
  revolution,'' \emph{Foreign Affairs}, vol.~91, no.~6, pp. 43--57, 2012.
  [Online]. Available: \url{http://www.jstor.org/stable/41720933}
\BIBentrySTDinterwordspacing

\bibitem{ulu2018coupling}
\BIBentryALTinterwordspacing
N.~G. Ulu, S.~Coros, and L.~B. Kara, ``Designing coupling behaviors using
  compliant shape optimization,'' \emph{Computer-Aided Design}, vol. 101, pp.
  57--71, 2018. [Online]. Available:
  \url{https://www.sciencedirect.com/science/article/pii/S0010448518301386}
\BIBentrySTDinterwordspacing

\bibitem{ulu2018crowdsourcing}
N.~G. Ulu, M.~Messersmith, K.~Goucher-Lambert, J.~Cagan, and L.~B. Kara,
  ``Wisdom of micro-crowds in evaluating solutions to esoteric engineering
  problems,'' \emph{submitted to Journal of Mechanical Design}, 2018.

\bibitem{ulu2015jvlc}
N.~G. Ulu and L.~B. Kara, ``Generative interface structure design for
  supporting existing objects,'' \emph{Journal of Visual Languages and
  Computing}, vol. 31, Part B, pp. 171--183, 2015.

\bibitem{ulu2015vlc}
------, ``Generative interface structure design for supporting existing
  objects,'' in \emph{International Conference on Distributed Multimedia
  Systems Workshop on Visual Languages and Computing}, 2015.

\bibitem{wang20184DMesh}
G.~Wang, H.~Yang, Z.~Yan, N.~G. Ulu, Y.~Tao, J.~Yu, L.~B. Kara, and L.~Yao,
  ``{4}d mesh inverse design tools for self-assembling non-developable mesh
  surfaces via 4d printing,'' in \emph{submitted to Proceedings of the ACM
  Symposium on User Interface Software and Technology (UIST)}.\hskip 1em plus
  0.5em minus 0.4em\relax ACM, 2018.

\bibitem{zhang2018high}
W.~Zhang, J.~Z. Yu, F.~Zhu, Y.~Zhu, N.~G. Ulu, B.~Arisoy, and L.~B. Kara,
  ``High degree of freedom hand pose tracking using limited strain sensing and
  optical training,'' in \emph{ASME 2018 International Design Engineering
  Technical Conferences and Computers and Information in Engineering
  Conference}, 2018.

\bibitem{arisoy2016data}
E.~B. Arisoy, G.~Ren, E.~Ulu, N.~G. Ulu, and S.~Musuvathy, ``A data-driven
  approach to predict hand positions for two-hand grasps of industrial
  objects,'' in \emph{ASME 2016 International Design Engineering Technical
  Conferences and Computers and Information in Engineering Conference}.\hskip
  1em plus 0.5em minus 0.4em\relax American Society of Mechanical Engineers,
  2016.

\bibitem{musuvathy2017methods}
S.~Musuvathy, G.~Allen, L.~Mirabella, L.~Komzsik, and N.~G. Ulu, ``System and
  method for modeling of parts with lattice structures,'' Patent
  WO2\,017\,088\,134 A1, 2017.

\bibitem{arisoy2017methods}
E.~B. Arisoy, S.~Musuvathy, E.~Ulu, and N.~G. Ulu, ``Methods and system to
  predict hand positions for multi-hand grasps of industrial objects,'' Patent
  WO2\,017\,132\,134 A1, 2017.

\bibitem{sketCHair_11}
\BIBentryALTinterwordspacing
G.~Saul, M.~Lau, J.~Mitani, and T.~Igarashi, ``Sketchchair: An all-in-one chair
  design system for end users,'' in \emph{Proceedings of the Fifth
  International Conference on Tangible, Embedded, and Embodied Interaction},
  ser. TEI '11.\hskip 1em plus 0.5em minus 0.4em\relax New York, NY, USA: ACM,
  2011, pp. 73--80. [Online]. Available:
  \url{http://doi.acm.org/10.1145/1935701.1935717}
\BIBentrySTDinterwordspacing

\bibitem{umetani_sigg12}
N.~Umetani, T.~Igarashi, and N.~J. Mitra, ``Guided exploration of physically
  valid shapes for furniture design,'' \emph{ACM Transactions on Graphics
  (Proceedings of SIGGRAPH 2012)}, vol.~31, no.~4, 2012.

\bibitem{Umetani:Pter:2014}
\BIBentryALTinterwordspacing
N.~Umetani, Y.~Koyama, R.~Schmidt, and T.~Igarashi, ``Pteromys: Interactive
  design and optimization of free-formed free-flight model airplanes,''
  \emph{ACM Trans. Graph.}, vol.~33, no.~4, pp. 65:1--65:10, Jul. 2014.
  [Online]. Available: \url{http://doi.acm.org/10.1145/2601097.2601129}
\BIBentrySTDinterwordspacing

\bibitem{Prevost:2013}
\BIBentryALTinterwordspacing
R.~Pr{\'e}vost, E.~Whiting, S.~Lefebvre, and O.~Sorkine-Hornung, ``Make it
  stand: Balancing shapes for 3d fabrication,'' \emph{ACM Trans. Graph.},
  vol.~32, no.~4, pp. 81:1--81:10, Jul. 2013. [Online]. Available:
  \url{http://doi.acm.org/10.1145/2461912.2461957}
\BIBentrySTDinterwordspacing

\bibitem{Bacher:2014}
\BIBentryALTinterwordspacing
M.~B\"{a}cher, E.~Whiting, B.~Bickel, and O.~Sorkine-Hornung, ``Spin-it:
  Optimizing moment of inertia for spinnable objects,'' \emph{ACM Trans.
  Graph.}, vol.~33, no.~4, pp. 96:1--96:10, Jul. 2014. [Online]. Available:
  \url{http://doi.acm.org/10.1145/2601097.2601157}
\BIBentrySTDinterwordspacing

\bibitem{Stava_2012}
\BIBentryALTinterwordspacing
O.~Stava, J.~Vanek, B.~Benes, N.~Carr, and R.~M\v{e}ch, ``Stress relief:
  Improving structural strength of 3d printable objects,'' \emph{ACM Trans.
  Graph.}, vol.~31, no.~4, pp. 48:1--48:11, Jul. 2012. [Online]. Available:
  \url{http://doi.acm.org/10.1145/2185520.2185544}
\BIBentrySTDinterwordspacing

\bibitem{Wang2002}
\BIBentryALTinterwordspacing
Y.~Wang and J.~P. Duarte, ``Automatic generation and fabrication of designs,''
  \emph{Automation in Construction}, vol.~11, no.~3, pp. 291 -- 302, 2002,
  rapid Prototyping. [Online]. Available:
  \url{http://www.sciencedirect.com/science/article/pii/S0926580500001126}
\BIBentrySTDinterwordspacing

\bibitem{Sass}
\BIBentryALTinterwordspacing
L.~Sass and R.~Oxman, ``Materializing design: the implications of rapid
  prototyping in digital design,'' \emph{Design Studies}, vol.~27, no.~3, pp.
  325 -- 355, 2006, digital Design Digital Design. [Online]. Available:
  \url{http://www.sciencedirect.com/science/article/pii/S0142694X05000864}
\BIBentrySTDinterwordspacing

\bibitem{Muller_2006}
\BIBentryALTinterwordspacing
P.~M\"{u}ller, P.~Wonka, S.~Haegler, A.~Ulmer, and L.~Van~Gool, ``Procedural
  modeling of buildings,'' \emph{ACM Trans. Graph.}, vol.~25, no.~3, pp.
  614--623, Jul. 2006. [Online]. Available:
  \url{http://doi.acm.org/10.1145/1141911.1141931}
\BIBentrySTDinterwordspacing

\bibitem{Parish_2001}
\BIBentryALTinterwordspacing
Y.~I.~H. Parish and P.~M\"{u}ller, ``Procedural modeling of cities,'' in
  \emph{Proceedings of the 28th Annual Conference on Computer Graphics and
  Interactive Techniques}, ser. SIGGRAPH '01.\hskip 1em plus 0.5em minus
  0.4em\relax New York, NY, USA: ACM, 2001, pp. 301--308. [Online]. Available:
  \url{http://doi.acm.org/10.1145/383259.383292}
\BIBentrySTDinterwordspacing

\bibitem{Prusinkiewicz_sigg94}
\BIBentryALTinterwordspacing
P.~Prusinkiewicz, M.~James, and R.~M\v{e}ch, ``Synthetic topiary,'' in
  \emph{Proceedings of the 21st Annual Conference on Computer Graphics and
  Interactive Techniques}, ser. SIGGRAPH '94.\hskip 1em plus 0.5em minus
  0.4em\relax New York, NY, USA: ACM, 1994, pp. 351--358. [Online]. Available:
  \url{http://doi.acm.org/10.1145/192161.192254}
\BIBentrySTDinterwordspacing

\bibitem{Runions_sigg2005}
\BIBentryALTinterwordspacing
A.~Runions, M.~Fuhrer, B.~Lane, P.~Federl, A.-G. Rolland-Lagan, and
  P.~Prusinkiewicz, ``Modeling and visualization of leaf venation patterns,''
  in \emph{ACM SIGGRAPH 2005 Papers}, ser. SIGGRAPH '05.\hskip 1em plus 0.5em
  minus 0.4em\relax New York, NY, USA: ACM, 2005, pp. 702--711. [Online].
  Available: \url{http://doi.acm.org/10.1145/1186822.1073251}
\BIBentrySTDinterwordspacing

\bibitem{Bendsoe:2004}
M.~P. Bends{\o}e and O.~Sigmund, \emph{\BIBforeignlanguage{en}{Topology
  Optimization: Theory, Methods and Applications}}.\hskip 1em plus 0.5em minus
  0.4em\relax Springer, Feb. 2004.

\bibitem{Runions_07}
\BIBentryALTinterwordspacing
A.~Runions, B.~Lane, and P.~Prusinkiewicz, ``Modeling trees with a space
  colonization algorithm,'' in \emph{Proceedings of the Third Eurographics
  Conference on Natural Phenomena}, ser. NPH'07.\hskip 1em plus 0.5em minus
  0.4em\relax Aire-la-Ville, Switzerland, Switzerland: Eurographics
  Association, 2007, pp. 63--70. [Online]. Available:
  \url{http://dx.doi.org/10.2312/NPH/NPH07/063-070}
\BIBentrySTDinterwordspacing

\bibitem{Palubicki_sigg09}
\BIBentryALTinterwordspacing
W.~Palubicki, K.~Horel, S.~Longay, A.~Runions, B.~Lane, R.~M\v{e}ch, and
  P.~Prusinkiewicz, ``Self-organizing tree models for image synthesis,'' in
  \emph{ACM SIGGRAPH 2009 Papers}, ser. SIGGRAPH '09.\hskip 1em plus 0.5em
  minus 0.4em\relax New York, NY, USA: ACM, 2009, pp. 58:1--58:10. [Online].
  Available: \url{http://doi.acm.org/10.1145/1576246.1531364}
\BIBentrySTDinterwordspacing

\bibitem{Longay2012}
\BIBentryALTinterwordspacing
S.~Longay, A.~Runions, F.~Boudon, and P.~Prusinkiewicz, ``Treesketch:
  Interactive procedural modeling of trees on a tablet,'' in \emph{Proceedings
  of the International Symposium on Sketch-Based Interfaces and Modeling}, ser.
  SBIM '12.\hskip 1em plus 0.5em minus 0.4em\relax Aire-la-Ville, Switzerland,
  Switzerland: Eurographics Association, 2012, pp. 107--120. [Online].
  Available: \url{http://dl.acm.org/citation.cfm?id=2331067.2331083}
\BIBentrySTDinterwordspacing

\bibitem{Pirk_sigg12}
\BIBentryALTinterwordspacing
S.~Pirk, O.~Stava, J.~Kratt, M.~A.~M. Said, B.~Neubert, R.~M\v{e}ch, B.~Benes,
  and O.~Deussen, ``Plastic trees: interactive self-adapting botanical tree
  models,'' \emph{ACM Trans. Graph.}, vol.~31, no.~4, pp. 50:1--50:10, Jul.
  2012. [Online]. Available: \url{http://doi.acm.org/10.1145/2185520.2185546}
\BIBentrySTDinterwordspacing

\bibitem{Bickel_sigg2010}
\BIBentryALTinterwordspacing
B.~Bickel, M.~B\"{a}cher, M.~A. Otaduy, H.~R. Lee, H.~Pfister, M.~Gross, and
  W.~Matusik, ``Design and fabrication of materials with desired deformation
  behavior,'' \emph{ACM Trans. Graph.}, vol.~29, no.~4, pp. 63:1--63:10, Jul.
  2010. [Online]. Available: \url{http://doi.acm.org/10.1145/1778765.1778800}
\BIBentrySTDinterwordspacing

\bibitem{Panetta_sigg2015}
\BIBentryALTinterwordspacing
J.~Panetta, Q.~Zhou, L.~Malomo, N.~Pietroni, P.~Cignoni, and D.~Zorin,
  ``Elastic textures for additive fabrication,'' \emph{ACM Trans. Graph.},
  vol.~34, no.~4, pp. 135:1--135:12, Jul. 2015. [Online]. Available:
  \url{http://doi.acm.org/10.1145/2766937}
\BIBentrySTDinterwordspacing

\bibitem{Schumacher_sigg2015}
\BIBentryALTinterwordspacing
C.~Schumacher, B.~Bickel, J.~Rys, S.~Marschner, C.~Daraio, and M.~Gross,
  ``Microstructures to control elasticity in 3d printing,'' \emph{ACM Trans.
  Graph.}, vol.~34, no.~4, pp. 136:1--136:13, Jul. 2015. [Online]. Available:
  \url{http://doi.acm.org/10.1145/2766926}
\BIBentrySTDinterwordspacing

\bibitem{Skouras_sigg2013}
\BIBentryALTinterwordspacing
M.~Skouras, B.~Thomaszewski, S.~Coros, B.~Bickel, and M.~Gross, ``Computational
  design of actuated deformable characters,'' \emph{ACM Trans. Graph.},
  vol.~32, no.~4, pp. 82:1--82:10, Jul. 2013. [Online]. Available:
  \url{http://doi.acm.org/10.1145/2461912.2461979}
\BIBentrySTDinterwordspacing

\bibitem{Perez_sigg2015}
\BIBentryALTinterwordspacing
J.~P{\'e}rez, B.~Thomaszewski, S.~Coros, B.~Bickel, J.~A. Canabal, R.~Sumner,
  and M.~A. Otaduy, ``Design and fabrication of flexible rod meshes,''
  \emph{ACM Trans. Graph.}, vol.~34, no.~4, pp. 138:1--138:12, Jul. 2015.
  [Online]. Available: \url{http://doi.acm.org/10.1145/2766998}
\BIBentrySTDinterwordspacing

\bibitem{Xu:2015:nonlinearMaterial}
\BIBentryALTinterwordspacing
H.~Xu, F.~Sin, Y.~Zhu, and J.~Barbi\v{c}, ``Nonlinear material design using
  principal stretches,'' \emph{ACM Trans. Graph.}, vol.~34, no.~4, pp.
  75:1--75:11, Jul. 2015. [Online]. Available:
  \url{http://doi.acm.org/10.1145/2766917}
\BIBentrySTDinterwordspacing

\bibitem{Xu:2015:materialmodes}
\BIBentryALTinterwordspacing
H.~Xu, Y.~Li, Y.~Chen, and J.~Barbi\v{c}, ``Interactive material design using
  model reduction,'' \emph{ACM Trans. Graph.}, vol.~34, no.~2, pp. 18:1--18:14,
  Mar. 2015. [Online]. Available: \url{http://doi.acm.org/10.1145/2699648}
\BIBentrySTDinterwordspacing

\bibitem{Chen:2014}
\BIBentryALTinterwordspacing
X.~Chen, C.~Zheng, W.~Xu, and K.~Zhou, ``An asymptotic numerical method for
  inverse elastic shape design,'' \emph{ACM Trans. Graph.}, vol.~33, no.~4, pp.
  95:1--95:11, Jul. 2014. [Online]. Available:
  \url{http://doi.acm.org/10.1145/2601097.2601189}
\BIBentrySTDinterwordspacing

\bibitem{Musialski:2015}
\BIBentryALTinterwordspacing
P.~Musialski, T.~Auzinger, M.~Birsak, M.~Wimmer, and L.~Kobbelt,
  ``Reduced-order shape optimization using offset surfaces,'' \emph{ACM Trans.
  Graph.}, vol.~34, no.~4, pp. 102:1--102:9, Jul. 2015. [Online]. Available:
  \url{http://doi.acm.org/10.1145/2766955}
\BIBentrySTDinterwordspacing

\bibitem{Stava:2012}
\BIBentryALTinterwordspacing
O.~Stava, J.~Vanek, B.~Benes, N.~Carr, and R.~M\v{e}ch, ``Stress relief:
  Improving structural strength of 3d printable objects,'' \emph{ACM Trans.
  Graph.}, vol.~31, no.~4, pp. 48:1--48:11, Jul. 2012. [Online]. Available:
  \url{http://doi.acm.org/10.1145/2185520.2185544}
\BIBentrySTDinterwordspacing

\bibitem{Wang:2013}
\BIBentryALTinterwordspacing
W.~Wang, T.~Y. Wang, Z.~Yang, L.~Liu, X.~Tong, W.~Tong, J.~Deng, F.~Chen, and
  X.~Liu, ``Cost-effective printing of 3d objects with skin-frame structures,''
  \emph{ACM Trans. Graph.}, vol.~32, no.~6, pp. 177:1--177:10, Nov. 2013.
  [Online]. Available: \url{http://doi.acm.org/10.1145/2508363.2508382}
\BIBentrySTDinterwordspacing

\bibitem{Lu:2014}
\BIBentryALTinterwordspacing
L.~Lu, A.~Sharf, H.~Zhao, Y.~Wei, Q.~Fan, X.~Chen, Y.~Savoye, C.~Tu,
  D.~Cohen-Or, and B.~Chen, ``Build-to-last: Strength to weight 3d printed
  objects,'' \emph{ACM Trans. Graph.}, vol.~33, no.~4, pp. 97:1--97:10, Jul.
  2014. [Online]. Available: \url{http://doi.acm.org/10.1145/2601097.2601168}
\BIBentrySTDinterwordspacing

\bibitem{ulu2017lightweight}
E.~Ulu, J.~McCann, and L.~B. Kara, ``Lightweight structure design under force
  location uncertainty,'' \emph{ACM Transactions on Graphics (Proc. of SIGGRAPH
  2017)}, vol.~36, no.~4, 2017.

\bibitem{Bharaj:2015}
\BIBentryALTinterwordspacing
G.~Bharaj, D.~I.~W. Levin, J.~Tompkin, Y.~Fei, H.~Pfister, W.~Matusik, and
  C.~Zheng, ``Computational design of metallophone contact sounds,'' \emph{ACM
  Trans. Graph.}, vol.~34, no.~6, pp. 223:1--223:13, Oct. 2015. [Online].
  Available: \url{http://doi.acm.org/10.1145/2816795.2818108}
\BIBentrySTDinterwordspacing

\bibitem{Umetani:duduk:2016}
\BIBentryALTinterwordspacing
N.~Umetani, A.~Panotopoulou, R.~Schmidt, and E.~Whiting, ``Printone:
  Interactive resonance simulation for free-form print-wind instrument
  design,'' \emph{ACM Trans. Graph.}, vol.~35, no.~6, pp. 184:1--184:14, Nov.
  2016. [Online]. Available: \url{http://doi.acm.org/10.1145/2980179.2980250}
\BIBentrySTDinterwordspacing

\bibitem{Koyama:2015}
\BIBentryALTinterwordspacing
Y.~Koyama, S.~Sueda, E.~Steinhardt, T.~Igarashi, A.~Shamir, and W.~Matusik,
  ``Autoconnect: Computational design of 3d-printable connectors,'' \emph{ACM
  Trans. Graph.}, vol.~34, no.~6, pp. 231:1--231:11, Oct. 2015. [Online].
  Available: \url{http://doi.acm.org/10.1145/2816795.2818060}
\BIBentrySTDinterwordspacing

\bibitem{Lu:2003}
\BIBentryALTinterwordspacing
K.-J. Lu and S.~Kota, ``Synthesis of shape morphing compliant mechanisms using
  a load path representation method,'' vol. 5049, 2003, pp. 337--348. [Online].
  Available: \url{http://dx.doi.org/10.1117/12.484020}
\BIBentrySTDinterwordspacing

\bibitem{Lu:2005}
------, ``Topology and dimensional synthesis of compliant mechanisms using
  discrete optimization,'' \emph{Journal of Mechanical Design}, vol. 128,
  no.~5, pp. 1080--1091, 2005.

\bibitem{pedersen2001}
\BIBentryALTinterwordspacing
C.~B.~W. Pedersen, T.~Buhl, and O.~Sigmund, ``Topology synthesis of
  large-displacement compliant mechanisms,'' \emph{International Journal for
  Numerical Methods in Engineering}, vol.~50, no.~12, pp. 2683--2705, 2001.
  [Online]. Available: \url{http://dx.doi.org/10.1002/nme.148}
\BIBentrySTDinterwordspacing

\bibitem{gripper2003topology}
M.~M. Shalaby, H.~A. Hegazi, A.~O. Nassef, and S.~M. Metwalli, ``Topology
  optimization of a compliant gripper using hybrid simulated annealing and
  direct search,'' in \emph{ASME 2003 International Design Engineering
  Technical Conferences and Computers and Information in Engineering
  Conference}.\hskip 1em plus 0.5em minus 0.4em\relax American Society of
  Mechanical Engineers, 2003, pp. 641--648.

\bibitem{gripperfreeform}
D.~Xu and G.~Ananthasuresh, ``Freeform skeletal shape optimization of compliant
  mechanisms,'' \emph{Journal of Mechanical Design}, vol. 125, no.~2, pp.
  253--261, 2003.

\bibitem{gripper98}
S.~Nishiwaki, M.~I. Frecker, S.~Min, and N.~Kikuchi, ``Topology optimization of
  compliant mechanisms using the homogenization method,'' \emph{International
  Journal for Numerical Methods in Engineering}, vol.~42, no.~3, pp. 535--559,
  1998.

\bibitem{Bruns2004}
T.~E. Bruns and O.~Sigmund, ``Toward the topology design of mechanisms that
  exhibit snap-through behavior,'' \emph{Computer Methods in Applied Mechanics
  and Engineering}, vol. 193, no. 36 - 38, pp. 3973 -- 4000, 2004.

\bibitem{Ohsaki2005}
\BIBentryALTinterwordspacing
M.~Ohsaki and S.~Nishiwaki, ``Shape design of pin-jointed multistable compliant
  mechanisms using snapthrough behavior,'' \emph{Structural and
  Multidisciplinary Optimization}, vol.~30, no.~4, pp. 327--334, 2005.
  [Online]. Available: \url{http://dx.doi.org/10.1007/s00158-005-0532-2}
\BIBentrySTDinterwordspacing

\bibitem{lawry2017level}
M.~Lawry and K.~Maute, ``Level set shape and topology optimization of finite
  strain bilateral contact problems,'' \emph{arXiv preprint arXiv:1701.06092},
  2017.

\bibitem{yunConnector}
Y.~Ling, ``Mating mechanics and stubbing of separable connectors,'' in
  \emph{1998 Proceedings. 48th Electronic Components and Technology Conference
  (Cat. No.98CH36206)}, May 1998, pp. 6--13.

\bibitem{hsu2000shape}
Y.-L. H. Y.-L. Hsu, Y.-C. H. Y.-C. Hsu, M.-S. H. M.-S. Hsu, and Y.~Hsu, ``Shape
  optimal design of the contact springs of a connector,'' in \emph{2000 PCB
  Manufacturing Technology Conference, Yuan Ze University, Chung-Li, Taiwan},
  2000.

\bibitem{Kloosterman:2002:contact}
G.~Kloosterman, ``Contact methods in finite element simulations,'' Ph.D.
  dissertation, University of Twente, 2002.

\bibitem{McNeely:1999:boeing}
\BIBentryALTinterwordspacing
W.~A. McNeely, K.~D. Puterbaugh, and J.~J. Troy, ``Six degree-of-freedom haptic
  rendering using voxel sampling,'' in \emph{Proceedings of the 26th Annual
  Conference on Computer Graphics and Interactive Techniques}, ser. SIGGRAPH
  '99.\hskip 1em plus 0.5em minus 0.4em\relax New York, NY, USA: ACM
  Press/Addison-Wesley Publishing Co., 1999, pp. 401--408. [Online]. Available:
  \url{http://dx.doi.org/10.1145/311535.311600}
\BIBentrySTDinterwordspacing

\bibitem{Barbic:2008}
J.~Barbi\v{c} and D.~L. James, ``Six-dof haptic rendering of contact between
  geometrically complex reduced deformable models,'' \emph{IEEE Transactions on
  Haptics}, vol.~1, no.~1, pp. 39--52, Jan 2008.

\bibitem{Kaufman:2008}
\BIBentryALTinterwordspacing
D.~M. Kaufman, S.~Sueda, D.~L. James, and D.~K. Pai, ``Staggered projections
  for frictional contact in multibody systems,'' in \emph{ACM SIGGRAPH Asia
  2008 Papers}, ser. SIGGRAPH Asia '08.\hskip 1em plus 0.5em minus 0.4em\relax
  New York, NY, USA: ACM, 2008, pp. 164:1--164:11. [Online]. Available:
  \url{http://doi.acm.org/10.1145/1457515.1409117}
\BIBentrySTDinterwordspacing

\bibitem{Harmon:2009}
\BIBentryALTinterwordspacing
D.~Harmon, E.~Vouga, B.~Smith, R.~Tamstorf, and E.~Grinspun, ``Asynchronous
  contact mechanics,'' \emph{ACM Trans. Graph.}, vol.~28, no.~3, pp.
  87:1--87:12, Jul. 2009. [Online]. Available:
  \url{http://doi.acm.org/10.1145/1531326.1531393}
\BIBentrySTDinterwordspacing

\bibitem{Tang:2012}
\BIBentryALTinterwordspacing
M.~Tang, D.~Manocha, M.~A. Otaduy, and R.~Tong, ``Continuous penalty forces,''
  \emph{ACM Trans. Graph.}, vol.~31, no.~4, pp. 107:1--107:9, Jul. 2012.
  [Online]. Available: \url{http://doi.acm.org/10.1145/2185520.2185603}
\BIBentrySTDinterwordspacing

\bibitem{AMT}
``{Amazon} mechanical turk,'' \url{https://www.mturk.com}, accessed:
  2017-01-30.

\bibitem{crowdflower}
``{CrowdFlower},'' \url{https://www.crowdflower.com}, accessed: 2017-01-30.

\bibitem{prolificAcademic}
``{Prolific} academic,'' \url{https://www.prolific.ac/}, accessed: 2017-01-30.

\bibitem{berinsky2012evaluating}
A.~J. Berinsky, G.~A. Huber, and G.~S. Lenz, ``Evaluating online labor markets
  for experimental research: Amazon. com's mechanical turk,'' \emph{Political
  Analysis}, vol.~20, no.~3, pp. 351--368, 2012.

\bibitem{kittur2008crowdsourcing}
A.~Kittur, E.~H. Chi, and B.~Suh, ``Crowdsourcing user studies with mechanical
  turk,'' in \emph{Proceedings of the SIGCHI conference on human factors in
  computing systems}.\hskip 1em plus 0.5em minus 0.4em\relax ACM, 2008, pp.
  453--456.

\bibitem{kittur2013future}
A.~Kittur, J.~V. Nickerson, M.~Bernstein, E.~Gerber, A.~Shaw, J.~Zimmerman,
  M.~Lease, and J.~Horton, ``The future of crowd work,'' in \emph{Proceedings
  of the 2013 conference on Computer supported cooperative work}.\hskip 1em
  plus 0.5em minus 0.4em\relax ACM, 2013, pp. 1301--1318.

\bibitem{rzeszotarski2011instrumenting}
J.~M. Rzeszotarski and A.~Kittur, ``Instrumenting the crowd: using implicit
  behavioral measures to predict task performance,'' in \emph{Proceedings of
  the 24th annual ACM symposium on User interface software and
  technology}.\hskip 1em plus 0.5em minus 0.4em\relax ACM, 2011, pp. 13--22.

\bibitem{summers2010mechanical}
J.~D. Summers and J.~J. Shah, ``Mechanical engineering design complexity
  metrics: size, coupling, and solvability,'' \emph{Journal of Mechanical
  Design}, vol. 132, no.~2, p. 021004, 2010.

\bibitem{takai2010game}
S.~Takai, ``A game-theoretic model of collaboration in engineering design,''
  \emph{Journal of Mechanical Design}, vol. 132, no.~5, p. 051005, 2010.

\bibitem{hong2004groups}
L.~Hong and S.~E. Page, ``Groups of diverse problem solvers can outperform
  groups of high-ability problem solvers,'' \emph{Proceedings of the National
  Academy of Sciences of the United States of America}, vol. 101, no.~46, pp.
  16\,385--16\,389, 2004.

\bibitem{surowiecki2005wisdom}
J.~Surowiecki, \emph{The wisdom of crowds}.\hskip 1em plus 0.5em minus
  0.4em\relax Anchor, 2005.

\bibitem{lorenz2011social}
J.~Lorenz, H.~Rauhut, F.~Schweitzer, and D.~Helbing, ``How social influence can
  undermine the wisdom of crowd effect,'' \emph{Proceedings of the National
  Academy of Sciences}, vol. 108, no.~22, pp. 9020--9025, 2011.

\bibitem{burnap2015crowdsourcing}
A.~Burnap, Y.~Ren, R.~Gerth, G.~Papazoglou, R.~Gonzalez, and P.~Y. Papalambros,
  ``When crowdsourcing fails: A study of expertise on crowdsourced design
  evaluation,'' \emph{Journal of Mechanical Design}, vol. 137, no.~3, p.
  031101, 2015.

\bibitem{burnap_identify}
\BIBentryALTinterwordspacing
A.~Burnap, R.~Gerth, R.~Gonzalez, and P.~Y. Papalambros, ``Identifying experts
  in the crowd for evaluation of engineering designs,'' \emph{Journal of
  Engineering Design}, vol.~28, no.~5, pp. 317--337, 2017. [Online]. Available:
  \url{http://dx.doi.org/10.1080/09544828.2017.1316013}
\BIBentrySTDinterwordspacing

\bibitem{burnap2015balancing}
A.~Burnap, J.~Hartley, Y.~Pan, R.~Gonzalez, and P.~Y. Papalambros, ``Balancing
  design freedom and brand recognition in the evolution of automotive brand
  styling,'' in \emph{ASME 2015 International Design Engineering Technical
  Conferences and Computers and Information in Engineering Conference}.\hskip
  1em plus 0.5em minus 0.4em\relax American Society of Mechanical Engineers,
  2015, pp. V007T06A047--V007T06A047.

\bibitem{orbay2015deciphering}
G.~Orbay, L.~Fu, and L.~B. Kara, ``Deciphering the influence of product shape
  on consumer judgments through geometric abstraction,'' \emph{Journal of
  Mechanical Design}, vol. 137, no.~8, p. 081103, 2015.

\bibitem{morgan2014ge}
H.~Morgan, H.~Levatti, J.~Sienz, A.~Gil, and D.~Bould, ``Ge jet engine bracket
  challenge: A case study in sustainable design,'' \emph{Sustainable Design and
  Manufacturing 2014 Part 1}, p.~95, 2014.

\bibitem{fu2013meaning}
K.~Fu, J.~Chan, J.~Cagan, K.~Kotovsky, C.~Schunn, and K.~Wood, ``The meaning of
  near and far?: the impact of structuring design databases and the effect of
  distance of analogy on design output,'' \emph{Journal of Mechanical Design},
  vol. 135, no.~2, p. 021007, 2013.

\bibitem{murphy2014function}
J.~Murphy, K.~Fu, K.~Otto, M.~Yang, D.~Jensen, and K.~Wood, ``Function based
  design-by-analogy: a functional vector approach to analogical search,''
  \emph{Journal of Mechanical Design}, vol. 136, no.~10, p. 101102, 2014.

\bibitem{moreno2014fundamental}
D.~P. Moreno, A.~A. Hernandez, M.~C. Yang, K.~N. Otto, K.~H{\"o}ltt{\"a}-Otto,
  J.~S. Linsey, K.~L. Wood, and A.~Linden, ``Fundamental studies in
  design-by-analogy: A focus on domain-knowledge experts and applications to
  transactional design problems,'' \emph{Design Studies}, vol.~35, no.~3, pp.
  232--272, 2014.

\bibitem{shah2000evaluation}
J.~J. Shah, S.~V. Kulkarni, and N.~Vargas-Hernandez, ``Evaluation of idea
  generation methods for conceptual design: effectiveness metrics and design of
  experiments,'' \emph{Journal of mechanical design}, vol. 122, no.~4, pp.
  377--384, 2000.

\bibitem{goucher2017using}
K.~Goucher-Lambert and J.~Cagan, ``Using crowdsourcing to provide analogies for
  designer ideation in a cognitive study,'' in \emph{DS 87-8 Proceedings of the
  21st International Conference on Engineering Design (ICED 17) Vol 8: Human
  Behaviour in Design, Vancouver, Canada}, 2017.

\bibitem{Mota_2011}
\BIBentryALTinterwordspacing
C.~Mota, ``The rise of personal fabrication,'' in \emph{Proceedings of the 8th
  ACM Conference on Creativity and Cognition}, ser. C\&C '11.\hskip 1em plus
  0.5em minus 0.4em\relax New York, NY, USA: ACM, 2011, pp. 279--288. [Online].
  Available: \url{http://doi.acm.org/10.1145/2069618.2069665}
\BIBentrySTDinterwordspacing

\bibitem{de2003design}
\BIBentryALTinterwordspacing
B.~De~Mozota, \emph{Design Management: Using Design to Build Brand Value and
  Corporate Innovation}.\hskip 1em plus 0.5em minus 0.4em\relax Skyhorse
  Publishing Company, Incorporated, 2003. [Online]. Available:
  \url{https://books.google.com/books?id=jpy\_JBhZ7nUC}
\BIBentrySTDinterwordspacing

\bibitem{ShirleyGraphicsBook}
P.~Shirley and S.~Marschner, \emph{Fundamentals of Computer Graphics},
  3rd~ed.\hskip 1em plus 0.5em minus 0.4em\relax Natick, MA, USA: A. K. Peters,
  Ltd., 2009.

\bibitem{Revelles00}
J.~Revelles, C.~Ureña, and M.~Lastra, ``An efficient parametric algorithm for
  octree traversal,'' in \emph{Journal of WSCG}, 2000, pp. 212--219.

\bibitem{mesh_processing_book}
M.~Botsch, L.~Kobbelt, M.~Pauly, P.~Alliez, and B.~uno Levy, \emph{Polygon Mesh
  Processing}.\hskip 1em plus 0.5em minus 0.4em\relax AK Peters, 2010.

\bibitem{arisoy2012}
E.~B. Arisoy, G.~Orbay, and L.~B. Kara, ``Free form surface skinning of 3d
  curve clouds for conceptual shape design,'' \emph{Journal of Computing and
  Information Science in Engineering}, vol.~12, p. 031005, 2012.

\bibitem{sketch2007}
L.~B. Kara and K.~Shimada, ``Sketch-based 3d-shape creation for industrial
  styling design,'' \emph{IEEE Computer Graphics and Applications}, vol.~27,
  pp. 60--71, 2007.

\bibitem{zoran2013hybrid}
A.~Zoran and L.~Buechley, ``Hybrid reassemblage: an exploration of craft,
  digital fabrication and artifact uniqueness,'' \emph{Leonardo}, vol.~46,
  no.~1, pp. 4--10, 2013.

\bibitem{Steif:2012}
P.~S. Steif, \emph{Mechanics of Materials}.\hskip 1em plus 0.5em minus
  0.4em\relax Pearson, 2012.

\bibitem{bendsoe1989optimal}
M.~P. Bends{\o}e, ``Optimal shape design as a material distribution problem,''
  \emph{Structural optimization}, vol.~1, no.~4, pp. 193--202, 1989.

\bibitem{casals2004dynamic}
J.~Casals-Terre and A.~Shkel, ``Dynamic analysis of a snap-action
  micromechanism,'' in \emph{Proceedings of IEEE Sensors, 2004.}, Oct 2004, pp.
  1245--1248 vol.3.

\bibitem{Sifakis:2012}
\BIBentryALTinterwordspacing
E.~Sifakis and J.~Barbic, ``Fem simulation of 3d deformable solids: A
  practitioner's guide to theory, discretization and model reduction,'' in
  \emph{ACM SIGGRAPH 2012 Courses}, ser. SIGGRAPH '12.\hskip 1em plus 0.5em
  minus 0.4em\relax New York, NY, USA: ACM, 2012, pp. 20:1--20:50. [Online].
  Available: \url{http://doi.acm.org/10.1145/2343483.2343501}
\BIBentrySTDinterwordspacing

\bibitem{Kolluri:2005}
\BIBentryALTinterwordspacing
R.~Kolluri, ``Provably good moving least squares,'' in \emph{Proceedings of the
  Sixteenth Annual ACM-SIAM Symposium on Discrete Algorithms}, ser. SODA
  '05.\hskip 1em plus 0.5em minus 0.4em\relax Philadelphia, PA, USA: Society
  for Industrial and Applied Mathematics, 2005, pp. 1008--1017. [Online].
  Available: \url{http://dl.acm.org/citation.cfm?id=1070432.1070578}
\BIBentrySTDinterwordspacing

\bibitem{Nocedal:2006}
J.~Nocedal and S.~J. Wright, \emph{Numerical Optimization, Second
  Edition}.\hskip 1em plus 0.5em minus 0.4em\relax Springer, 2006.

\bibitem{mosek2015}
\BIBentryALTinterwordspacing
M.~ApS, \emph{The MOSEK optimization software}, 2015. [Online]. Available:
  \url{http://www.mosek.com (2015)}
\BIBentrySTDinterwordspacing

\bibitem{Deuflhard:2008}
\BIBentryALTinterwordspacing
P.~Deuflhard, R.~Krause, and S.~Ertel, ``A contact-stabilized newmark method
  for dynamical contact problems,'' \emph{International Journal for Numerical
  Methods in Engineering}, vol.~73, no.~9, pp. 1274--1290, 2008. [Online].
  Available: \url{http://dx.doi.org/10.1002/nme.2119}
\BIBentrySTDinterwordspacing

\bibitem{Jacobson:2011:BBW}
\BIBentryALTinterwordspacing
A.~Jacobson, I.~Baran, J.~Popovi\'{c}, and O.~Sorkine, ``Bounded biharmonic
  weights for real-time deformation,'' in \emph{ACM SIGGRAPH 2011 Papers}, ser.
  SIGGRAPH '11.\hskip 1em plus 0.5em minus 0.4em\relax New York, NY, USA: ACM,
  2011, pp. 78:1--78:8. [Online]. Available:
  \url{http://doi.acm.org/10.1145/1964921.1964973}
\BIBentrySTDinterwordspacing

\bibitem{Frey:2007}
B.~J. Frey and D.~Dueck, ``Clustering by passing messages between data
  points,'' \emph{Science}, vol. 315, no. 5814, pp. 972--976, 2007.

\bibitem{Sorkine:2004}
\BIBentryALTinterwordspacing
O.~Sorkine, D.~Cohen-Or, Y.~Lipman, M.~Alexa, C.~R{\"o}ssl, and H.~P. Seidel,
  ``Laplacian surface editing,'' in \emph{Proceedings of the 2004
  Eurographics/ACM SIGGRAPH Symposium on Geometry Processing}, ser. SGP
  '04.\hskip 1em plus 0.5em minus 0.4em\relax New York, NY, USA: ACM, 2004, pp.
  175--184. [Online]. Available:
  \url{http://doi.acm.org/10.1145/1057432.1057456}
\BIBentrySTDinterwordspacing

\bibitem{astin1999finger}
A.~D. Astin, ``Finger force capability: measurement and prediction using
  anthropometric and myoelectric measures,'' Ph.D. dissertation, Virginia
  Polytechnic Institute and State University, 1999.

\bibitem{Kirkpatrick:1983}
\BIBentryALTinterwordspacing
S.~Kirkpatrick, C.~D. Gelatt, and M.~P. Vecchi, ``Optimization by simulated
  annealing,'' \emph{Science}, vol. 220, no. 4598, pp. 671--680, 1983.
  [Online]. Available: \url{http://science.sciencemag.org/content/220/4598/671}
\BIBentrySTDinterwordspacing

\bibitem{triangleSoftware}
J.~R. Shewchuk, ``Triangle: Engineering a 2d quality mesh generator and
  delaunay triangulator,'' in \emph{Applied Computational Geometry Towards
  Geometric Engineering}, M.~C. Lin and D.~Manocha, Eds.\hskip 1em plus 0.5em
  minus 0.4em\relax Berlin, Heidelberg: Springer Berlin Heidelberg, 1996, pp.
  203--222.

\bibitem{TetGen}
\BIBentryALTinterwordspacing
H.~Si, ``Tetgen, a delaunay-based quality tetrahedral mesh generator,''
  \emph{ACM Trans. Math. Softw.}, vol.~41, no.~2, pp. 11:1--11:36, Feb. 2015.
  [Online]. Available: \url{http://doi.acm.org/10.1145/2629697}
\BIBentrySTDinterwordspacing

\bibitem{konakovic2016beyond}
M.~Konakovi{\'c}, K.~Crane, B.~Deng, S.~Bouaziz, D.~Piker, and M.~Pauly,
  ``Beyond developable: computational design and fabrication with auxetic
  materials,'' \emph{ACM Transactions on Graphics (TOG)}, vol.~35, no.~4,
  p.~89, 2016.

\bibitem{guseinov2017curveups}
R.~Guseinov, E.~Miguel, and B.~Bickel, ``Curveups: shaping objects from flat
  plates with tension-actuated curvature,'' \emph{ACM Transactions on Graphics
  (TOG)}, vol.~36, no.~4, p.~64, 2017.

\bibitem{perez2017computational}
J.~P{\'e}rez, M.~A. Otaduy, and B.~Thomaszewski, ``Computational design and
  automated fabrication of kirchhoff-plateau surfaces,'' \emph{ACM Transactions
  on Graphics (TOG)}, vol.~36, no.~4, p.~62, 2017.

\bibitem{harnett2013digital}
C.~Harnett and C.~Kimmer, ``Digital origami from geometrically frustrated
  tiles,'' in \emph{International Design Engineering Technical Conferences and
  Computers and Information in Engineering Conference, Portland, Oregon}, vol.
  2103, 2013.

\bibitem{Springborn:2008:CETM}
\BIBentryALTinterwordspacing
B.~Springborn, P.~Schr\"{o}der, and U.~Pinkall, ``Conformal equivalence of
  triangle meshes,'' \emph{ACM Trans. Graph.}, vol.~27, no.~3, pp. 77:1--77:11,
  Aug. 2008. [Online]. Available:
  \url{http://doi.acm.org/10.1145/1360612.1360676}
\BIBentrySTDinterwordspacing

\bibitem{Sawhney:2017:BFF}
\BIBentryALTinterwordspacing
R.~Sawhney and K.~Crane, ``Boundary first flattening,'' \emph{ACM Trans.
  Graph.}, vol.~37, no.~1, pp. 5:1--5:14, Dec. 2017. [Online]. Available:
  \url{http://doi.acm.org/10.1145/3132705}
\BIBentrySTDinterwordspacing

\bibitem{shoup1971use}
T.~E. Shoup and C.~W. McLarnan, ``On the use of the undulating elastica for the
  analysis of flexible link mechanisms,'' \emph{Journal of Engineering for
  Industry}, vol.~93, no.~1, pp. 263--267, 1971.

\bibitem{bergou2008discrete}
M.~Bergou, M.~Wardetzky, S.~Robinson, B.~Audoly, and E.~Grinspun, ``Discrete
  elastic rods,'' in \emph{ACM transactions on graphics (TOG)}, vol.~27,
  no.~3.\hskip 1em plus 0.5em minus 0.4em\relax ACM, 2008, p.~63.

\bibitem{ode2008}
\BIBentryALTinterwordspacing
R.~Smith, \emph{Open Dynamics Engine}, 2008. [Online]. Available:
  \url{http://ode.org}
\BIBentrySTDinterwordspacing

\bibitem{galton1907ballot}
F.~Galton, ``The ballot-box,'' \emph{Nature}, vol.~75, no. 1952, p. 509, 1907.

\bibitem{hooker1907mean}
R.~H. Hooker, ``Mean or median,'' \emph{Nature}, vol.~75, pp. 487--488, 1907.

\bibitem{wah2006crowdsourcing}
C.~Wah, ``Crowdsourcing and its applications in computer vision,''
  \emph{University of California, San Diego}, 2006.

\bibitem{yang2010consensus}
M.~C. Yang, ``Consensus and single leader decision-making in teams using
  structured design methods,'' \emph{Design Studies}, vol.~31, no.~4, pp.
  345--362, 2010.

\bibitem{gurnani2008collaborative}
A.~Gurnani and K.~Lewis, ``Collaborative, decentralized engineering design at
  the edge of rationality,'' \emph{Journal of Mechanical Design}, vol. 130,
  no.~12, p. 121101, 2008.

\bibitem{cabrerizo2014building}
F.~J. Cabrerizo, R.~Ure{\~n}a, W.~Pedrycz, and E.~Herrera-Viedma, ``Building
  consensus in group decision making with an allocation of information
  granularity,'' \emph{Fuzzy Sets and Systems}, vol. 255, pp. 115--127, 2014.

\bibitem{nobel2016improving}
M.~Nobel-J{\o}rgensen, D.~Malmgren-Hansen, J.~A. B{\ae}rentzen, O.~Sigmund, and
  N.~Aage, ``Improving topology optimization intuition through games,''
  \emph{Structural and Multidisciplinary Optimization}, vol.~54, no.~4, pp.
  775--781, 2016.

\bibitem{whitehill2009whose}
J.~Whitehill, T.-f. Wu, J.~Bergsma, J.~R. Movellan, and P.~L. Ruvolo, ``Whose
  vote should count more: Optimal integration of labels from labelers of
  unknown expertise,'' in \emph{Advances in neural information processing
  systems}, 2009, pp. 2035--2043.

\bibitem{bachrach2012grade}
Y.~Bachrach, T.~Graepel, T.~Minka, and J.~Guiver, ``How to grade a test without
  knowing the answers---a bayesian graphical model for adaptive crowdsourcing
  and aptitude testing,'' \emph{arXiv preprint arXiv:1206.6386}, 2012.

\bibitem{welinder2010multidimensional}
P.~Welinder, S.~Branson, S.~J. Belongie, and P.~Perona, ``The multidimensional
  wisdom of crowds.'' in \emph{NIPS}, vol.~23, 2010, pp. 2424--2432.

\bibitem{lakshminarayanan2013inferring}
B.~Lakshminarayanan and Y.~W. Teh, ``Inferring ground truth from
  multi-annotator ordinal data: a probabilistic approach,'' \emph{arXiv
  preprint arXiv:1305.0015}, 2013.

\bibitem{wauthier2011bayesian}
F.~L. Wauthier and M.~I. Jordan, ``Bayesian bias mitigation for
  crowdsourcing,'' in \emph{Advances in neural information processing systems},
  2011, pp. 1800--1808.

\bibitem{vul2008measuring}
E.~Vul and H.~Pashler, ``Measuring the crowd within probabilistic
  representations within individuals,'' \emph{Psychological Science}, vol.~19,
  no.~7, pp. 645--647, 2008.

\bibitem{Linsey2008}
\BIBentryALTinterwordspacing
J.~s. Linsey, K.~l. Wood, and A.~b. Markman, ``Modality and representation in
  analogy,'' \emph{Artif. Intell. Eng. Des. Anal. Manuf.}, vol.~22, no.~2, pp.
  85--100, Jan. 2008. [Online]. Available:
  \url{http://dx.doi.org/10.1017/S0890060408000061}
\BIBentrySTDinterwordspacing

\bibitem{Miller2014}
\BIBentryALTinterwordspacing
S.~R. Miller, B.~P. Bailey, and A.~Kirlik, ``Exploring the utility of bayesian
  truth serum for assessing design knowledge,'' \emph{Hum.-Comput. Interact.},
  vol.~29, no. 5-6, pp. 487--515, Aug. 2014. [Online]. Available:
  \url{http://dx.doi.org/10.1080/07370024.2013.870393}
\BIBentrySTDinterwordspacing

\bibitem{wilson2010}
\BIBentryALTinterwordspacing
J.~O. Wilson, D.~Rosen, B.~A. Nelson, and J.~Yen, ``The effects of biological
  examples in idea generation,'' \emph{Design Studies}, vol.~31, no.~2, pp. 169
  -- 186, 2010. [Online]. Available:
  \url{http://www.sciencedirect.com/science/article/pii/S0142694X0900074X}
\BIBentrySTDinterwordspacing

\bibitem{viswanathan2013design}
V.~K. Viswanathan and J.~S. Linsey, ``Design fixation and its mitigation: a
  study on the role of expertise,'' \emph{Journal of Mechanical Design}, vol.
  135, no.~5, p. 051008, 2013.

\end{thebibliography}

\end{document}